\documentclass[useAMS,usenatbib]{mn2e}
\usepackage{graphics,latexsym,epsfig,graphicx,amssymb,natbib,lscape,aas_macros}     
\renewcommand{\d}[0]{{\rm d}}
\newcommand{\ave}[1]{\langle #1 \rangle}
\newcommand{\Ave}[1]{\big\langle #1\big\rangle}
\newcommand{\Ref}[1]{(\ref{#1})}

\renewcommand{\vec}[1]{{\bmath{#1}}}
\newcommand{\mat}[1]{\mathbfss{#1}}
\newcommand{\be}{\begin{equation}}
\newcommand{\ee}{\end{equation}}

\newcommand{\myarcsec}{\hbox{$.\!\!^{\prime\prime}$}}

\newcommand{\combo}{\mbox{COMBO-17}}

\newcommand{\msol}[0]{{\rm M}_\odot}
\newcommand{\pprime}[0]{{\prime\prime}\!}
\newcommand{\caln}[0]{{\cal N}}

\newcommand{\changed}[1]{#1} 

\title[Clustering and joint HOD of red and blue galaxies in \combo]
      {Relative clustering and the joint halo occupation distribution 
	of red-sequence and blue-cloud galaxies in \combo}

\author[P. Simon et al.]{P. Simon$^1$\thanks{E-mail: psimon@roe.ac.uk},
  M. Hetterscheidt$^2$, C. Wolf$^3$, K. Meisenheimer$^4$,
  H. Hildebrandt$^{5,2}$,\newauthor P. Schneider$^2$, M. Schirmer$^2$ and
  T. Erben$^2$\\ $^1$The Scottish Universities Physics Alliance
  (SUPA), Institute for Astronomy, School of Physics, University of
  Edinburgh,\\ Royal Observatory, Blackford Hill, Edinburgh EH9 3HJ,
  UK\\ $^2$Argelander-Institut f\"ur Astronomie, Auf dem H\"ugel 71,
  53121 Bonn, Germany\\ $^3$ Department of Physics, Denys Wilkinson
  Bldg., University of Oxford, Keble Road, Oxford OX1 3RH, UK\\ $^4$
  Max-Planck-Institut f\"ur Astronomie, K\"onigsstuhl 17, 69117
  Heidelberg, Germany\\ $^5$ Leiden Observatory, Leiden University,
  Niels Bohrweg 2, 2333 CA Leiden , The Netherlands}

\def\LaTeX{L\kern-.36em\raise.3ex\hbox{a}\kern-.15em
    T\kern-.1667em\lower.7ex\hbox{E}\kern-.125emX}

\begin{document}

\date{Released 2007 Xxxxx XX}

\pagerange{\pageref{firstpage}--\pageref{lastpage}} \pubyear{2007}

\maketitle

\label{firstpage}

\begin{abstract}
 This paper studies the relative spatial distribution of red-sequence
 and blue-cloud galaxies, and their relation to the dark matter
 distribution in the \mbox{COMBO-17} survey as function of scale down
 to \mbox{$z\sim1$}.  We measure the $2^{\rm nd}$-order auto- and
 cross-correlation functions of galaxy clustering and express the
 relative biasing by using aperture statistics. Also estimated is the
 relation between the galaxies and the dark matter distribution
 exploiting galaxy-galaxy lensing (GGL).  All observables are further
 interpreted in terms of a halo model. To fully explain the galaxy
 clustering cross-correlation function with a halo model, we introduce
 a new parameter, $R$, that describes the statistical correlation
 between numbers of red and blue galaxies within the same halo.

 We find that red and blue galaxies are clearly differently clustered,
 a significant evolution of the relative clustering with redshift is
 not found. There is evidence for a scale-dependence of relative
 biasing: The linear relative bias factor varies slightly between
 \mbox{$b\sim1.7\pm0.5$} and \mbox{$b\sim2.2\pm0.1$} on spatial scales
 between roughly $100\,h^{-1}\rm kpc$ and $7\,h^{-1}\rm Mpc$,
 respectively. The linear correlation coefficient of galaxy number
 densities drops from a value near unity on large scales to
 \mbox{$r\sim0.6\pm0.15$}. Both biasing trends, the GGL and
 \changed{with some tension} the galaxy numbers can be explained
 consistently within a halo model.  \changed{Red galaxies typically
 start to populate haloes with masses starting from
 \mbox{$\gtrsim10^{12.1\pm0.2}\,h^{-1}\msol$}, blue galaxies from
 \mbox{$\gtrsim10^{11.2\pm0.1}\,h^{-1}\msol$}.}  For the
 cross-correlation function one requires a HOD variance that becomes
 Poisson even for relatively small occupancy numbers. \changed{This
 rules out for our samples with high confidence a ``Poisson
 satellite'' scenario, as found in semi-analytical models.}
 \changed{We compare different model flavours, with and without
 galaxies at the halo centres, using Bayesian evidence. The result is
 inconclusive. However, red galaxies have to be concentrated towards
 the halo centre, either by a red central galaxy or by a concentration
 parameter above that of dark matter.}  \changed{The value of $R$
 depends on the presence or absence of central galaxies: If no central
 galaxies or only red central galaxies are allowed, $R$ is consistent
 with zero, whereas a positive correlation $R=+0.5\pm0.2$ is needed if
 both blue and red galaxies can have central galaxies.}
\end{abstract}



\section{Introduction}

Today a confusing wealth of different galaxy populations is known,
which yet is thought to have arisen from a fairly simple early
Universe.  Morphologically, local galaxies fall into two broad
classes: early-type galaxies, with almost spheroidal appearance and
none or only a very small disk component, and late-type galaxies, with
a small central bulge and a dominating stellar disk exhibiting
different degrees of spiral structure and star formation. Within the
context of the cold dark matter paradigm for cosmological structure
formation \citep{2008arXiv0803.4003C,2005Natur.435..629S} galaxies
merge, grow and interact with the ambient intergalactic medium by
participating in a hierarchical merging process
\citep{2000MNRAS.319..168C}. The ongoing research is trying to test
whether the today's known variety of galaxies can indeed be explained
within this paradigm.

To trace the evolution of galaxy populations with time, a
morphological identification of a large sample of galaxies down to
higher redshifts has proven to be difficult. The most practical
solution to this problem is to exploit the bimodal distribution of
galaxies in a colour-magnitude diagram (CMD). In such a diagram early-
and late-type galaxies can roughly be separated down to redshifts of
$z\sim1$, possibly even beyond that \citep[see e.g.][and references
therein]{2008arXiv0802.3004L,2007ApJ...665..265F,2004ApJ...608..752B}.
The red mode in the CMD is the well-known colour-magnitude relation
(CMR), or red-sequence, of early-type galaxies. The blue mode is often
referred to as the blue cloud galaxies. To distinguish between a red
and a blue galaxy population we proceed according to
\citet{2004ApJ...608..752B}, using a (rest-frame) $U-V$ vs. $M_V$ CMD
and cut the galaxy sample along the CMR to obtain a red-sequence and
blue-cloud sample.  In adopting this division line about $80\%$ of the
selected red galaxies have morphologies earlier than or equal Hubble
type Sa, while the blue-cloud galaxies are mainly late-type, star
forming galaxies. \changed{A better morphological separation of
galaxies, not pursued for this paper though, may be achieved by
applying inclination corrections as discussed in
\citet{2008arXiv0801.3286M} that have been tested for low-$z$ galaxies
from SDSS and 2dF.}

The so far strongest observational clues about the emergence of the
red-sequence from the blue cloud come from careful number counts in
CMDs and estimates of the galaxy luminosity functions for different
redshifts. \citet{2007ApJ...665..265F} have found strong evidence that
the red-sequence has been built up by a mixture of dry mergers between
red-sequence galaxies, wet mergers between blue-cloud galaxies and
quenching of star formation with subsequent aging of the stellar
populations of blue-cloud galaxies.

Another important source of information hinting to the nature of
galaxies is their spatial distribution.  For example, early-type
galaxies are preferentially found in the cores of rich galaxy clusters
where their fraction is about $90$ percent, whereas outside of galaxy
clusters about $70$ percent of the field galaxies are late-type
galaxies \citep{1980ApJ...236..351D}. As another example, it has also
been found by modelling stellar populations of local early-types that
the star-formation history of early-type galaxies depends on the
galaxy density of the environment \citep{2005ApJ...621..673T}.

One traditional way to study the spatial distribution of galaxies is
to look at correlations in the galaxy distribution, in particular the
two-point correlation function
\citep{peebles80,1969PASJ...21..221T}. Analyses revealed that galaxy
clustering depends on the properties of the galaxy population like
morphology, colour, luminosity or spectral type \citep[e.g.][and
references therein]{2008ApJ...672..153C,2005ApJ...630....1Z,
2003MNRAS.344..847M}. Therefore, different galaxy populations are
differently clustered -- \emph{biased} -- with respect to the total
matter component and with respect to each other. The detailed
dependence of spatial clustering on galaxy characteristics, scale and
redshift is a opportunity to learn more about the formation and
evolution history of galaxies, see for example
\citet{2007ApJ...655L..69W} or \citet{2005A&A...430..827S}.

Along with this motivation, one aim of this paper is to measure the
relative clustering of red (early-type) and blue (late-type) galaxies
for different epochs in terms of the linear stochastic biasing
parameters. These biasing parameters require $2^{\rm nd}$-order
clustering statistics \citep{1999ApJ...520...24D}. That
parametrisation is a completely model-independent, albeit in a
statistical sense for non-Gaussian random fields incomplete, measure
for comparing two random distributions. It quantifies as function of
angular scale the relative clustering strength of two galaxy types and
the correlation of their number densities.

The machinery that is applied here to study galaxy biasing is the
aperture statistics as formalised in \citet{1998MNRAS.296..873S} and
\citet{1998A&A...334....1V}. It is convenient for analysing weak
lensing data and, in particular, to measure the linear stochastic
galaxy bias as a function of scale
\citep{2007A&A...461..861S,hvg02}. In order to have a compatible
statistical measure that quantifies the relative bias between red and
blue galaxies the formalism is slightly extended.

Another aim of this paper is to give a physical interpretation of the
relative clustering. For that purpose, we use the measurements for
setting constraints on parameters of galaxies within the framework of
a halo model \changed{\citep{2007MNRAS.376..841V, 2005ApJ...633..791Z,
2002ApJ...575..587B, 2001ApJ...546...20S, 2000MNRAS.318..203S,
2000MNRAS.318.1144P}}. In this context, we introduce and discuss a new
parameter -- the correlation factor of the joint HOD -- that regulates
the likelihood to find a certain number of red and blue galaxies
within the same halo. This allows us to investigate whether two galaxy
populations avoid or \changed{attract} each other inside/inside the
same dark matter halo. \changed{In \citet{2003MNRAS.339..410S,
2002MNRAS.332..697S} a similar modelling is carried out to explain the
relative clustering of red and blue galaxy samples, however assuming
for simplicity uncorrelated galaxy numbers inside same haloes. In
\citet{2005MNRAS.361..415C} also the clustering of red and blue
galaxies was studied by looking at the projected galaxy density profiles
of groups.}

For the scope of this analysis, the \mbox{COMBO-17} Survey
\citep{2004A&A...421..913W,2001A&A...377..442W} offers an unique
opportunity. It provides one of the so far largest deep galaxy samples
in the redshift regime $0.2\le z\le 1.1$ covering an area of
$\sim0.78~\rm deg^2$, observed in five broad-band and twelve
narrow-band filters. Based on the photometry, photometric redshifts of
galaxies brighter than $R\le24~\rm mag$ have been derived within a few
percent accuracy as well as absolute rest-frame luminosities and
colours. We are analysing the data from three \mbox{COMBO-17} patches
which are known as S11, A901 and CDFS (also known as AXAF).

The survey has also been designed to fit the requirements of
gravitational lensing applications \citep{2006A&A...455..441K, btb03,
2002ApJ...568..141G}.  The coherent shear distortions of images of
background galaxies can therefore be used to infer the relation
between galaxy and matter distribution as well \citep{bas01}. We use
this additional piece of information to further constrain parameters
of the halo-model by cross-correlating the \mbox{COMBO-17} galaxies
with the corresponding shear catalogues taken from the Garching-Bonn
Deep Survey (GaBoDS) \citep{hss07,esd05}.

\changed{The structure of this paper is as
follows. Sect. \ref{sect:formalism} outlines the quantities that are
used to measure the angular galaxy clustering, the aperture
$\caln$-statistics. Sect. \ref{sect:data} introduces the \combo-survey
and \mbox{GaBoDS} which are the sources of the galaxy samples, red and
blue galaxies, and shear catalogues for this study,
respectively. Sect. 4 is the place where we describe and present the
details of our clustering analysis and compare the results to the
literature. The cosmic shear information is harnessed for the GGL,
Sect. \ref{sect:ggl}, quantifying the typical matter distribution
about the galaxies in our $\combo$ samples.  Sect. 6 outlines the halo
model which is then used to interpret the (relative) clustering and
the GGL signal of the blue and red galaxy samples. In particular, we
introduce and discuss the correlation factor of the joint HOD of two
galaxy populations. We finish with a summary in the last section.}

Unless stated otherwise we use as fiducial cosmology a $\Lambda$CDM
model (adiabatic fluctuations) with \mbox{$\Omega_{\rm m}=0.24$},
\mbox{$\Omega_{\rm b}=0.0416$}, \mbox{$\Omega_{\Lambda}=1-\Omega_{\rm
m}$} and \mbox{$H_{0}=h\,100\,{\rm km}\,{\rm s}^{-1}\,{\rm Mpc}^{-1}$}
with \mbox{$h=0.732$}. The normalisation of the matter fluctuations
within a sphere of radius $8\,h^{-1}\rm Mpc$ at redshift zero is
assumed to be $\sigma_8=0.76$. For the spectral index of the
primordial matter power spectrum we use $n_{\rm s}=0.96$. These values
are consistent with the third-year WMAP results
\citep{2007ApJS..170..377S}.

\section{Clustering quantifiers}
\label{sect:formalism}

\subsection{Aperture statistics}

The statistics used in this paper to quantify galaxy clustering is the
so-called aperture number count statistics. It originally stems from
the gravitational lensing literature.  As shown in
\citet{2007A&A...461..861S} the aperture number count statistics is
useful for studying galaxy clustering even outside the context of
gravitational lensing. Its advantage is that no correction for the
integral-constraint of the angular correlation function
\citep{peebles80} is needed.

The aperture number count, $\caln(\vec{\theta},\theta_{\rm ap})$,
measures the fluctuations, excluding shot-noise from discrete
galaxies, of the galaxy number density by smoothing the density with a
compensated filter $u$, i.e. \mbox{$\int\d x\,x\,u(x)=0$}:
\begin{equation}\label{aperturecountdef}
  \caln(\vec{\theta},\theta_{\rm ap})= 
  \frac{1}{\bar{\eta}\theta^2_{\rm ap}}
  \int {\rm d}^2\vec{\theta}^\prime~
  u\left(\frac{|\vec{\theta}-\vec{\theta}^\prime|} {\theta_{\rm
      ap}}\right) \eta(\vec{\theta}^\prime) \; ,
 \end{equation}
where $\eta(\vec{\theta})$ and $\bar{\eta}$ denote the (projected)
number density of galaxies in some direction $\vec{\theta}$ and the
mean number density of galaxies, respectively. The variable
$\theta_{\rm ap}$ defines the smoothing radius of the aperture.

One focus of our analysis is the $2^{\rm nd}$-order statistics of
galaxy clustering. All information on the $2^{\rm nd}$-order
statistics is comprised by \changed{second moments of the aperture
number counts as function of $\theta_{\rm ap}$}:
\begin{equation}\label{nstat}
  \Ave{\caln_i\left(\theta_{\rm ap}\right)\caln_j\left(\theta_{\rm ap}\right)}=
  2\pi\int_0^\infty {\rm d}\ell\,\ell\,P_{ij}\left(\ell\right)
  \left[{I}\left(\ell\theta_{\rm ap}\right)\right]^2 ,
 \end{equation}
where $I(x)$ is a filter kernel
\begin{equation}
  {I}\left(x\right)\equiv
  \int_0^\infty {\rm d}s\,s\,u\left(s\right)
  J_0\left(s\,x\right)
\end{equation}
to the angular power spectrum $P_{ij}(\ell)$ defined by
\begin{equation}
  \frac{\Ave{\tilde{\eta}_i(\vec{\ell}_1)\tilde{\eta}_j(\vec{\ell}_2)}}
   {\bar{\eta}_i\bar{\eta}_j}=
   (2\pi)^2P_{ij}(|\vec{\ell}|)\delta_{\rm
     D}(\vec{\ell}_1+\vec{\ell}_2)\;.
\end{equation}
In this equation, \mbox{$\bar{\eta}_i$} is the mean number density of
galaxies on the sky, for possibly different galaxy samples. We use
$J_0(x)$ for the $0^{\rm th}$-order Bessel function of the first kind
and $\delta_{\rm D}(\vec{x})$ for the Delta-function. The tilde on top
of the galaxy number density $\tilde{\eta}$ denotes the angular
Fourier transform of $\eta$ assuming a flat sky with Cartesian
coordinates which we do throughout this paper:
\begin{equation}\label{fouriertransf}
  \tilde{\eta}(\vec{\ell})=
  \int\d^2\vec{\theta}\,\eta(\vec{\theta})\,{\rm e}^{+{\rm
      i}\vec{\theta}\cdot\vec{\ell}}~;~ \eta(\vec{\theta})=
  \int\frac{\d^2\vec{\ell}}{(2\pi)^2}\,\tilde{\eta}(\vec{\ell})\,{\rm
    e}^{-{\rm i}\vec{\theta}\cdot\vec{\ell}} \;.
\end{equation}
For \mbox{$i=j$}, $P_{ij}(\ell)$ is an auto-correlation power
spectrum, a cross-power spectrum otherwise.

To weigh density fluctuations inside apertures we use a compensated
polynomial filter \citep{1998ApJ...498...43S}
\begin{equation}
  \label{filter}
  u\left(x\right)=
  \frac{9}{\pi}\left(1-x^2\right)\left(\frac{1}{3}-x^2\right)\,{\rm H}(1-x)\;,
\end{equation} 
which by definition vanishes for $x\ge 1$; ${\rm H}(x)$ denotes the
Heaviside step function.  The filter has the effect that only galaxy
number density fluctuations from a small range of angular scales
contribute to the signal; it acts as a narrow-band filter for the
angular modes, $\ell$, with highest sensitivity to $\ell_{\rm
c}\approx 1.5\times10^4\,\frac{1^\prime}{\theta_{\rm ap}}$. Therefore,
the $2^{\rm nd}$-order $\caln$-statistics are essentially a probe for
the (band) power spectrum $P_{ij}(\ell_{\rm c})$. 

In practice, the $\caln$-statistics are easily derived from the
two-point correlation functions, $\omega(\theta)$, of galaxy
clustering \citep[e.g.][]{2005PhDT........41S} by a weighted
integral. \changed{As an aside, this is very similar to the approach
recently advocated in \citet{2007MNRAS.376.1702P} which points out that
weighting $\omega(\theta)$ with a compensated filter can also be
useful for deprojecting $\omega(\theta)$ to obtain the 3D-correlation
function on small cosmological scales.}

\subsection{Linear stochastic biasing parameters}
  
 The linear stochastic biasing parameters
 \citep{1999ApJ...520...24D,1998ApJ...500L..79T}, expressed here in
 terms of the $\caln$-statistics, quantify the relative clustering of
 two random fields, which are in our case the number density of blue
 and red galaxies:
  \begin{eqnarray}\label{eq:linbias}
    b(\theta_{\rm ap})&=&
    \sqrt{\frac{\Ave{\caln^2_{\rm red}(\theta_{\rm ap})}}{\Ave{\caln^2_{\rm
	    blue}(\theta_{\rm ap})}}}\;,\\
    \label{eq:lincorr}
    r(\theta_{\rm ap})&=&
    \frac{\Ave{\caln_{\rm red}(\theta_{\rm ap})\caln_{\rm blue}(\theta_{\rm
	  ap})}}{\sqrt{\Ave{\caln^2_{\rm red}(\theta_{\rm
    ap})}\Ave{\caln^2_{\rm blue}(\theta_{\rm ap})}}}\;.   
  \end{eqnarray}
 \changed{Note that the aperture number count vanishes on average,
 $\ave{{\cal N}}=0$, due to the compensated filter used.}  Galaxy
 samples unbiased with respect to each other have \mbox{$r(\theta_{\rm
 ap})=b(\theta_{\rm ap})=1$}.  The parameters are complete for
 Gaussian random fields, which is only approximately true for large
 scales but clearly wrong for non-linear scales, effective scale
 smaller than \mbox{$\sim8{\rm Mpc}/h$}.  Owing to the incomplete
 picture those biasing parameters convey for the non-Gaussian regime,
 one is unable to distinguish stochasticity from non-linearity in the
 relation between the two random fields. Therefore a bias factor
 $r\ne1$ can mean a stochastic scatter between galaxy number densities
 or a non-linear but deterministic mapping
 \changed{\citep{1993ApJ...413..447F}} between number densities -- or
 both. Higher-order statistics or non-Gaussian models for the
 clustering are required to make this distinction
 \citep{2005MNRAS.356..247W,2000ApJ...544...63B,1999ApJ...520...24D}.
 
 Note that the parameter $r(\theta_{\rm ap})$ \emph{can be larger than
 unity} because shot-noise contributions to the variances in the
 aperture galaxy number count are subtracted, or put another way,
 spatial shot-noise in the fluctuation power of the galaxy number
 density fields is \changed{automatically subtracted} as in
 \citet{2001MNRAS.321..439G} or
 \citet{2000MNRAS.318..203S}. \changed{The underlying assumption is
 that galaxies trace a general galaxy number density field by a
 Poisson sampling process (Poisson shot-noise), which is widely
 assumed in large-scale structure studies and, in fact, in the
 definition of the clustering correlation function $\omega(\theta)$.}

 We employ the linear stochastic biasing parameters here in order to
 quantify, without too many assumptions, the relative biasing of our
 blue and red sample as function of scale, $\theta_{\rm ap}$. A more
 sophisticated and physical, albeit very model-dependent,
 interpretation of the relative biasing is given within the framework
 of the halo-model\changed{, see Sect. 6.}

\section{Data set}
 \label{sect:data}

This study is based on three fields: the S11, A901 and CDFS. The
observations of the fields were obtained with the Wide Field Imager
(WFI) of the MPG/ESO 2.2m telescope on La Silla, Chile.  The camera
consists of eight $2\,{\rm k}\times 4\,{\rm k}$ CCDs with a pixel size
of $15\,\mu {\rm m}$, corresponding to a pixel scale of $0\myarcsec
238$ in the sky.  The field-of-view in the sky is $34^\prime \times
33^\prime$.

The data from two surveys, carried out with the same instrument, are
used. We select blue and red galaxies, possessing photometric
redshifts, from the \combo-survey. These data sets are further
subdivided into four redshifts bins covering the range between $z=0.2$
and $z=1.0$. Shear catalogues from another survey, GaBoDS, covering
the same patches on the sky as \combo, are utilised to quantify the
relation between the total (dark) matter density and the galaxy
positions by using the gravitational lensing technique.  The following
sections describe the details of the two surveys and the extracted
galaxy and shear catalogues.

\subsection{Red and blue galaxy samples: \combo}

\begin{figure*}
  \begin{center}
    \epsfig{file=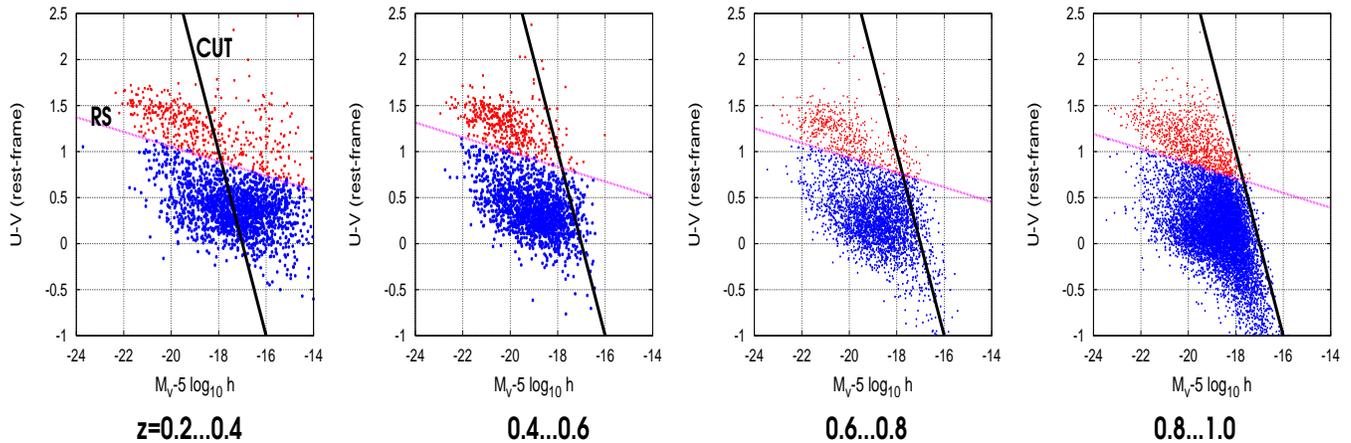,width=18cm,height=6cm,angle=0}
  \end{center}
  \caption{Rest-frame U-V colour vs. absolute magnitude in V-band for
    the S11 galaxies inside four different redshift ranges (numbers at
    bottom). Analogous colour-magnitude diagrams for the fields CDFS
    and A901 look similar. The galaxy sample is split into a red and
    blue-cloud population by the red-sequence division line ``RS''.
    The steep line ``CUT'' is an additional cut applied to obtain
    comparable absolute magnitude-limits in all
    bins. \label{combocolmag}}
\end{figure*}

The observations and data reduction of the \mbox{COMBO-17} survey are
described in detail in \citet{2001A&A...377..442W} and
\citet{2003A&A...401...73W}.  Overall the total survey consists of
four different, non-contiguous fields observed in 17 optical
filters\footnote{The filters include UBVRI and 12 medium-band
filters.}.

\begin{figure}
  \begin{center}
    \epsfig{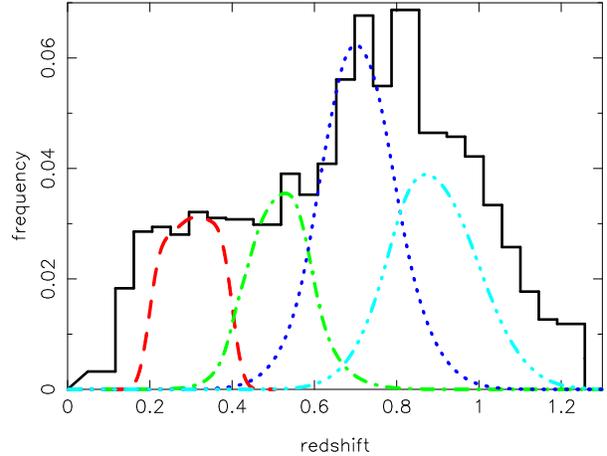}
  \end{center}
  \caption{\label{fig:pzlenses} \changed{Histogram (arbitrary
  normalisation) of photometric redshifts of galaxies with $m_{\rm
  R}\le24\,\rm mag$ in \mbox{COMBO-17} (solid line). The mean
  photometric redshift is $\bar{z}=0.68$. The sample is split into
  four distinct photo-z bins. The photo-z distribution inside the bins
  is further convolved with the photo-z error of the individual
  galaxies to estimate the true redshift distribution inside the
  photo-z bins (dashed: $\bar{z}=0.3$; dashed-dotted: $\bar{z}=0.5$;
  dotted: $\bar{z}=0.7$; dashed-dotted-dotted: $\bar{z}=0.9$).}}
\end{figure}

\changed{The photometric information was used to derive photometric
redshifts of galaxies with $m_R\lesssim 24~\rm mag$, based on a set of
galaxy spectrum templates \citep[see references
in][]{2004A&A...421..913W}.  The quality of the estimate depends
primarily on the apparent magnitude of the object. As estimator for
the redshift uncertainty we use Eq. (5) of
\citet{2004A&A...421..913W}:
\begin{equation}\label{eq:photozerror}
  \sigma_z=0.007(1+z)\sqrt{1+10^{0.8(m_R-21.6\,{\rm mag})}}\;,
\end{equation}
where $\sigma_z$ is the 1-$\sigma$ standard deviation of the object
redshift. Fig. \ref{fig:pzlenses} shows the frequency distribution of
the photometric redshifts (solid line) of the full sample.}

Based on photometry, rest-frame colours with accuracy $\delta m\sim
0.1~\rm mag$ and absolute luminosities with accuracies $\delta m\sim
0.1~\rm mag$ ($0.2~\rm mag$) for redshifts $z\gtrsim0.5$ ($\sim0.3$)
were calculated.

Our object catalogue consists of galaxies with reliable photometric
redshifts. Galaxies are only contained in the object catalogue if both
spectral classification and estimation of the photometric redshift has
been successful. Therefore there is a certain probability with which a
galaxy of some absolute magnitude, redshift and template spectrum
(SED) cannot be identified. This means that the galaxy sample is
incomplete. The completeness of \mbox{COMBO-17} has been studied using
extensive Monte Carlo simulations \citep{2003A&A...401...73W} and has
been found to be a complex function of galaxy type and
redshift. Roughly, the completeness is about $90\%$ for $m_R\lesssim
23~\rm mag$ and about $50\%$ near $m_R\approx 23.8~\rm mag$ (blue,
late-type galaxies) or near $m_R\approx 23.5~\rm mag$ (red, early-type
galaxies). 

\begin{table*}
  \caption{\changed{Number of galaxies for all redshift bins and
    survey fields. Only the mean redshifts of the redshift bins
    $0.2\le z<0.4$, $0.4\le z<0.6$, $0.6\le z<0.8$ and $0.8\le z<1.0$
    are quoted. The quoted fractions are red and blue galaxies (from
    \mbox{COMBO-17}) relative to the total number of galaxies in the
    corresponding redshift bin. The comoving number densities derived
    from the galaxy numbers and the estimated comoving volume of a
    z-bin and field are corrected for incompleteness. The statistical
    errors for the individual fields ($2\sigma$) are estimates
    assuming (uncorrelated) Poisson errors for the absolute galaxy
    numbers; errors do hence not include cosmic variance for
    individual fields.  The last column contains the number of
    galaxies (sources) in the shear catalogue (from \mbox{GaBoDS}),
    with $21.5\le R<24.5~\rm mag$, used for the lensing analysis. The
    source galaxies have a mean redshift of $\bar{z}=0.78$. The bottom
    block of the table contains values averaged over all three
    fields. The errors ($1\sigma$) are \emph{now} Jackknife-estimates,
    reflecting cosmic variance derived from the field-to-field
    variances.}
    \label{redbluefig}}
  \begin{center}
  \begin{tabular}{cc|ccc|ccc|c}
    Field&$\bar{z}$&
    \#RED&Fraction&density&
    \#BLUE&Fraction&density&\#Sources\\
    &&&&$[10^{-3}h^3\rm Mpc^{-3}]$&&&$[10^{-3}h^3\rm Mpc^{-3}]$&(GaBoDS)\\
    \hline\hline&&&&&&&&\\
    A901&$0.31$&$297$&$0.22\pm0.02$&$10.7\pm0.6$&$1045$&$0.78\pm0.02$&$38.3\pm1.2$&$17084$\\
    &$0.50$&$397$&$0.24\pm0.02$&$6.2\pm0.3$&$1264$&$0.76\pm0.02$&$20.0\pm0.6$\\
    &$0.71$&$413$&$0.17\pm0.01$&$3.5\pm0.2$&$2046$&$0.83\pm0.01$&$18.2\pm0.4$\\
    &$0.89$&$376$&$0.13\pm0.01$&$2.1\pm0.1$&$2421$&$0.87\pm0.01$&$12.8\pm0.3$\\
    \hline&&&&&&&&\\
    CDFS&$0.31$&$118$&$0.16\pm0.02$&$4.5\pm0.4$&$621$&$0.84\pm0.02$&$22.4\pm0.9$&$20487$\\
    &$0.50$&$338$&$0.18\pm0.01$&$5.2\pm0.3$&$1516$&$0.82\pm0.01$&$23.5\pm0.6$\\
    &$0.71$&$433$&$0.16\pm0.01$&$3.8\pm0.2$&$2323$&$0.84\pm0.01$&$20.5\pm0.4$\\
    &$0.89$&$154$&$0.11\pm0.01$&$0.8\pm0.1$&$1263$&$0.89\pm0.01$&$6.8\pm0.2$\\
    \hline&&&&&&&&\\
    S11&$0.31$&$280$&$0.25\pm0.02$&$10.4\pm0.6$&$843$&$0.75\pm0.02$&$30.3\pm1.1$&$18996$\\
    &$0.50$&$402$&$0.21\pm0.01$&$6.4\pm0.3$&$1518$&$0.79\pm0.01$&$23.7\pm0.6$\\
    &$0.71$&$393$&$0.18\pm0.01$&$3.5\pm0.2$&$1832$&$0.82\pm0.01$&$16.3\pm0.4$\\
    &$0.89$&$367$&$0.14\pm0.01$&$2.0\pm0.1$&$2334$&$0.86\pm0.01$&$12.4\pm0.3$\\
    \hline\hline&&&&&&&&\\
    COMBINED&$0.31$&.&$0.21\pm0.08$&$8.5\pm3.0$&.&$0.79\pm0.08$&$30.3\pm6.9$\\
    &$0.50$&.&$0.21\pm0.05$&$5.9\pm0.6$&.&$0.79\pm0.05$&$22.4\pm1.8$\\
    &$0.70$&.&$0.17\pm0.02$&$3.6\pm0.2$&.&$0.83\pm0.02$&$18.3\pm2.8$\\
    &$0.89$&.&$0.13\pm0.03$&$1.6\pm0.6$&.&$0.87\pm0.03$&$10.7\pm2.9$
  \end{tabular}
  \end{center}
\end{table*}

We split the total object catalogue into four distinct
photo-z bins, namely a) \mbox{$0.2\le z<0.4$}, b) \mbox{$0.4\le
z<0.6$}, c) \mbox{$0.6\le z<0.8$} and d) \mbox{$0.8\le z<1.0$}. The
mean redshifts of galaxies belonging to a)-d) are
\mbox{$\bar{z}=0.3,0.5,0.7,0.9$}, respectively. The sizes of the
samples are listed in Table \ref{redbluefig}.  

\changed{In order to have a better estimate for the true redshift
distribution, the photo-z distribution of every bin is convolved with
the photo-z error (Gaussian errors) of the individual galaxies,
Eq. \Ref{eq:photozerror}. The average redshift uncertainties are
\mbox{$\sigma_z=0.02,0.04,0.06,0.08$} for the samples a)-d),
respectively. See Fig. \ref{fig:pzlenses} for the resulting
distributions. Obviously, the true redshift distribution is wider than
the photo-z distribution. Ignoring this effect would lead to a
systematic under-estimation of the galaxy clustering amplitude by
$\sim20\%$ \citep{2008arXiv0804.2293B}.}

The galaxy samples are further subdivided by applying a cut in the
rest-frame $U-V$ vs. $M_V-5\,\log_{10} h$ CMD (Johnson filter) along
the line
\begin{eqnarray}\label{eq:cmrcut}
  &&(U-V)(M_V,z)=\\
  &&1.15-0.31\,z-0.08(M_V-5\log_{10}h+20)\nonumber
  \; . 
\end{eqnarray}
This model-independent, empirical cut has been chosen by
\citet{2004ApJ...608..752B} to study red galaxies near the galaxy
red-sequence. It slices the bimodal distribution of galaxies in the
CMD between the two modes.  Galaxies redder than $(U-V)(M_V,z)$ are
dubbed ``red galaxies'', ``blue galaxies'' otherwise.  For the
redshifts considered here, most of the red galaxies selected this way
are morphologically early-type with dominant old stellar populations,
while blue galaxies are mainly late-type, star-forming galaxies.   

The redshift dependence of the CMR zero-point, in Eq. \Ref{eq:cmrcut},
was fitted to match the colour evolution of \combo~early-type galaxies
\emph{and} to be consistent with the SDSS CMR zero-point at low
redshift. From the viewpoint of the \combo~early-types the zero-point
for redshifts \mbox{$z\lesssim0.2$} is slightly too low, giving a
small contamination of the red sample with blue cloud galaxies. Since
we consider only galaxies starting from \mbox{$z\ge0.2$}, this
contamination is negligible for this study, though.

\begin{figure*}
  \begin{center}
    \epsfig{file=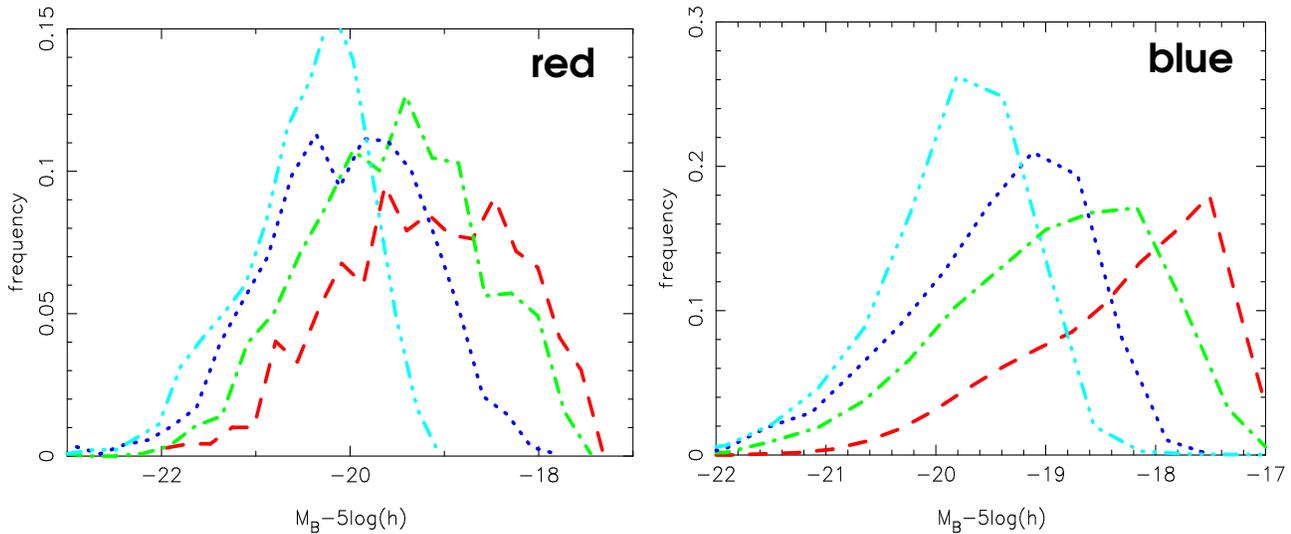,width=170mm,angle=0}
  \end{center}
  \caption{\label{fig:mbhist}\changed{Frequency distribution of
      (rest-frame) B-band magnitudes, separated for red (left) and
      blue (right) galaxies, for the four different photo-z bins
      (dashed lines: $\bar{z}=0.3$; dashed-dotted: $\bar{z}=0.5$;
      dotted: $\bar{z}=0.7$; dashed-dotted-dotted: $\bar{z}=0.9$). The
      corresponding mean magnitudes and standard deviations in order
      of increasing mean redshifts are ($h=1$): $\ave{M_{\rm
      B}}=-19.20\pm0.94\,{\rm mag}, -19.56\pm0.91\,{\rm
      mag},-20.06\pm0.84\,{\rm mag}, -20.49\pm0.68\,{\rm mag}$ (red)
      and $\ave{M_{\rm B}}=-18.34\pm0.88\,{\rm mag},
      -18.96\pm0.94\,{\rm mag},-19.46\pm0.83\,{\rm mag},
      -19.83\pm0.69\,{\rm mag}$ (blue).}}
\end{figure*}

As we took only galaxies with reliable photometric redshifts, we have
as further selection rule $m_{\rm R}\le24\,\rm mag$. The
distribution of our samples in a rest-frame CMD is plotted in Fig.
\ref{combocolmag}.  Obviously, in the lowest redshift bin CMD galaxies
populate faint regions in the diagram that are excluded in the other
redshift bins due to the survey flux-limit.  We estimate that in the
three deeper redshift bins galaxies have roughly to be brighter than
\changed{(rest-frame)}
\begin{equation}\label{eq:extracut}
  M_{\rm V}-5\log_{10} h\approx-17\,\rm mag-(U-V)
\end{equation}
 in order to be included (see steep black lines in Fig.
\ref{combocolmag}).  To acquire comparable galaxy samples at all
redshifts we artificially apply this limit as cut to all redshift
bins. After applying this cut, the galaxy samples of all redshift bins
have comparable absolute \changed{rest-frame} $M_V$ luminosities. The
red sample has an average of $\ave{M_V}=-20.0\pm0.1\,\rm mag$, the
blue sample $\ave{M_V}=-18.8\pm0.3\,\rm mag$ (for $h=1$).

\changed{In contrast to the red-sequence cut, this luminosity cut does
  not take into account the colour/luminosity evolution of the samples
  but is placed at the same position of the rest-frame CMD-diagram at
  all redshifts. We therefore expect the selected galaxy populations
  of all redshifts not to be totally equivalent at the faint end.}
\changed{For an easier comparison with the literature,
  e.g. \citet{2008arXiv0804.2293B}, Fig. \ref{fig:mbhist} shows also
  the distribution of rest-frame $M_{\rm B}$-magnitudes of the various
  samples. The samples become bluer with increasing redshift which is,
  at least partially, explained by the passive evolution of the
  stellar populations.}

\changed{We determine the absolute number density, $\phi_{\rm type}$,
  of our galaxy samples by using the $\rm V_{max}$-estimator
  \citep[e.g.][]{2001A&A...367..788F}. This estimator needs to be
  slightly modified since we are not selecting the galaxies from a
  top-hat redshift range. Instead we are selecting, for every photo-z
  bin, galaxies in redshift with a probability proportional to $p(z)$,
  which is the aforementioned photo-z distribution convolved with the
  photo-z error; $p(z)$ resembles only for the lowest z-bin roughly a
  top-hat selection window, see Fig. \ref{fig:pzlenses}. Due to the
  uncertainty in redshift we also select galaxies from a redshift
  range (volume) larger than the photo-z window would imply. Not
  taking this effect into account would mean to over-estimate the
  galaxy number density. \emph{Assuming} that we are selecting, apart
  from the incompleteness expressed by the incompleteness function
  $C(z,{\rm SED},U-V,R)$ \citep{2003A&A...401...73W}, all galaxies at
  the redshift of maximum $p(z)$, $p_{\rm max}$ say, the $V_{\rm
  max}$-estimator is:
  \begin{eqnarray}
    &&\!\!\!\phi_{\rm type}=\sum_{i=1}^{N_{\rm gal}}\frac{1}{V_i}~;~
    \sigma^2(\phi_{\rm type})=\sum_{i=1}^{N_{\rm
    gal}}\frac{1}{V^2_i}\;,\\ 
    &&\!\!\!V_i\equiv\\
    &&\frac{\Omega}{p_{\rm
    max}}\int_0^\infty\d z\,p(z)\, \frac{dV}{d\Omega\d z}\,C(z,{\rm
    type},U_i-V_i,R_i)\;.
  \end{eqnarray}
  Therefore, $p(z)$ is used here to correct the incompleteness
  function $C$ for our additional galaxy redshift selection
  criterion. By $\Omega$ we denote the survey area of a
  \mbox{COMBO-17}-patch which is $\Omega=f\times(39.7\,\rm arcmin)^2$
  with a filling factor $f$, estimated from the patch masks, of
  $f=0.56,0.54,0.55$ for A901, CDFS and S11, respectively. The
  estimator $\sigma^2(\phi_{\rm type})$ is used for the Poisson
  shot-noise error of $\phi_{\rm type}$. For a top-hat selection
  function $p(z)$, one obtains the estimator mentioned in Fried et
  al.}

\subsection{Cosmic shear data: GaBoDS}

The data reduction of GaBoDS was performed with a nearly fully
automatic, stand-alone pipeline which we had developed to reduce
optical and near infrared images, especially those taken with
multi-chip cameras.  Since weak gravitational lensing was our main
science driver, the pipeline algorithms were optimised to produce deep
co-added mosaics from individual exposures obtained from empty field
observations.  Special care was taken to achieve an accurate
astrometry to reduce possible artificial PSF patterns in the final
co-added images.  For the co-addition we used the programme {\it
EISdrizzle}.  A detailed description of the pipeline can be found in
\citet{esd05}.\footnote{The \texttt{THELI} pipeline is freely
  available under \mbox{\texttt{ftp://ftp.ing.iac.es/mischa/THELI}}.}

The shape of galaxies is influenced by the anisotropic PSF.  In order
to obtain unbiased shear estimates from observed source galaxy
ellipticities we use the so-called KSB algorithm \citep{ksb95}.  For a
detailed description of our implementation of the KSB-algorithm and
catalogue creation we refer the reader to \citet{hss07}.  Our
KSB-algorithm pipeline was blind-tested with simulated data within the
STEP project \citep{mhb07,hvb06}.

For the PSF-anisotropy correction we utilise stars which are
point-like and unaffected by lensing.  By using a sample of bright,
unsaturated stars, we measure the anisotropic PSF with a Gaussian
filter scale matched to the size of the galaxy image to be corrected
\citep{hfk98}.  In the case of the WFI@2.2m instrument the PSF of the
co-added images varies smoothly over the total field-of-view.
Therefore we perform a third-order two-dimensional polynomial fit to
the PSF anisotropy with $3.5\,\sigma$-clipping as a function of
position over the entire field-of-view.  With this fit it is possible
to estimate the PSF anisotropy at the position of the galaxies.

All objects for which problems concerning the determination of shape
or centroid position occur are rejected (e.g. objects near the border,
with negative total flux, with negative semi major and/or semi major
axis).  In addition, we only use those objects with a half-light
radius which is larger than that measured for stars and an ellipticity
(after PSF correction) of less than $1.0$.  Additionally, we only use
those galaxies with a signal-to-noise ratio larger than five, since a
comparison of ground- and space-based data showed that galaxies with a
lower S/N do not contain any shear information (Hetterscheidt et al.,
in prep.).  We adopt the scheme in \citet{ewb01} and \citet{het05} to
estimate uncertainties for each galaxy ellipticities. In the following
lensing analysis galaxies are weighted with the inverse of the square
of the estimates variances.
%
%
%
%
\begin{figure}
\centering
\epsfig{file=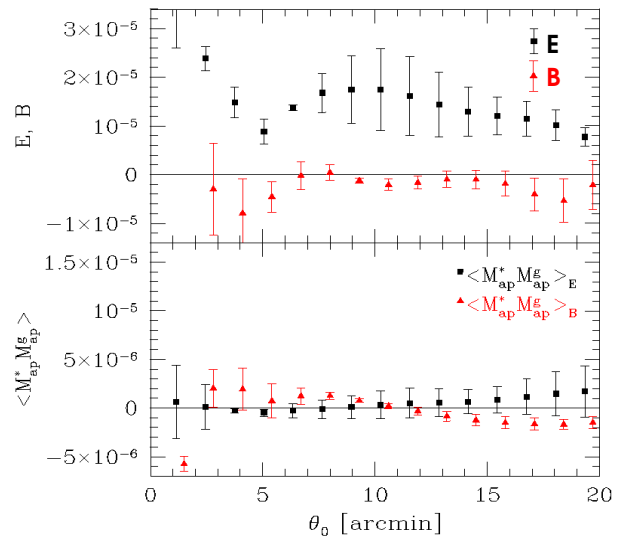,width=8cm,angle=0}
\caption{Various measurements of the aperture mass statistics of the
three \combo-fields in $R$-band ($\approx 0.75\,{\rm deg}^2$).  All
galaxies in the magnitude range $R \in[21.5,24.5]$ are used for the
cosmic shear analysis.  The effective number density of galaxies is
$16\,{\rm arcmin}^{-2}$. The error-bars denote the field-to-field
variance between the three fields. {\it Upper panel: } E- and B-mode
decomposition of the $M_{\rm ap}$-signal.  {\it Lower panel: }
Cross-correlation between uncorrected stars- and corrected galaxies
for E- and B-modes.}
\label{fig:ecorrpol_shear}
\end{figure}

A powerful way to reveal possible systematic errors in the
PSF-correction is the application of the aperture mass statistics as
it provides an unambiguous splitting of E- and B-modes:
\begin{eqnarray}\label{mapdef}
  &&\Ave{M_{\rm ap,\perp}^2(\theta_{\rm ap})}=\\\nonumber
  &&\frac{1}{2}\int_0^{2\theta_{\rm
  ap}}\frac{\d\theta\,\theta}{\theta_{\rm ap}^2} \left[
  \xi_+(\theta)\,T_+\left(\frac{\theta}{\theta_{\rm ap}}\right)\pm
  \xi_-(\theta)\,T_-\left(\frac{\theta}{\theta_{\rm ap}}\right)
  \right]\;.
\end{eqnarray}
The $M_{\rm ap}$-statistics quantify the fluctuations of the shear
signal (E-mode: plus sign on r.h.s, B-mode: minus sign) within an
aperture of radius $\theta_{\rm ap}$. They can be obtained by
transforming the shear-shear correlation functions $\xi_\pm$
\citep[e.g.][]{hss07}. We use $T_\pm$ as derived in
\citet{2002A&A...396....1S}. The presence of non-vanishing B-modes is
a good indicator for systematics arising, for instance, from an
imperfect anisotropy correction.

However, there are several possible astronomical sources of B-modes,
like the intrinsic alignment of galaxies \citep[e.g.][]{hrh00,heh03},
the intrinsic shape-shear correlation
\citep[e.g.][]{2004PhRvD..70f3526H,hwh06} and the redshift clustering
of source galaxies \citep{svm02}.  Furthermore, \citet{kse06} found in
their work a mixing of E- and B-modes due to a cut off in $\xi_\pm$ on
small angular scales.  However, all those B-mode sources are expected
to be much smaller than the statistical errors of the three fields and
are therefore irrelevant in the following analyses.
  
A further method to check for systematics is the cross-correlation
between PSF-uncorrected stars and anisotropy corrected galaxies
\citep[e.g.][]{bmr03}. For that purpose, shear-shear
cross-correlations, $\xi_\pm$, between star-ellipticities and
galaxy-ellipticities are computed and transformed according to
Eq. \Ref{mapdef}. We denote the thereby obtained $M_{\rm
ap}$-variances as $M_{\rm cross;E}$ and $M_{\rm cross;B}$ for the E-
and B-modes, respectively.

In Fig. \ref{fig:ecorrpol_shear} the average E- and B-mode signal and
the average signal of the cross-correlation between
anisotropy-corrected stars and uncorrected galaxies, $M_{\rm cross;E}$
and $M_{\rm cross;B}$, are displayed \citep[further details on $M_{\rm
cross;E}$ and $M_{\rm cross;B}$ are given in][]{hss07}.  The measured
B-mode signal is consistent with zero within the $1\,\sigma$-range for
\mbox{$\theta_0 > 2^\prime$}, and the cross-correlation between
uncorrected stars and corrected galaxies, $M_{\rm cross;B}$, is
consistent with, or close to zero.  Hence the B-mode signal does
not indicate an imperfect anisotropy correction.  Additionally, the
cross-correlation signal, $M_{\rm cross;E}$ is consistent with zero.

Taking the B-modes and the cross-correlation signals $M_{\rm cross;E}$
and $M_{\rm cross;B}$ into account we conclude that the influence of
systematics on the calculated E-mode signal is negligible compared to
statistical errors.

\section{Relative biasing of red and blue galaxies}
\label{sect:analysis}

\subsection{Combining measurements}
\label{sect:combining}

We outline here how measurements of the same quantity in the
\mbox{$N_{\rm p}=3$} different fields were combined, and how the
covariance of the combined value was estimated.

The quantities estimated from the data, as function of galaxy-galaxy
separation on the sky, are the angular clustering of the galaxy
samples, in terms of the aperture statistics (auto- and
cross-correlations), and later on the mean tangential shear around the
galaxies. The measurements are binned into five logarithmic
\changed{angular} bins and compiled as a data vector $\vec{d}_i$,
where \mbox{$i=1\ldots3$} is an index for the survey field (either
A901, S11 or CDFS).

A commonly applied technique for combining all measurements and
estimating the covariance of the combination is by looking at the
field-to-field variance of $\vec{d}_i$ \citep[e.g.][]{hss07}. Applying
this technique to mere three fields, however, poses problems that are
unsolved so far \citep{2007A&A...464..399H} and would bias the final
results. We therefore use a different approach here.

For each field $i$ individually, the measurement is repeated
\mbox{$N_{\rm b}=100$} times on bootstrapped data, which is acquired
by randomly drawing galaxies (\changed{with replacement}) from the
original samples. \changed{The size of the bootstrap samples equals
the the size of the original sample. For a recent paper on this and
related statistical tools for error estimation see
\citet{2008arXiv0810.1885N}, which points bootstrapping out as
appropriate, albeit conservative (errors are overestimated by
$\sim40\%$), method for uncertainties in two-point correlation
functions.}. The data vector of the $j$-th bootstrapped sample of the
$i$-th field is denoted by $\vec{d}_{ij}$. The variance of $\vec{d}_i$
among the bootstrap samples yields an estimate for the covariance of
the statistical errors in $\vec{d}_i$ due to galaxy shot noise:
\begin{equation}\label{eq:covi}
  \mat{C}_i=\frac{1}{N_{\rm b}-1}\sum_{j=1}^{N_{\rm b}}
  \left(\vec{d}_{ij}-\vec{d}_i\right)
  \left(\vec{d}_{ij}-\vec{d}_i\right)^{\rm t}\;.
\end{equation}

The most likely value of a combined $\vec{d}$ of all fields,
constrained by all individual $\vec{d}_i$ and their covariances
$\mat{C}_i$, is
\begin{equation}\label{eq:dest}
  \vec{d}=\mat{C}\sum_{i=1}^{N_{\rm
  p}}\left[\mat{C}_i\right]^{-1}\vec{d}_i~;~
  \mat{C}\equiv\left[\sum_{i=1}^{N_{\rm p}}\left(\mat{C}_i\right)^{-1}\right]^{-1}\;,
\end{equation}
obtained by finding the minimum of $\vec{d}$ in the negative
log-likelihood (assuming Gaussian errors):
\begin{equation}
  \frac{\partial\chi^2(\vec{d})}{\partial\vec{d}}=0~;~
  \chi^2(\vec{d})\equiv\sum_{i=1}^{N_{\rm p}}
  \left(\vec{d}_i-\vec{d}\right)^{\rm t}\left[\mat{C}_i\right]^{-1}
  \left(\vec{d}_i-\vec{d}\right)\;.
\end{equation}
This is the generalisation of the well-known rule to combine
measurements by inversely weighting with their statistical
error. \changed{We consider the assumption of Gaussian statistics for
the likelihood function as valid approximation for the following
reasons:
\begin{itemize}
\item The statistical errors of the angular clustering estimator,
outlined below, are known to be Poisson, hence closely Gaussian for a
not too small number of galaxy pairs inside a bin \citep{ls93}. Since
we linearly combine different, little correlated, angular bins of the
clustering estimator to obtain the $\cal N$-statistics, the $\cal N$-estimates are even more Gaussianly distributed according to the
central limit theorem of statistics. 
\item As for GGL, the complex
ellipticities of the lensed sources obey roughly a bivariate Gaussian
distribution which makes the estimate for GGL inside an angular bin
also approximately Gaussianly distributed.
\end{itemize}
Even if the noise in the data does not obey Gaussian statistics, the
l.h.s. estimator Eq. \Ref{eq:dest} yields an unbiased however not
optimal (maximum likelihood) estimate of $\vec{d}$, with $\mat{C}$ as
covariance, as any weighted average of unbiased estimates $\vec{d}_i$
is itself an unbiased estimator.}

As pointed out by \citet{2007A&A...464..399H} taking the inverse of
the (bootstrapped) $\mat{C}_i$ gives a biased estimate of the
inverse. To obtain an unbiased estimator of the inverse we multiply
$\mat{C}_i$ in Eq. \Ref{eq:covi} by the factor:
\begin{equation}
  \frac{N_{\rm b}}{N_{\rm b}-p-1}\;,
\end{equation}
where $p$ is the size of the vector $\vec{d}$ \changed{(here: $p=15$,
      five angular bins for $\ave{{\cal N}_{\rm red}^2}$, $\ave{{\cal
      N}_{\rm blue}^2}$, $\ave{{\cal N}_{\rm red}{\cal N}_{\rm
      blue}}$, respectively; for the halo-model fit later on we will
      extend $\vec{d}_i$ by ten more components comprising the GGL
      signal of the samples)}.

The covariance of the combined mean $\vec{d}$ is simply
\begin{equation}
  \ave{\vec{d}\vec{d}^{\rm t}}-\ave{\vec{d}}\ave{\vec{d}}^{\rm
  t}=\mat{C}\;.
\end{equation}
This covariance does not contain an estimate of the cosmic variance,
though, because it is solely based on bootstrapping using individual
fields (galaxy shot-noise) and does not include a field-to-field
variance between the fields. 

This can be seen by considering a toy example which has just one bin,
$d_i$, for the field data vectors and equal (co)variance for all
fields, \mbox{$C_i=\sigma^2$} say. The above equations tell us for
that case that $d$ is the arithmetic mean of all $d_i$, and the
combined variance is \mbox{$C=\sigma^2N_{\rm p}^{-1}$}. Based on this,
if we had an infinite number of galaxies within each field,
i.e. bootstrapped \mbox{$\sigma\rightarrow0$}, we would expect a
``perfect'' measurement with \mbox{$C=0$}. Hence, $C$ completely
ignores the possibility that \mbox{$d_i\ne d_j$} for $i\ne j$ even
with a hypothetically infinite number of galaxies inside each patch
(cosmic variance).

We expect the actual uncertainty therefore to be somewhat higher than
expressed by $\mat{C}$, especially for the larger separation bins
(larger aperture radii) where cosmic variance errors are becoming
larger with respect to shot-noise errors. \changed{The fact that the
bootstrap variances overestimate the covariance presumably compensates
this deficit to some extend.}

\subsection{Clustering of red and blue galaxies}
\label{sect:clustering}

\begin{figure}
  \begin{center}
    \epsfig{file=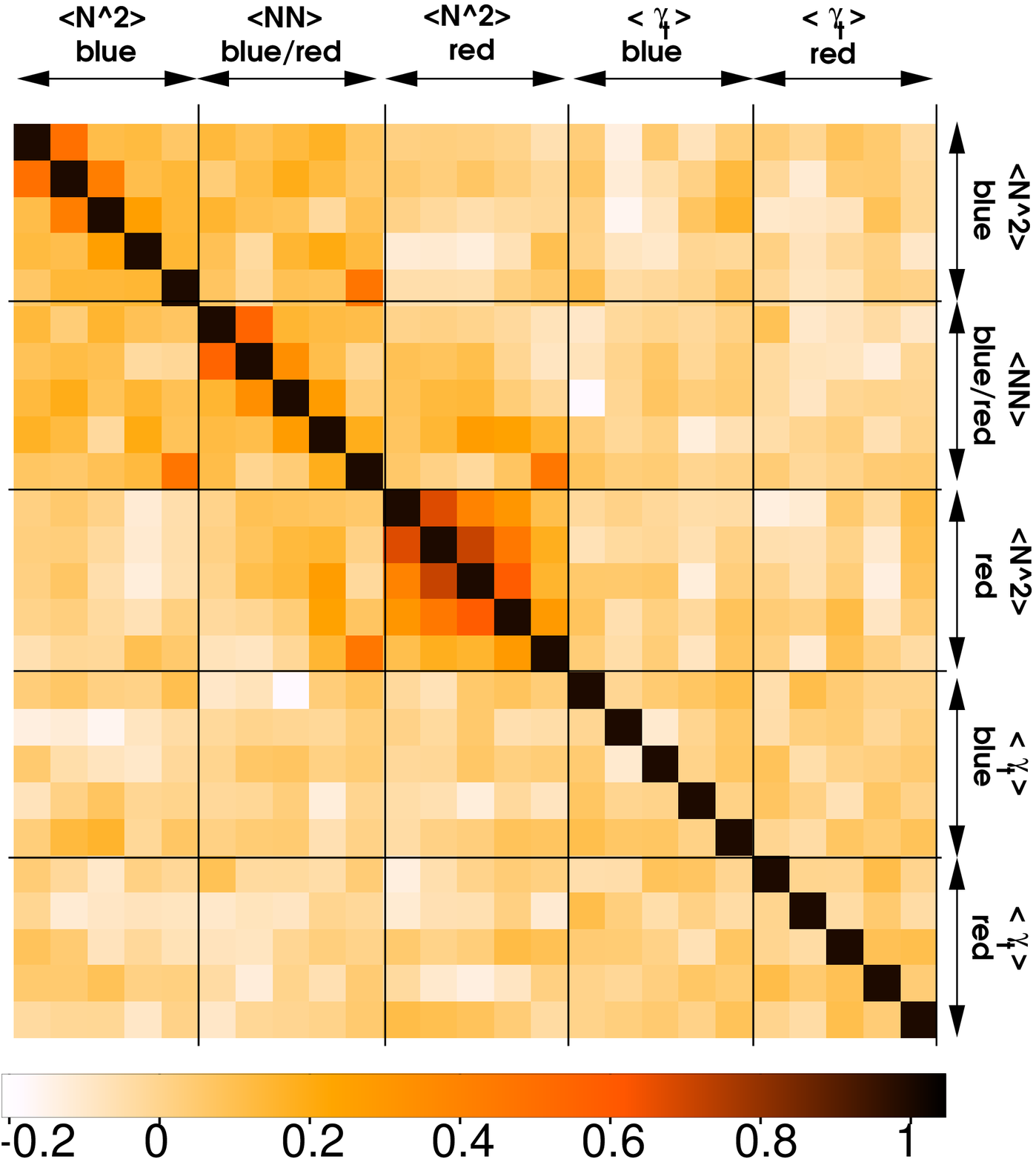,width=8.5cm,angle=0}
  \end{center}
  \caption{Correlation of statistical errors of the aperture number
  count statistics -- $\ave{\caln^2_{\rm blue}}$, $\ave{\caln_{\rm
  blue}\caln_{\rm red}}$ and $\ave{\caln^2_{\rm red}}$ for five
  aperture radii each -- and the GGL signal -- $\ave{\gamma_{\rm t}}$
  for the blue and red sample -- for a mean redshift of \mbox{$z=0.7$}
  as obtained from bootstrapping the data. The correlation matrix for
  the other redshift bins look similar. \label{correlation}}
\end{figure}

\subsubsection{Method}
The $\caln$-statistics is derived from the angular clustering of the
galaxy samples, $\omega(\theta)$ \citep{peebles80}, by a linear
transformation \citep{2007A&A...461..861S}
\begin{equation}
  \ave{\caln_i(\theta_{\rm ap})\caln_j(\theta_{\rm ap})} = \int_0^2\d
  x\,x\,\omega_{ij}(\theta_{\rm ap}\,x)\,T_+(x)\;,
\end{equation}
where
\begin{equation}
  T_+\left(x\right)\equiv\left(2\pi\right)^2\int_0^\infty
  {\rm d}s~s~\left[\rm I\left(s\right)\right]^2J_0\left(sx\right)\;.
\end{equation}
The indices $i$ and $j$ are used to denote the different galaxy
samples.  For estimating the angular clustering of a single sample,
\mbox{$i=j$}, we use the standard method of counting the number of
galaxy pair within a certain separation, namely pairs of galaxies from
the same sample, \mbox{$\ave{DD}$}, pairs of galaxies from a random
mock sample and the \mbox{COMBO-sample}, \mbox{$\ave{DR}$}, and pairs
between galaxies from the same mock sample, \mbox{$\ave{RR}$}
\citep{ls93}. The number of pairs involving random mock samples is
averaged over $50$ mock realisations for each correlation function.

\changed{For the random catalogue, we assume that the completeness of
the galaxies inside a redshift is homogeneous, so that the only
relevant parameters for the random mocks is the number of galaxies and
the masking, which is applied for all galaxies equally.}

The cross-correlation function, $i\ne j$, is computed implementing the
estimator \citep{1997astro.ph..4241S}:
\begin{equation}
  \omega_{ij}(\theta)=
  \frac{\ave{D_iD_j}}{\ave{R_iR_j}}-\frac{\ave{D_iR_j}+\ave{D_jR_i}}{\ave{R_iR_j}}+1
  \;,
\end{equation}
for which counting the number of pairs between different galaxy
samples, $D_{i/j}$, and different mock samples, $R_{i/j}$, is
required. For example, \mbox{$\ave{R_iD_j}$} denotes the number of
pairs between galaxies from a $i$-mock catalogue and galaxies $j$
within a certain $\theta$ separation interval. Note that the size of
the mock sample $R_k$ is the same as the size of the galaxy sample
$D_k$.

Traditionally, galaxy clustering is studied using the angular
correlation function $\omega(\theta)$. For a comparison of our results
for the two-point statistics of galaxy clustering in \mbox{COMBO-17}
with the literature, we infer the angular correlation function from
the $\caln$-statistics by applying the method outlined in
\citet{2007A&A...461..861S}. This method allows us to be ignorant
about the integral constraint which offsets the estimates of
$\omega(\theta)$ obtained from the aforementioned estimators. We
parameterise $\omega(\theta)$ as a simple power-law
\begin{equation}
  \omega(\theta)=A_\omega\,\left(\frac{\theta}{1^\prime}\right)^{-\delta}\;;
\end{equation}
$A_\omega$ and $\delta$ are constants. Moreover, we deproject
$\omega(\theta)$ in order to obtain an average 3D-correlation function
for the clustering of the samples,
\begin{equation}
  \xi(r)=\left(\frac{r}{r_0}\right)^{-\delta-1}\;,
\end{equation}
after having made sure that the Limber approximation
\citep{peebles80,limber53} was valid here \citep{2006astro.ph..9165S};
the constant $r_0$ is the correlation length.

\changed{As the 2D-correlation functions are not exactly power-laws
\citep{2004ApJ...608...16Z}, the foregoing procedure will yield
parameters for the 3D-clustering which are biased to some extent. We
do not discuss this effect further but point out here that this bias
could, in principle, be estimated from the halo-model fit, which also
predicts the 3D-correlation function.}

\begin{figure*}
  \begin{center}
    \epsfig{file=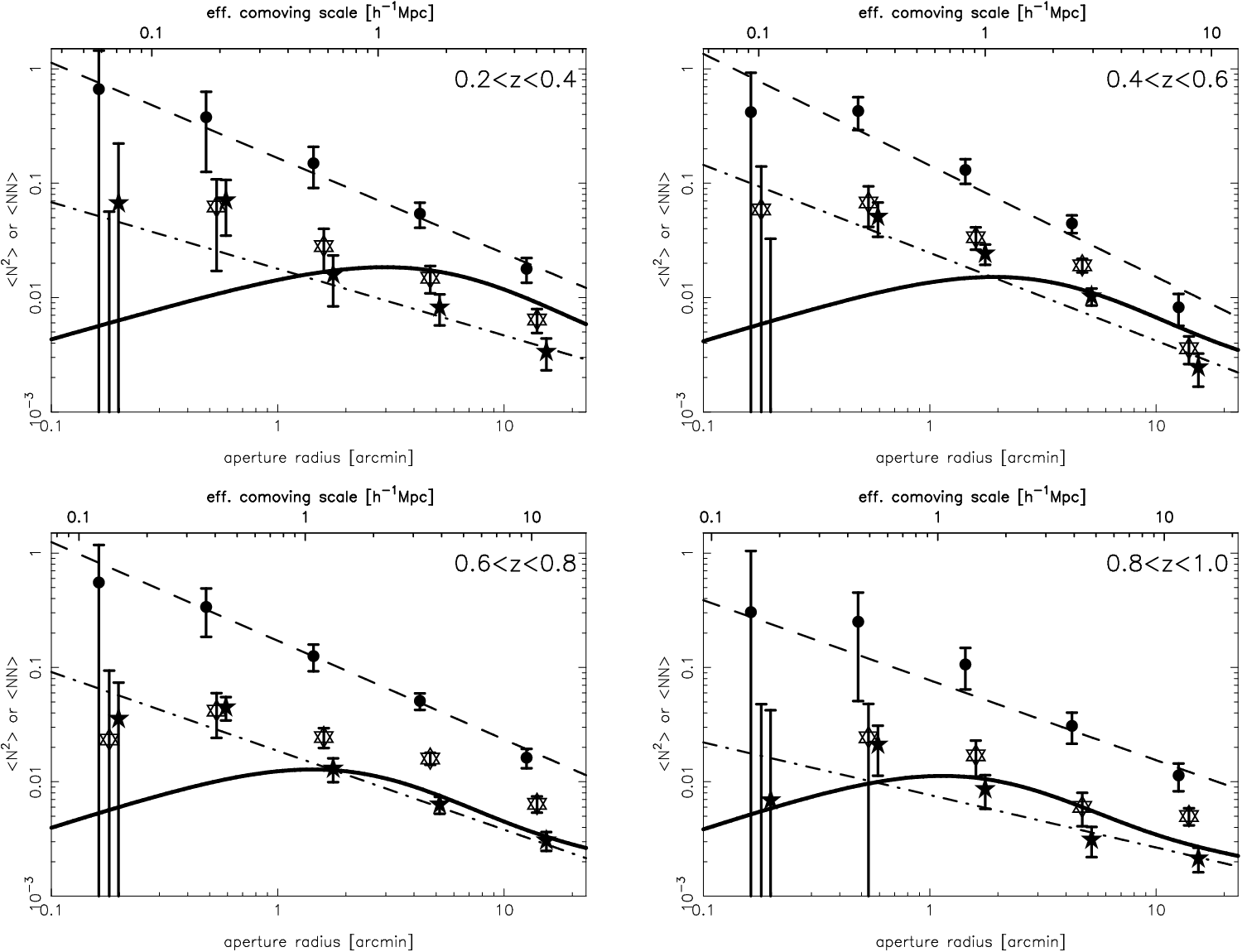,width=13.4cm,angle=-90}
  \end{center}
  \caption{Variance of the aperture number count, $\ave{\caln_{\rm
    red}^2}$ and $\ave{\caln_{\rm blue}^2}$, for red (filled circles)
    and blue (filled stars) galaxies. The straight lines are the
    best-fitting power laws to the measurements. The thick solid line
    denotes the theoretical variance for galaxies unbiased to the dark
    matter \citep[as in][]{spj03}. Also plotted is the aperture number
    count cross-correlation, $\ave{\caln_{\rm red}\caln_{\rm blue}}$,
    between blue and red galaxies (open stars). The data points are
    slightly shifted \changed{along the abscissa} to increase the
    visibility. \label{combonsqd}}
\end{figure*}

\begin{table*}
  \caption{\changed{The results are for $\Omega_{\rm m}=0.3$,
    $\Omega_\Lambda=0.7$. \emph{Upper half of table}: summary of
    results (median and $68\%$ confidence levels) obtained considering
    values for $\ave{\caln^2}$ out of
    $\theta\in[0.1^\prime,23^\prime]$ for red and blue galaxies for
    different mean redshifts, $\bar{z}$.  Note that the statistical
    errors of $\delta$ and $A_\omega$ are strongly anti-correlated
    with a correlation coefficient of about \mbox{$-0.6$}. \emph{Lower
    half of table}: the fit for $\omega(\theta)$ has been transformed
    into the 3D-galaxy clustering function $\xi(r)=(r/r_0)^{-\gamma}$
    using Limber's equation.  The clustering length $r_0$ is in units
    of $h^{-1}\rm Mpc$; $\xi_{1\rm Mpc}$ denotes $\xi(r)$ at a
    comoving length of $r=1\,h^{-1}\rm Mpc$. Parameters with a
    ``$|\delta$''-suffix are fits for a fixed slope:
    $\delta=0.85,0.65$ for red and blue,
    respectively.} \label{comboclustering}}
  \begin{center}\small
  \begin{tabular}{l|ccc|cc||ccc|cc}
    &\multicolumn{5}{c|}{\textbf{RED SAMPLE}}&\multicolumn{5}{c}{\textbf{BLUE SAMPLE}}\\ $\bar{z}$&
    $[A_\omega$&$\delta$&$\chi^2/n$&$A_\omega|\delta$&$\chi^2/n$]&
    $[A_\omega$&$\delta$&$\chi^2/n$&$A_\omega|\delta$&$\chi^2/n]$\\
    \hline\hline &&&&&&&&\\ 0.31&
    $0.46_{-0.14}^{+0.15}$&$0.81_{-0.27}^{+0.19}$&$0.3$&$0.47^{+0.12}_{-0.12}$&$0.2$&
    $0.11_{-0.04}^{+0.20}$&$0.49_{-0.39}^{+0.35}$&$1.2$&$0.09^{+0.02}_{-0.02}$&$0.8$\\
    0.50&
    $0.27_{-0.07}^{+0.07}$&$0.97_{-0.16}^{+0.13}$&$2.4$&$0.29^{+0.06}_{-0.06}$&$3.0$&
    $0.08_{-0.02}^{+0.02}$&$0.76_{-0.13}^{+0.12}$&$3.8$&$0.09^{+0.02}_{-0.02}$&$4.5$\\
    0.70&
    $0.44_{-0.09}^{+0.09}$&$0.85_{-0.17}^{+0.14}$&$0.6$&$0.45^{+0.07}_{-0.07}$&$0.6$&
    $0.08_{-0.02}^{+0.02}$&$0.67_{-0.14}^{+0.12}$&$2.3$&$0.08^{+0.01}_{-0.01}$&$2.3$\\
    0.90&
    $0.31_{-0.11}^{+0.16}$&$0.69_{-0.44}^{+0.31}$&$0.7$&$0.28^{+0.08}_{-0.08}$&$0.4$&
    $0.08_{-0.03}^{+0.27}$&$0.33_{-0.27}^{+0.30}$&$1.3$&$0.05^{+0.01}_{-0.01}$&$0.8$\\
    \hline &&&&&&&&\\ $\bar{z}$&[$r_0$&$\gamma$&$\xi_{1\rm
    Mpc}$&$r_0|\delta$&$\xi_{1\rm
    Mpc}|\delta$]&[$r_0$&$\gamma$&$\xi_{1\rm
    Mpc}$&$r_0|\delta$&$\xi_{1\rm Mpc}|\delta$]\\ \hline&&&&&&&&\\
    0.31&
    $5.37_{-1.54}^{+2.72}$&$1.81_{-0.27}^{+0.19}$&$20.6_{-6.9}^{+7.4}$&$5.19^{+0.67}_{-0.74}$&$21.0^{+5.3}_{-5.2}$&
    $2.98_{-1.12}^{+2.74}$&$1.49_{-0.39}^{+0.35}$&$5.0_{-2.0}^{+2.5}$&$2.50^{+0.38}_{-0.42}$&$4.6^{+1.2}_{-1.2}$\\
    0.50&
    $4.37_{-0.86}^{+1.06}$&$1.97_{-0.16}^{+0.13}$&$17.8_{-4.7}^{+4.9}$&$5.04^{+0.56}_{-0.62}$&$19.9^{+4.3}_{-4.3}$&
    $2.66_{-0.51}^{+0.57}$&$1.76_{-0.13}^{+0.12}$&$5.6_{-1.4}^{+1.5}$&$3.08^{+0.35}_{-0.38}$&$6.4^{+1.2}_{-1.2}$\\
    0.70&
    $7.43_{-1.42}^{+2.00}$&$1.85_{-0.17}^{+0.14}$&$40.3_{-7.7}^{+7.9}$&$7.51^{+0.66}_{-0.71}$&$41.7^{+7.0}_{-7.0}$&
    $3.20_{-0.51}^{+0.59}$&$1.67_{-0.14}^{+0.12}$&$6.9_{-0.6}^{+1.3}$&$3.30^{+0.29}_{-0.31}$&$7.2^{+1.1}_{-1.1}$\\
    0.90&
    $7.60_{-2.41}^{+5.20}$&$1.69_{-0.44}^{+0.31}$&$28.7_{-9.25}^{+10.3}$&$6.58^{+0.94}_{-1.07}$&$32.7^{+9.19}_{-9.17}$&
    $3.52_{-0.97}^{+1.82}$&$1.33_{-0.27}^{+0.30}$&$5.5_{-1.4}^{+1.4}$&$2.70^{+0.33}_{-0.36}$&$5.1^{+1.1}_{-1.1}$
  \end{tabular}
  \end{center}
\end{table*}

\subsubsection{Results}
The combined measurements of the $2^{\rm nd}$-order $\caln$-statistics
for blue and red galaxies can be found in Fig. \ref{combonsqd}. The
$\caln$-statistics is binned between \mbox{$0^\prime.1\le\theta_{\rm
ap}<23^\prime$} using five logarithmic bins. The statistical errors
are somewhat correlated which can be seen in
Fig.\ref{correlation}. The best fits of the angular clustering and
3D-clustering parameters are listed in Table
\ref{comboclustering}. \changed{Averaging over all redshift bins, we
find for the red sample $A_\omega=0.35\pm0.06$, $\delta=0.85\pm0.10$
and $r_0=5.5\pm0.9\,h^{-1}\rm Mpc$. The corresponding values of the
blue sample are $A_\omega=0.08\pm0.02$, $\delta=0.65\pm0.08$ and
$r_0=3.0\pm0.4\,h^{-1}\rm Mpc$. If we combine the red and blue
sample, we find as average over all redshifts for all galaxies
$A_\omega=0.11\pm0.01$, $\delta=0.79\pm0.03$ and
$r_0=3.8\pm0.3\,h^{-1}\rm Mpc$.}

In the following, we would like to compare the best-fit parameters for
the galaxy clustering to the results of other papers.  Since the data
sample selections between different surveys are in general not equal,
we can only make a crude comparison to other results. Typical values
for the clustering of galaxies, regardless of their colour, at low
redshifts are $r_0=4-6\,h^{-1}\rm Mpc$ and $\delta=0.6-0.9$
\changed{\citep[cf.][]{2008A&A...479..321M, 2003MNRAS.346...78H,
2005ApJ...630....1Z, 2002MNRAS.332..827N,
2002ApJ...571..172Z}}. Compared to these values our results are
compatible, although we may have somewhat lower values for $r_0$. A
lower clustering amplitude may be explained by a different mean
luminosity of our sample, though. See in particular
\citet{2008ApJ...672..153C} for the dependence of galaxy clustering
parameters on absolute luminosity.

Subdividing the galaxy sample of \mbox{COMBO-17} into red and blue
galaxies yields different clustering properties: red galaxies are more
strongly clustered than blue galaxies and red galaxies have steeper
slopes $\gamma$ than blue galaxies.  Similar cuts have been done in
\citet{2008A&A...479..321M} and \citet{2008ApJ...672..153C} who find
clustering properties in good agreement with our measurements up to
redshifts of \mbox{$z\sim1$}. Beyond the statistical uncertainties of
our measurements we do not find a change in clustering of our samples
as, for example, reported by the more accurate measurements of Coil et
al.

\citet{2006A&A...457..145P} measured the clustering of red and blue
galaxies, examining the clustering in redshift space within the range
of \mbox{$0.4\le z\le 0.8$}, with the same data set as we do and
applying the same red-sequence cut. They quote values for the red
sample which are comparable to our result. The blue sample is somewhat
different, though, with a marginally higher
$r_0=3.65\pm0.25\,h^{-1}\rm Mpc$ and a shallower
$\delta=0.45\pm0.03$. We suspect that the difference is due to a
different magnitude cut in addition to the red-sequence cut: Phleps et
al. selected only galaxies brighter than $M_B=-18$, whereas our
magnitude limits are colour dependent, Eq. \Ref{eq:extracut}.

\subsection{Relative linear stochastic bias}
\label{sect:relbiasresults}

\subsubsection{Method}
The aim of this section is to constrain the relative linear stochastic
bias of the red and blue galaxy sample using the measurements of the
aperture number count statistics. Simply applying the definitions,
Eq. \Ref{eq:linbias} and \Ref{eq:lincorr}, to the measurements would
probably result in a biased estimate of the bias parameters due to the
relatively large uncertainty in the aperture statistics. A more
reliable, but also more elaborate, approach consists in employing
Bayesian statistics, see Appendix \ref{sect:bayesian}.

Our combined measurement of the aperture statistics for different
aperture radii, $\theta_i$, is compiled inside the vector:
\begin{eqnarray}\nonumber
  \vec{d}&=& \Big( \ave{\caln^2_{\rm red}(\theta_1)},\ave{\caln^2_{\rm
  red}(\theta_2)},\ldots,\ave{\caln^2_{\rm
  red}(\theta_i)},\ldots\\\nonumber &&\ave{\caln_{\rm
  red}(\theta_1)\caln_{\rm blue}(\theta_1)},\ave{\caln_{\rm
  red}(\theta_2)\caln_{\rm blue}(\theta_2)},\ldots,\\
  &&\ave{\caln^2_{\rm blue}(\theta_1)},\ave{\caln^2_{\rm
  blue}(\theta_2)},\ldots\Big)^{\rm t}\;.
\end{eqnarray}
The covariance of $\vec{d}$, $\mat{C}$, is worked out according to
what has been outlined in Sect. \ref{sect:combining}. The measurement
$\vec{d}$ is an estimate of the true, underlying aperture
statistics. Let $\ave{\caln^2_{\rm blue}(\theta_i)|_{\rm true}}$ be
the true aperture number count dispersion of blue
galaxies.\footnote{Without further assumptions, the best estimate for
the true $\caln$-dispersion is $\ave{\caln_{\rm blue}^2(\theta)}$
itself. By putting constraints on the relation between the aperture
statistics of red and blue galaxies, as we are doing by constraining
the parameters of the linear stochastic biasing, this can change.} For
given linear stochastic bias parameters and the true aperture number
count dispersion of blue galaxies,
\begin{equation}
  \vec{p}\equiv\left(\ldots,\ave{\caln^2_{\rm blue}(\theta_i)|_{\rm
    true}},\ldots,b(\theta_i),\ldots,r(\theta_i),\ldots\right)^{\rm t}\;,    
\end{equation}
the expected $\caln$-statistics of blue and red galaxies, ``fitted''
to the data $\vec{d}$, is:
\begin{eqnarray}\nonumber
  &&\!\!\!\!\!\!\!\!\!\!\!\!\vec{m}(\vec{p})=\\\nonumber
  &&\Big(
  b^2(\theta_1)\ave{\caln^2_{\rm blue}(\theta_1)|_{\rm true}},b^2(\theta_2)\ave{\caln^2_{\rm
      blue}(\theta_2)|_{\rm true}},\ldots,\\\nonumber
  &&r(\theta_1)b(\theta_1)\ave{\caln_{\rm blue}(\theta_1)|_{\rm true}},
  r(\theta_2)b(\theta_2)\ave{\caln_{\rm blue}(\theta_2)|_{\rm true}},\ldots,\\
  &&\ave{\caln^2_{\rm blue}(\theta_1)|_{\rm true}},\ave{\caln^2_{\rm
      blue}(\theta_2)|_{\rm true}},\ldots\Big)^{\rm t}\;.
\end{eqnarray}
The $\caln$-statistics involving the red galaxy population is
expressed in terms of the blue population statistics and the linear
stochastic bias parameters.

Now, the likelihood of the parameters $\vec{p}$ given the data
$\vec{d}$ is for Gaussian errors:
\begin{equation}
  P(\vec{d}|\vec{p})\propto\exp{\left(
    -\frac{1}{2}\left[\vec{d}-\vec{m}(\vec{p})\right]^{\rm
      t}\mat{C}^{-1}\left[\vec{d}-\vec{m}(\vec{p})\right]\right)}\;.
\end{equation}
The posterior likelihood, up to a constant factor, of $b(\theta_i)$
and $r(\theta_i)$ given $\vec{d}$ and marginalised over $\ave{N^2_{\rm
blue}(\theta_i)|_{\rm true}}$ is
\begin{eqnarray}
  &&P\left(b(\theta_i),r(\theta_i)|\vec{d}\right)\propto\\\nonumber
  &&\int\left[\prod_{i=1}^{N_{\rm bin}}\d\ave{\caln^2_{\rm blue}(\theta_i)|_{\rm true}}\,\,
  P\left(b(\theta_i)\right)P\left(r(\theta_i)\right)\right]P(\vec{d}|\vec{p})\;.
\end{eqnarray}
\changed{The probabilities $P\left(b(\theta_i)\right)$ and
$P\left(r(\theta_i)\right)$ are priors on the bias parameters which we
chose to be flat within \mbox{$b(\theta_i)\in[0,4]$},
\mbox{$r(\theta_i)\in[0,1.3]$} and zero otherwise. The upper limits of
the priors were chosen to be well above crude estimates for $b$ and
$r$, obtained by blindly applying Eqs. \Ref{eq:linbias} and
\Ref{eq:lincorr} to the data.} The number of aperture angular radii
bins is $N_{\rm bin}$.

\changed{The marginalised posterior likelihood
$P\left(b(\theta_i),r(\theta_i)|\vec{d}\right)$ of the overall ten
variables (ten bias parameters for five aperture radii)} is most
conveniently sampled employing the Monte-Carlo Markov Chain (MCMC)
technique \citep[e.g.][]{2005A&A...429..383T}. Especially, the
marginalisation is trivial within this framework. \changed{Remember
that the size of the data vector is $p=15$.}

\begin{figure*}
  \begin{center}
    \epsfig{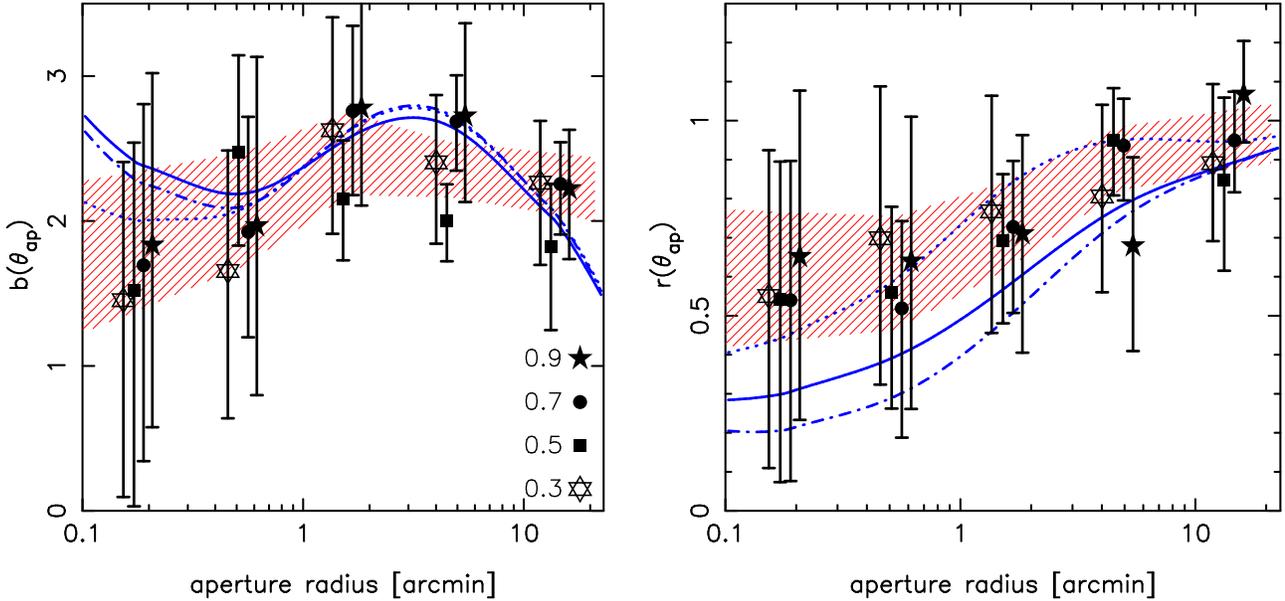}
  \end{center}
  \caption{Relative linear stochastic bias between the red and blue
    galaxy sample for four mean redshifts, $b(\theta_{\rm ap})$ is in
    the left panel and $r(\theta_{\rm ap})$ in the right right
    panel. The shaded area denotes the $1\sigma$-confidence region
    about the mean of all redshifts taken together. The effective
    proper spatial scale for an aperture radius of $10^\prime$ at
    redshift \mbox{$z=0.3,0.5,0.7,0.9$} is
    \mbox{$2.8,3.8,4.5,4.8\,h^{-1}\rm Mpc$}.  The lines show the
    best-fitting models for $\bar{z}=0.7$ discussed later on in
    Sect. \ref{sect:jhodresults} (solid: no central galaxies,
    dot-dashed: red central galaxies, dotted: red and blue central
    galaxies). The data points are slightly shifted to increase the
    visibility.\label{redbluebias}}
\end{figure*}

\subsubsection{Results}
The Fig. \ref{redbluebias} shows the inferred constraints on the
relative linear stochastic bias of red and blue galaxies.  Owing to
relatively large remaining statistical uncertainties, which makes all
redshift bins indistinguishable, we have combined the signal from all
redshift bins (shaded areas).

As already seen in Fig. \ref{combonsqd}, red and blue galaxies are
differently clustered with respect to the dark matter and therefore
have also to be biased relative to each other. This difference is
equivalent to a relative bias factor of about $b\sim1.7-2.2$ with some
evidence for a rise towards radii of about $\theta_{\rm
ap}\sim2^\prime$ and a subsequent decline for even smaller radii. The
evidence for this scale-dependence is, however, weak.  A rise is
expected, though, because of the different power-law slopes of
$\langle \caln^2\rangle$ for the two samples.

For the lowest redshift bin, this is in good agreement with
\citet{2003MNRAS.344..847M}, their Fig. 4.  Furthermore, the observed
scale-dependence explains why we find an overall larger value for the
red/blue bias than other authors who determined the relative bias with
various different methods on larger scales,
e.g. \citet{2005MNRAS.356..456C}, $b\approx1.3$ at $15\,h^{-1}\rm
Mpc$, \citet{2005MNRAS.356..247W}, $b\approx 1.8$ at $10\,h^{-1}\rm
Mpc$, \citet{1998AJ....115..869W}, $b\approx 1.2$ at $8\,h^{-1}\rm
Mpc$, \citet{1997ApJ...489...37G}, $b\approx1.7$. \changed{According
to our halo model (Fig. \ref{fig:split}), which will be discussed in
the following sections,} we should expect a steep decline of the bias
factor down to \mbox{$b\sim1.5$} beyond our largest aperture radius
which would reconcile our $b\sim2-3$ with other studies.

Within the errors we do not see an evolution of the relative bias with
redshift as, for example, has been reported by
\citet{1996ApJ...461..534L}.  This would support the finding of
\citet{2003A&A...407..855P}.

Recently, \citet{2008ApJ...672..153C} have reported for DEEP2 a
relative bias of \mbox{$b\sim1.28$} averaged over spatial scales
between \mbox{$1-15\,h^{-1}\rm Mpc$} and \mbox{$b\sim1.44$} for
\mbox{$0.1-15\,h^{-1}\rm Mpc$} which implies an increase of the bias
for scales smaller than \mbox{$1\,h^{-1}\rm Mpc$}. This also agrees
with our finding.

For the bias parameter $r(\theta_{\rm ap})$ we can make out a trend of
decorrelation, $r\ne1$, between the two samples towards small scales
starting from about $10^\prime$, which corresponds to (proper)
$\sim3-5\,h^{-1}\rm Mpc$ depending on the mean redshift. We estimate
that the correlation factor drops to \mbox{$r\sim0.60\pm0.15$} on the
smallest measured scale.  An evolution with redshift exceeding the
statistical errors is not visible.

Therefore, as other authors, we find a correlation close to unity on
large scales \citep[e.g.][]{2005MNRAS.356..247W, 2005MNRAS.356..456C,
2000ApJ...544...63B, 1999ApJ...518L..69T} that is decreasing towards
smaller scales. We expect our data points to become eventually
consistent with $r\sim 1$ not far beyond $\theta_{\rm
ap}\gtrsim20^\prime$. Note, again, that we are probing smaller scales
than the cited authors due to different methods.

\citet{2007ApJ...664..608W} studied the cross-correlation statistics
of galaxy samples with different luminosities and colours in
SDSS. Their results for the cross-correlation between the faint red
and faint blue sample is consistent with our results. In particular,
they also find a decrease of $r(\theta_{\rm ap})$ towards smaller
scales (see their Fig. 16, bottom panels).\footnote{Note that Wang et
al. are essentially plotting $r^{-2}(\theta_{\rm ap})$.}
\citet{2008ApJ...672..153C} pointed out that below a scale of
\mbox{$1\,h^{-1}\rm Mpc$} the cross-correlation function of their blue
and red sample drops below the geometric mean of the separate
auto-correlation functions. This is to say that their correlation
factor becomes less than unity on these scales. In this context, see
also Fig. 11 of \citet{2008MNRAS.385.1635S} where a decorrelation
towards smaller scales is found for splitting the data set into red
and blue galaxies.

\section{Galaxy-galaxy lensing}
\label{sect:ggl}

\begin{figure*}
  \begin{center}
    \epsfig{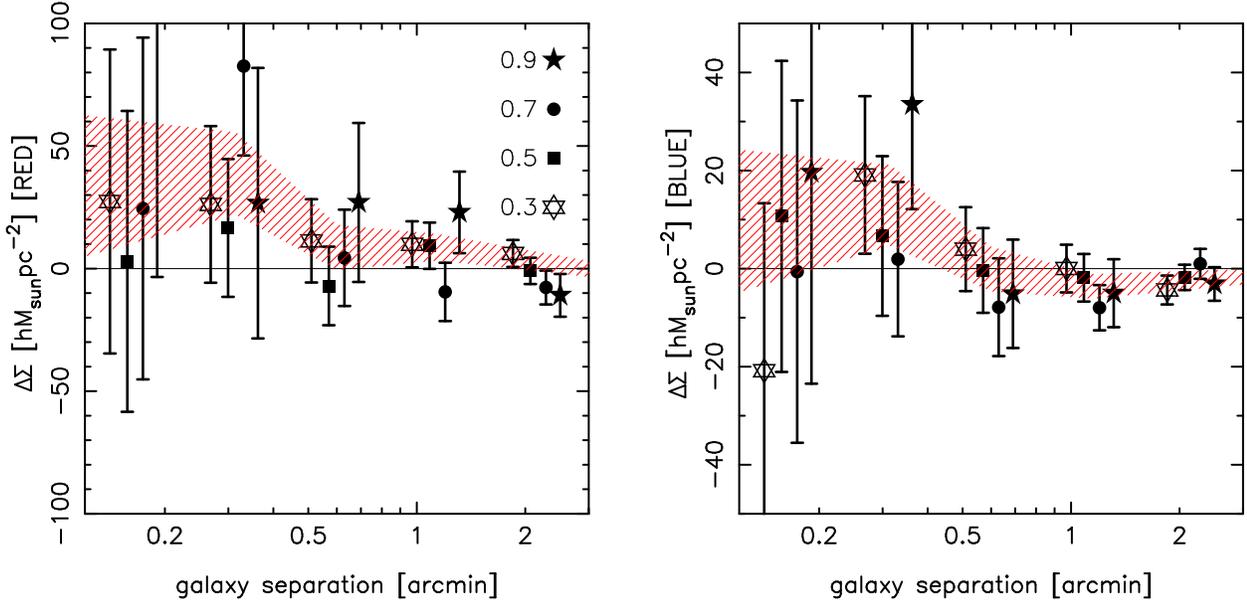}
  \end{center}
  \caption{\label{lambda}Differential surface mass density,
  $\Delta\Sigma$, as function of angular separation about lenses from
  the red (left panel) and blue (right panel) galaxy samples as
  inferred from GGL. The shaded area denotes the $1\sigma$-confidence
  region about the mean of all redshifts taken together. The data
  points are slightly shifted to increase the visibility. From the
  lowest to the highest redshift, a separation of $0^\prime\!.5$
  corresponds to a projected comoving distance of $125\,h^{-1}\rm kpc$,
  $195\,h^{-1}\rm kpc$, $265\,h^{-1}\rm kpc$ or $320\,h^{-1}\rm kpc$,
  respectively.}
\end{figure*}

\subsection{Method}
To impose further constraints on the following halo-model analysis, we
additionally measure the mean tangential shear of source galaxies
(\mbox{GaBoDS}), $\ave{\gamma_{\rm t}}$, as function of separation
$\theta$ about the red and blue lens-galaxies (\mbox{COMBO-17}) for
different redshifts, i.e.
\begin{equation}
  \ave{\gamma_{\rm t}(\theta)}=
  -\Ave{\epsilon(\vec{\theta}_1){\rm e}^{-2{\rm i}\angle\vec{\theta}_1,\vec{\theta}_2}}\;,
\end{equation}
where the average tangential ellipticity over all source-lens pairs
with separation \mbox{$\theta=|\vec{\theta}_1-\vec{\theta}_2|$} has to
be taken \citep[cf.][]{2006A&A...455..441K};
\mbox{$\angle\vec{\theta}_1,\vec{\theta}_2$} is the angle that is
spanned by $\vec{\theta}_2-\vec{\theta}_1$, the difference vector
between source and lens position, and the $x$-axis.

The mean tangential shear about lenses can be related to the
differential projected matter over-density about lenses, in excess to
the cosmic mean \citep[e.g.][]{2001astro.ph..8013M},
\begin{equation}
  \Delta\Sigma(\theta)\equiv\overline{\Sigma}(<\theta)-\Sigma(\theta)=
  \frac{\ave{\gamma_{\rm t}}(\theta)}{\Sigma_{\rm crit}^{-1}}
\end{equation}
with
\begin{equation}
  \Sigma_{\rm crit}^{-1}=\frac{4\pi G}{c^2}\left\langle\frac{f_{\rm
      k}(w_{\rm l})f_{\rm k}(w_{\rm s}-w_{\rm l})}{f_{\rm
      k}(w_{\rm s})}\right\rangle\;,
\end{equation}
if we specify a fiducial cosmology and the distribution of sources
(lenses) in distance from the observer, $w_{\rm s}$ ($w_{\rm l}$);
$\ave{\ldots}$ denotes the average over the lens and source
distribution.  The function $f_{\rm k}(w)$ is the comoving angular
diameter distance as function of the comoving radial distance $w$ and
the curvature, $k$, of the fiducial cosmological model. By
$\overline{\Sigma}(<\theta)$ we denote the average line-of-sight
over-density within a radius of $\theta$, the lens is at the disk
centre, whereas $\Sigma(\theta)$ is the average over-density over an
annulus with radius $\theta$. We rescale our measurements of
$\ave{\gamma_{\rm t}}$ with $\Sigma_{\rm crit}$ in order to get rid of
the influence of the lensing efficiency. This gives us comparable
quantities for lenses of all redshift bins.

\begin{figure}
  \begin{center}
    \epsfig{file=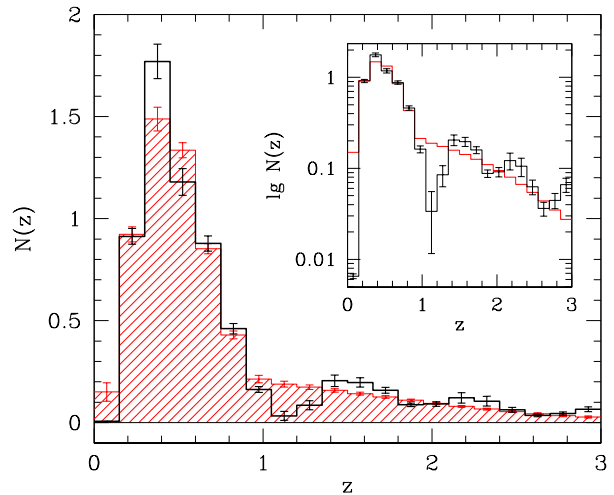,width=8.5cm,angle=0}
  \end{center}
  \caption{\label{fig:pofz} Distribution of sources belonging to the
  cosmic shear catalogue, as estimated by the photometric redshifts of
  galaxies inside the magnitude interval $R\in[21.5,24.5]$ in the
  DPS. The black histogram shows the empirical distribution, the red
  histogram is the fit to the data. The empirical data points around
  $z=1.1$ (dip) are down-weighted in the fit due to systematics in the
  photometric redshifts. The error bars denote the field-to-field
  variance. The mean redshift of the sources is $\bar{z}=0.78$. The
  figure is taken from \citet{hss07}.}
\end{figure}

For the redshift distribution of the sources we use the fit of
\citet{hss07} to the empirical distribution of photometric redshifts as
seen in the Deep-Public Survey (DPS) \citep{2006A&A...452.1121H}, see
Fig. \ref{fig:pofz}. Our shear catalogue is a sub-sample of the shear
catalogue used in that study.

A representative example for the correlation of statistical errors in
the GGL estimate, and the cross-correlation between $\cal
N$-statistics errors and GGL errors, is given by
Fig. \ref{correlation}.

\subsection{Results}
Our measurements for of $\Delta\Sigma(\theta)$ are shown in Fig.
\ref{lambda}. The data is binned between
\mbox{$6^\pprime\le\theta<3^\prime$} into five logarithmic bins
\changed{in angular separation.} The remaining statistical
uncertainties are high so that almost all measurements are at
$1\sigma$ consistent with zero. However, when combining all redshifts
bins, we find a slight signal for both the red and the blue galaxy
sample. The combined red signal is higher than the blue signal,
roughly by a factor of two or three. This is consistent with what was
found by \citet{2006A&A...455..441K}, using the same data,
\citet{2004AJ....127.2544S}, \citet{2001astro.ph..8013M} and
\citet{2001MNRAS.321..439G}.  It implies that the environment of red
galaxies (or the galaxy itself) contains more mass than the
environment of a typical blue galaxy, \changed{or that the typical
size of a blue lens halo is smaller than that of a red lens, although
it might have the same mass than the red lens halo.}

When fitting HOD parameters to the data, we use the GGL signal of the
different redshift bins to constrain the allowed regime of lensing
predictions by our model. Due to the large statistical uncertainties
of the measurement, the GGL mainly serves as an upper limit.

\section{Interpretation within a halo model framework}

\changed{In this section,} we translate the clustering statistics and
GGL-signal of the galaxy samples into halo-model parameters. The
halo-model description is utilised to describe the 3D-distribution of
galaxies and dark matter. We only briefly summarise the halo-model
formalism here and refer the reader to the literature for the
details. How the 3D-distributions, expressed by the halo-model,
relates to the observed projected angular distribution on the sky is
discussed later on in Sect. 6.3.  The reader may find the details on
the dark matter halo properties used for this paper in Appendix
\ref{sect:dmhaloes}. Note that the notation used therein is introduced
in this section.  The Fourier transform of a function $f(\vec{r})$,
denoted by a tilde $\tilde{f}(\vec{k})$, is defined in analogy to
Eq. \Ref{fouriertransf} but now for three dimensions.

\subsection{Halo-model description}

\subsubsection{General halo-model formalism} 

The halo model is an analytical prescription for the clustering of
dark matter that was motivated by N-body simulations of the cosmic
structure formation \citep[for a review:][]{2002PhR...372....1C}. All
matter is enclosed inside typical haloes with a given mass
spectrum. Galaxy mock catalogues can be generated within this
framework by populating virialised dark matter haloes with galaxies
according to a prescription taken directly from semi-analytic models
of galaxy formation or hydrodynamic simulation that include recipes
for the formation (evolution) of galaxies
\changed{\citep{2002ApJ...575..587B,2000MNRAS.311..793B}}. The
parameters for populating the haloes with galaxies depend solely on
the halo mass.  The halo-model description has been quite successful
in describing or fitting observational data, although doubts about the
strict validity of the basic model assumptions have been cast recently
by \changed{\citet{2007MNRAS.377L...5G} and
\citet{2004MNRAS.350.1385S}} (assembly bias).

One key assumption of the halo model is that the dark matter density
or galaxy number density are linear superpositions of in total $N_{\rm
h}$ haloes with a typical radial profile\footnote{Actually, this
expansion assumes that galaxies or dark matter are smoothly spread out
over the halo. This is a good approximation if the number of particles
is large. For \emph{discrete particles}, like galaxies, that are only
on average distributed according to the halo profile, the more
accurate expansion would be:
\begin{equation}\label{eq:discrete}
  n_i(\vec{r}) = \sum_{k=1}^{N_{\rm
  h}}\sum_{l=1}^{N_i(m_k)}\delta_{\rm D}(\vec{r}-\vec{r}_k-\vec{\Delta
  r}_{kl})\;,
\end{equation}
where $\vec{\Delta r}_{kl}$ is the position of the $l$-th galaxy
relative to the $k$-th halo centre. The discreteness makes a
difference for the one-halo term, which is taken into account for the
following results.}:
\begin{equation}
  n_i(\vec{r}) = \sum_{j=1}^{N_{\rm
  h}}N_i(m_j)u_i(\vec{r}-\vec{r}_j,m_j)\;.
\end{equation}
The halo profiles, $u_i$, describe the spatial distribution of
galaxies, \mbox{$i>0$}, or dark matter, \mbox{$i=0$}, within haloes.
All profiles considered here are normalised to unity.

The profiles depend on an additional parameter, $m$, which is the
total mass of the dark matter halo, $\vec{r}$ is a position within the
comoving frame and $\vec{r}_i$ is the centre of the $i$-th halo. By
$N_i(m_j)$ we denote either the number of galaxies, type $i$,
populating the $j$-th halo (if $i>0$), or if $i=0$ the dark matter
mass, $m_j$, that is attached to the $j$-th halo. In the latter case,
$n_0(\vec{r})$ is simply the dark matter density as function of
position. \changed{In particular, $N_0$ and $N_i$ with $i>0$ have, for
convenience within the formalism, different units\footnote{\changed{The proper
normalisation for the power spectra is done by $\overline{N}_i$.}}
(halo mass versus number of galaxies).}

In general, the halo-occupation number \mbox{$N_i(m_j)$} and the halo
centre, $\vec{r}_j$, are \emph{random numbers}. The conditional
probability distribution of the halo-occupation number,
\mbox{$P(N_i|m_j)$}, is a function of the halo mass only. In the case
of dark matter, $i=0$, one has simply \mbox{$N_0(m_j)\equiv m_j$},
i.e. \mbox{$P(N_0|m_j)=\delta_{\rm D}(N_0-m_j)$}, \changed{thus the
mass associated with a ``dark matter halo $m_i$'' is always $m_i$,
whereas the number of galaxies living inside haloes of same mass may
vary from halo to halo.}

Both \mbox{$N_i(m_j)$} and \mbox{$u_i(\vec{r},m_j)$} depend solely on
$m_j$ and not on the mass or position of any other halo. Moreover, the
position of a halo, $\vec{r}_j$, and its mass, $m_j$, are postulated
to be statistically independent. However, the halo-occupation numbers
of different galaxies $i$ and $j$ inside the same halo may be
statistically dependent on each other. For example, inside the
\emph{same} halo the number of red galaxies may be related to the
number of blue galaxies and vice versa. \changed{This is the idea
  behind the concept of the joint halo-occupation distribution.}

Based on these assumptions, the halo model predicts the 3D-power
spectra of galaxy number density correlations, galaxy-mass density
correlations and mass-mass correlations, \mbox{$P_{ij}(k)$}, as
function of a halo mass spectrum, halo bias parameters, typical
density profiles and a HOD of galaxies:
\begin{equation}
  \frac{\Ave{\tilde{n}_i(\vec{k})\tilde{n}_j(\vec{k}^\prime)}}
{\bar{n}_i\bar{n}_j}=
(2\pi)^3P_{ij}(|\vec{k}|)\delta_{\rm
D}(\vec{k}+\vec{k}^\prime)\;.
\end{equation}
Here and in the following, $\ave{\ldots}$ is the statistical average
over all possible haloes, and $\bar{n}_i$ means the average number of
galaxies per unit volume. This paper only considers isotropic halo
density profiles $u_i(\vec{r},m)$ independent of the direction of
$\vec{r}$, so that:
\begin{equation}
  \tilde{u}_i(k,m)= \frac{\int_0^\infty\d r\,rk^{-1}\,u_i(r,m)
  \sin{(kr)}} {\int_0^\infty\d r\,r^2\,u_i(r,m)} \;.
\end{equation}
This equation includes a normalisation of the density profile.

Following the calculations of \citet{1991ApJ...381..349S} one finds
for the power spectra (\mbox{$k>0$} and \mbox{$N_{\rm
h}\gg1$}):
\begin{equation}\label{eq:modelpower}
  P_{ij}(k)=P_{ij}^{\rm
    1h}(k)+P_{ij}^{\rm 2h}(k)\;,
\end{equation}
where the so-called one-halo term is
\begin{equation}\label{eq:onehalo}
  P_{ij}^{\rm 1h}(k)=
  \frac{\overline{m}}{\bar{\rho}\overline{N}_i\overline{N}_j}
  \int_0^\infty\!\!\!\!\!\!\d m\,n(m)K^{\rm 1h}_{ij}(k,m)\;,
\end{equation}
the two-halo term
\begin{eqnarray}\label{eq:twohalo}
  P_{ij}^{\rm 2h}(k)&=&
  \frac{1}{\overline{N}_i\overline{N}_j}\int_0^\infty\!\!\!\d m_1\,n(m_1)\int_0^\infty\!\!\!\d
  m_2\,n(m_2)\times\\\nonumber
  &&P(k,m_1,m_2)K^{\rm 2h}_{ij}(k,m_1,m_2)
\end{eqnarray}
\changed{and} 
\begin{equation}
  \overline{N}_i=\int_0^\infty\d m\,n(m)\,\Ave{N_i(m)}~;~
  \overline{m}=\int_0^\infty\d m\,n(m)\,m\;.
\end{equation}
The mean number density of galaxies is
\begin{equation}\label{eq:halonbar}
  \bar{n}_i=\Ave{n_i(\vec{r})}=
  \frac{\bar{\rho}}{\overline{m}}\,\overline{N}_i\;.
\end{equation}

\changed{The function $n(m)\d m$ is usually the mean number density of
haloes within the mass interval $[m,m+\d m]$. Note, however, that in
the above equations, and for the following $\bar{n}_i$, $n(m)$ may be
rescaled by any arbitrary constant without changing the results.}  The
average \mbox{$\Ave{N_i(m)}$} is the mean number of galaxies found
within a halo of mass $m$ (for \mbox{$i>0$}), or the mass of the halo
itself, if \mbox{$i=0$}. \changed{Notice that after the statistical
average performed for the power spectra we have shifted the notation
from $N_i(m_k)$ (number of $i$-galaxies inside halo $k$) to $N_i(m)$
(number of $i$-galaxies inside \emph{a} halo of mass $m$).}

By $\bar{\rho}=\rho_{\rm crit}\Omega_{\rm m}$ we denote the mean
\emph{comoving} matter density of the dark matter that is included
inside the dark matter haloes.  The constant $\Omega_{\rm m}$ is the
matter density parameter and $\rho_{\rm crit}$ the critical density.

The different cases of power spectra (galaxy auto-power spectra,
galaxy cross-power spectra, dark matter/galaxy cross-power spectra)
give different results for the above integrals.  \changed{To save
space, all variants are encapsulated into the integral
kernels\footnote{\changed{The spatial distributions of galaxies inside the same
halo are postulated to be statistically independent, the kernels
follow from Eq. \Ref{eq:discrete}.}}
\begin{eqnarray}\label{eq:k1h}
  K^{\rm 1h}_{ij}(k,m)&=&
  \Ave{
    \sum_{q=1}^{N_i(m)}\tilde{u}^{(q)}_i(k,m)
    \sum_{r=1}^{N_j(m)}\tilde{u}^{(r)}_j(k,m)}-\\
  \nonumber&&
  -\delta^{\rm K}_{ij}
  \Ave{\sum_{q=1}^{N_j(m)}[\tilde{u}^{(q)}_j(k,m)]^2}\;,\\
  \label{eq:k2h}
  K^{\rm 2h}_{ij}(k,m)&=&
  \Ave{
    \sum_{q=1}^{N_i(m)}\tilde{u}^{(q)}_i(k,m)}
  \Ave{\sum_{r=1}^{N_j(m)}\tilde{u}^{(r)}_j(k,m)}\;,
\end{eqnarray}
where $\ave{\ldots}$ denotes the average over all haloes of mass $m$
and $\tilde{u}_i^{(q)}(k,m)$ expresses the spatial probability
distribution of the $q$-th, out of $q\in[1,N_i(m)]$, galaxies belonging
to the sample $i$ and a particular halo of mass $m$.}

\changed{We need this further distinction into various spatial
distributions since we may split a galaxy sample into one central
galaxy -- sitting at the halo centre hence having $\tilde{u}(k,m)=1$
-- and $(N_i(m)-1)$ satellite galaxies with a different
distribution. The term with the Kronecker pre-factor, $\delta^{\rm
K}_{ij}$, in Eq. \Ref{eq:k1h} is only applied if both samples $i$ and
$j$ are discrete (galaxies). It accounts for the subtraction of white
shot-noise contribution, $1/\bar{n}_i$, to the clustering power that
is not measured due to the definition of the clustering correlation
function, $\omega(\theta)$ (excess of pairs over a uniform
distribution), and the aperture statistics. For the smooth dark matter,
$i=0$, one has to substitute in the Eqs. \Ref{eq:k1h} and \Ref{eq:k2h}
\mbox{$\sum_{q=1}^{N_i(m)}\tilde{u}^{(q)}_i(k,m)\equiv
m\,\tilde{u}_0(k,m)$}, which simplifies the equations.}

\changed{As an example, if all galaxies of a same sample $i$ have
identical spatial distributions, $\tilde{u}_i(k,m)$, one will find
for $i\ne j$ and $i,j>0$:
\begin{eqnarray}\label{eq:k1}
  \sum_{q=1}^{N_i(m)}\tilde{u}_i^{(q)}(k,m)&=&\!\!N_i(m)\tilde{u}_i(k,m)\;,\\
  \nonumber K^{\rm 1h}_{ij}(k,m)\!\!\!&=&
  \tilde{u}_i(k,m)\tilde{u}_j(k,m)\Ave{N_i(m)N_j(m)}\;,\\
  \nonumber K^{\rm 2h}_{ij}(k,m)\!\!\!&=&
  \tilde{u}_i(k,m)\tilde{u}_j(k,m)\Ave{N_i(m)}\Ave{N_j(m)}\;.
\end{eqnarray}
A galaxy sample with $N_i^{\rm cen}(m)\in\{0,1\}$ central galaxies and
$N_i^{\rm sat}(m)$ satellites of same in-halo distribution
$\tilde{u}_i(k,m)$ has 
\begin{equation}\label{eq:k2}
  \sum_{q=1}^{N_i(m)}\tilde{u}_i^{(q)}(k,m)=
  N_i^{\rm cen}(m)+N^{\rm sat}_i(m)\tilde{u}_i(k,m)\;.
\end{equation}
From Eq. \Ref{eq:k1} and \Ref{eq:k2} all kernels relevant for this paper
follow. They are listed in Table \ref{kernels}.}

\begin{table*}
\caption{\label{kernels}\changed{Integration kernels for the one- and
    two-halo term, Eqs. \Ref{eq:onehalo} and \Ref{eq:twohalo}. In the
    simple model, all galaxies of same type are distributed over the
    halo the same way. For the central-galaxy scenario, the halo
    occupation, \mbox{$N_i(m)=N_i^{\rm cen}(m)+N_i^{\rm sat}(m)$}, of
    every galaxy type, $i$, is split into a central galaxy and
    satellite galaxies. Here, it is explicitly assumed that $N_i^{\rm
      sat}(m)>0$ requires $N_i^{\rm cen}(m)=1$, i.e. $\ave{N_i^{\rm
        cen}(m)N_i^{\rm sat}(m)}=\ave{N_i^{\rm sat}(m)}$. The mixed
    model assumes one type of galaxies, $i$, to possess a central
    galaxy and another, $j$, having no central galaxy as in the simple
    model. For that case, $i$ is described by the central model, $j$
    by the simple model; only for the cross-correlation of both a new
    kernel is needed. Note that \mbox{$N_0(m)\equiv m$}, dark matter
    is always ``simple''.}}
\begin{center}
\begin{tabular}{ll|llr}
  $i$&$j$&&integral kernel $K_{ij}^{\rm 1h}(k,m)$&model type\\
  \hline\hline&&\\ $0$&$0$&&$[\tilde{u}_0(k,m)]^2m^2$&simple\\&&&&.\\
  $i>0$&$j=i$&&$[\tilde{u}_i(k,m)]^2\Ave{N_i(m)\left(N_i(m)-1\right)}$&.\\&&&&.\\
  $i>0$&$j\ne
  i$&&$\tilde{u}_i(k,m)\tilde{u}_j(k,m)\Ave{N_i(m)N_j(m)}$&.\\
  &&\\\hline&&\\ $i>0$&$0$&&$\tilde{u}_0(k,m)m\Ave{N_i^{\rm
  cen}(m)}+\tilde{u}_i(k,m)\tilde{u}_0(k,m)m\Ave{N_i^{\rm
  sat}(m)}$&central\\&&&&.\\
  $i>0$&$j=i$&&$2\tilde{u}_i(k,m)\Ave{N_i^{\rm
  sat}(m)}+[\tilde{u}_i(k,m)]^2\Ave{N_i^{\rm sat}(m)\left(N_i^{\rm
  sat}(m)-1\right)}$&.\\&&&&.\\ $i>0$&$j\ne i>0$&&
  $\tilde{u}_i(k,m)\Ave{N_i^{\rm sat}(m)N_j^{\rm cen}(m)}+
  \tilde{u}_j(k,m)\Ave{N_j^{\rm sat}(m)N_i^{\rm cen}(m)}+$ &.\\&&&&.\\
  &&&$\tilde{u}_i(k,m)\tilde{u}_j(k,m)\Ave{N_i^{\rm sat}(m)N_j^{\rm
  sat}(m)}+\Ave{N_i^{\rm cen}(m)N_j^{\rm
  cen}(m)}$&.\\&&\\\hline&&\\$i>0$&$j\ne i$&&
  $\tilde{u}_j(k,m)\Ave{N_i^{\rm cen}(m)N_j(m)}
  +\tilde{u}_i(k,m)\tilde{u}_j(k,m)\Ave{N_i^{\rm
  sat}(m)N_j(m)}$&mixed\\&&\\\\
  $i$&$j$&&integral kernel $K_{ij}^{\rm
  2h}(k,m_1,m_2)$&model type\\\hline\hline&&\\ $i$&$j$&&
  $\tilde{u}_i(k,m_1)\tilde{u}_j(k,m_2)\Ave{N_i(m_1)}
  \Ave{N_j(m_2)}$&simple\\&&\\\hline&&\\
  $i>0$&$0$&&$\tilde{u}_0(k,m_2)m_2
  \left[\Ave{N_i^{\rm
  cen}(m_1)}+\tilde{u}_i(k,m_1)\Ave{N_i^{\rm
  sat}(m_1)}\right]$&central\\&&&&.\\ $i>0$&$j>0$&&
  $\left[\tilde{u}_j(k,m_1)\Ave{N_i^{\rm
  cen}(m_1)}+\tilde{u}_i(k,m_1)\tilde{u}_j(k,m_1)\Ave{N_i^{\rm
  sat}(m_1)}\right]\times$
  &.\\&&&&.\\
  &&&$\left[\tilde{u}_i(k,m_2)\Ave{N_j^{\rm
  cen}(m_2)}+\tilde{u}_i(k,m_2)\tilde{u}_j(k,m_2)\Ave{N_j^{\rm
  sat}(m_2)}\right]$&.\\&&\\\hline&&\\
  $i>0$&$j\ne i$&&$\tilde{u}_j(k,m_2)\Ave{N_j(m_2)}\left[\Ave{N_i^{\rm
  cen}(m_1)}+\tilde{u}_i(k,m_1)\Ave{N_i^{\rm
  sat}(m_1)}\right]$&mixed
\end{tabular}
\end{center}
\end{table*}

The function $P(k,m_1,m_2)$ means the cross-power spectrum of the
number densities of haloes with masses $m_1$ and $m_2$. It is common
practice to assume a linear deterministic biasing between the halo
number density and the linear dark matter density
\citep{2002PhR...372....1C}:
\begin{equation}
  P(k,m_1,m_2)\approx P_{\rm lin}(k)\,b(m_1)\,b(m_2)
\end{equation}
with $b(m)$ being the linear bias factor of haloes with mass $m$ and
$P_{\rm lin}(k)$ the linear dark matter power spectrum.  This reduces
the 2D-integral in Eq. \Ref{eq:twohalo} to a simpler product of
1D-integrals. 

\begin{figure*}
  \begin{center}
    \epsfig{file=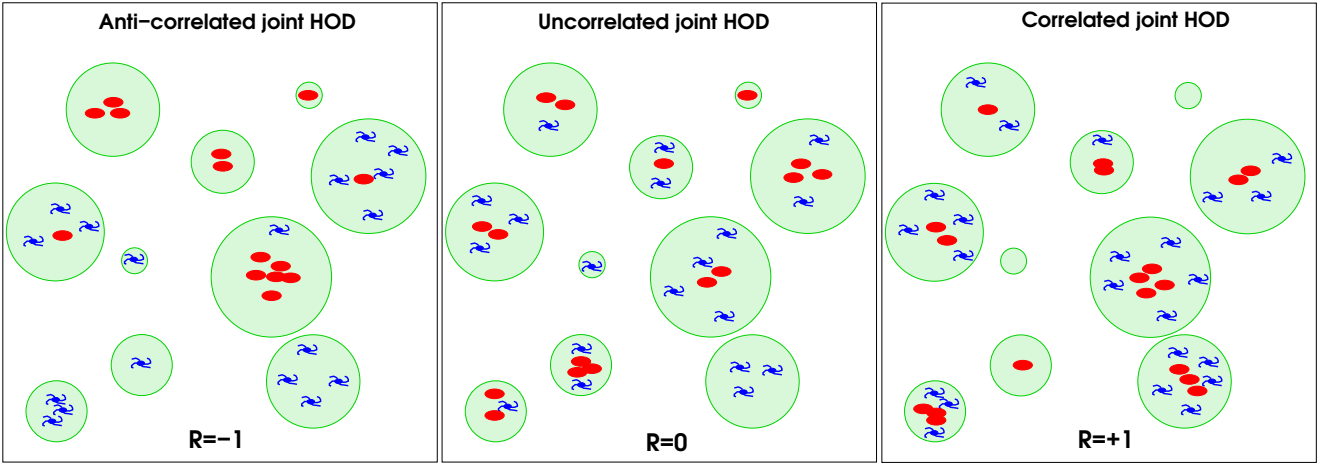,width=17.5cm,angle=0}
  \end{center}
  \caption{\label{jointhodsketch} Simplified illustration of the
  concept of the joint halo occupation distribution for three
  cases. \emph{Left panel:} the dark matter haloes (shady disks) are
  filled with different galaxy types that tend to avoid each other. If
  the number of red, elliptical, galaxies is increased for a halo of
  given size, the number of blue spiral, galaxies
  decreases. \emph{Right panel:} over- and under densities of the two
  galaxy populations are highly correlated. If the number of red
  galaxies for a halo of a given size increases, the number of blue
  galaxies increases as well. \emph{Middle panel:} the halo-occupation
  numbers are not correlated. }
\end{figure*}

\subsubsection{Joint halo-occupation distribution}

Within the framework of the halo model (see Eqs. \Ref{eq:onehalo} and
\Ref{eq:twohalo}), one works out the clustering statistics of galaxies
by specifying the mean number of galaxies for a halo of certain mass
$m$, \mbox{$\ave{N_i(m)}$}, and the mean number of galaxy pairs,
either pairs of the same galaxy type (auto-power),
\mbox{$\ave{N_i(m)(N_i(m)-1)}$}, or pairs between different galaxy
types (cross-power), \mbox{$\ave{N_i(m)N_j(m)}$}. Things become
slightly more difficult, though, if we distinguish between central
galaxies and satellite galaxies as can be seen in Table \ref{kernels}.

In general, the number of galaxies or galaxy pairs for a fixed halo
mass are first and $2^{\rm nd}$-order moments of a joint halo
occupation distribution (JHOD) of two galaxy populations $i$ and $j$.
The JHOD, \mbox{$P(N_i,N_j|m)$}, determines the probability to find a
certain number of ``galaxies $i$'' and ``galaxies $j$'' inside the
same halo of mass $m$. The $2^{\rm nd}$-order moment,
\begin{equation}
 \ave{N_i(m)N_j(m)}= \sum_{N_i,N_j=0}^\infty P(N_i,N_j|m)\,
 N_iN_j\;,
\end{equation}
of the JHOD can conveniently be parameterised in terms of the mean
halo-occupation number,
\begin{equation}
  \Ave{N_i(m)}=\sum_{N_i,N_j=0}^\infty P(N_i,N_j|m)\,N_i\;,
\end{equation}
the (central) variance of the HOD,
\begin{equation}
  \sigma_i(m)\equiv \sqrt{\Ave{N^2_i(m)}-\Ave{N_i(m)}^2}\;,
\end{equation}
and the JHOD-correlation factor:
\begin{equation}
  R_{ij}(m)\equiv
  \frac{\Ave{N_i(m)N_j(m)}-\Ave{N_i(m)}\Ave{N_j(m)}}
       {\sigma_i(m)\sigma_j(m)}\;.
\end{equation}

The JHOD correlation factor expresses the tendency of two populations
to avoid or \changed{attract} each other inside the same halo.  See
Fig. \ref{jointhodsketch} for a simplified illustration. This
tendency, however, can only be seen in the one-halo term of the
cross-power spectrum, Eq. \Ref{eq:onehalo}, which is observable when
considering the cross-power (cross-correlation function) of two galaxy
populations for small separations.

The effect of the JHOD correlation factor $R_{ij}(m)$ on the
cross-moments becomes negligible, i.e.
\begin{equation}
  \Ave{N_i(m)N_j(m)}\approx\Ave{N_i(m)}\Ave{N_j(m)}\;,
\end{equation}
if the relative fluctuations in galaxy numbers inside haloes become
small, i.e. if \mbox{$\sigma_i(m)/\Ave{N_i(m)}\ll1$}. Since haloes
with mean galaxy occupation numbers of more than roughly one are
expected to have a Poisson variance, one can expect that the JHOD
correlation factor loses significance for massive haloes. For large
$m$ the relative variance becomes $\propto1/\sqrt{\ave{N_i(m)}}$.

In order to avoid confusion, we would like to stress again that
$R_{ij}$ expresses the correlation of galaxy numbers inside single
haloes, whereas $r(\theta_{\rm ap})$ is a correlation of galaxy number
densities as seen in the angular clustering of galaxies.

\subsection{Adopted model for JHOD of red and blue galaxies}
\label{sect:jhodmodel}

\changed{In the following we describe the details of the HOD of the
  red and blue galaxy sample. A galaxy sample (blue or red) can either
  have a central galaxy and satellites or solely consists of
  satellites. Satellites are distributed according to the dark matter
  in a halo, but possibly with a different concentration
  parameters. Central galaxies sit at or close to the centre of a
  halo, for the latter of which our ``central models'' are an
  approximation. We will distinguish three different model flavours:
  \begin{enumerate}
    \item A scenario in which we have only red and blue satellites
    populating a dark matter halo.
    \item A scenario in which red galaxies are both central and
    satellite galaxies, blue galaxies are \emph{only satellites}.
    \item A scenario in which we have red and blue central galaxies
    \emph{and} red and blue satellites.
  \end{enumerate}}

\changed{Case ii) is motivated by the observation that galaxy clusters
  often have red galaxies as central galaxies. Case iii) is motivated
  by the possibility that the observed galaxy-galaxy lensing signal of
  the blue galaxies and the power-law clustering may also require a
  central blue galaxy.}

\changed{We start off with the description of the HOD of a single
  sample. The interconnection of the red and blue sample is discussed
  later on where we address the problem of red/blue galaxy pairs.}

For modelling the HOD of our red and blue galaxy sample we use, with
some minor modifications, the parametrisations that were discussed in
\citet{2007ApJ...667..760Z}, \citet{2007ApJ...659....1Z} and
\citet{2005ApJ...633..791Z}.  

\subsubsection{Mean galaxy numbers and distribution inside haloes}

The HOD (\changed{red and blue}) for haloes with mass $m$,
\begin{equation}
  N(m)=N^{\rm cen}(m)+N^{\rm sat}(m)\;,
\end{equation}
is split into one central galaxy, $N^{\rm cen}(m)\in\{0,1\}$, and
satellite galaxies, $N^{\rm sat}(m)$.  A central galaxy is placed at
the centre of the halo. If there is a galaxy \changed{inhabiting} a
halo, there is always one central galaxy. \changed{This is used in the
following relations.} Satellite galaxies are distributed according to
a NFW profile with concentration parameter $c^\prime=f c$, where $c$
is the concentration of the dark matter (Appendix
\ref{sect:dmhaloes}). For $f>1$ galaxies populating the haloes are
more concentrated than the dark matter, while for $f<1$ galaxies are
less concentrated than the dark matter.

The halo mass dependence of the mean HOD is assumed to be
\begin{eqnarray}\label{eq:nmean}
  \Ave{N(m)}&=&\Ave{N^{\rm cen}(m)}+\Ave{N^{\rm sat}(m)}\;,\\
  \Ave{N^{\rm cen}(m)}&=& \frac{1}{2}\left[1+{\rm erf}{
  \left(\frac{\log{m}-\log{m_{\rm
  min}}}{\sigma_{\log{m}}}\right)}\right]\;,\\ \Ave{N^{\rm
  sat}(m)}&=&\Ave{N^{\rm cen}(m)}
  \left(\frac{m-m_0}{m^\prime}\right)^\epsilon {\rm H}(m-m_0)\;,
\end{eqnarray}
where 
\begin{equation}
  {\rm erf}(x)=\frac{2}{\sqrt{\pi}}\int_0^x\d t\,{\rm e}^{-t^2}
\end{equation}
is the error function. 

We would like to keep the number of parameters, required to explain
the data, as small as possible. For that reason, we set $m_0=m_{\rm
min}$ because we found that a free $m_{\rm min}$ does not
significantly improve our fits. An additional parameter $f\ne1$, on
the other hand, that gives some freedom in the shape of the density
profiles of blue and red galaxies yields improved fits and is
therefore included.

Note that due to the previous definition of $\ave{N(m)}$ we can still
have (central) galaxies for \mbox{$m\le m_{\rm min}$}, as
\mbox{$1+{\rm erf}(x)\ne0$} for \mbox{$x<0$}. Other authors prefer to
define a hard cut-off for $N(m)$, as for instance in
\citet{2006A&A...457..145P}. 

\subsubsection{Number of galaxy pairs \changed{of same sample}}
\label{sect:lambdadef}
\changed{In the original model, the fluctuation in the number of
satellites is assumed to be Poisson \citep[referred to as ``Poisson
satellite'' model hereafter;][]{2004ApJ...609...35K}},
i.e. \mbox{$\ave{[N^{\rm sat}]^2}=\ave{N^{\rm sat}}^2+\ave{N^{\rm
sat}}$}. Their assumption completely fixes the variance of the
\emph{total number} of galaxies inside a halo to (we skip in the
following the arguments ``$(m)$'' to save space):
\begin{eqnarray}\nonumber
  \sigma^2(N)\!\!\!&=&\ave{N^2}-\ave{N}^2\\\label{eq:varpair}
  &=&\ave{N(N-1)}+\ave{N}(1-\ave{N})\\\nonumber
  &=&3\ave{N^{\rm sat}}+
  \ave{N^{\rm cen}}\left(1-2\ave{N^{\rm sat}}\right)- \ave{N^{\rm
  cen}}^2\;,
\end{eqnarray}
because the variance of the number of central galaxies is always
\begin{equation}
  \sigma^2(N^{\rm cen})=
  \ave{N^{\rm cen}}[1-\ave{N^{\rm cen}}]\;,
\end{equation}
owing to the fact that $N^{\rm cen}$ is only zero or one
\changed{(Bernoulli distribution}). In particular we have
\mbox{$\ave{[N^{\rm cen}]^2}=\ave{N^{\rm cen}}$}.

\begin{figure*}
  \begin{center}
    \epsfig{file=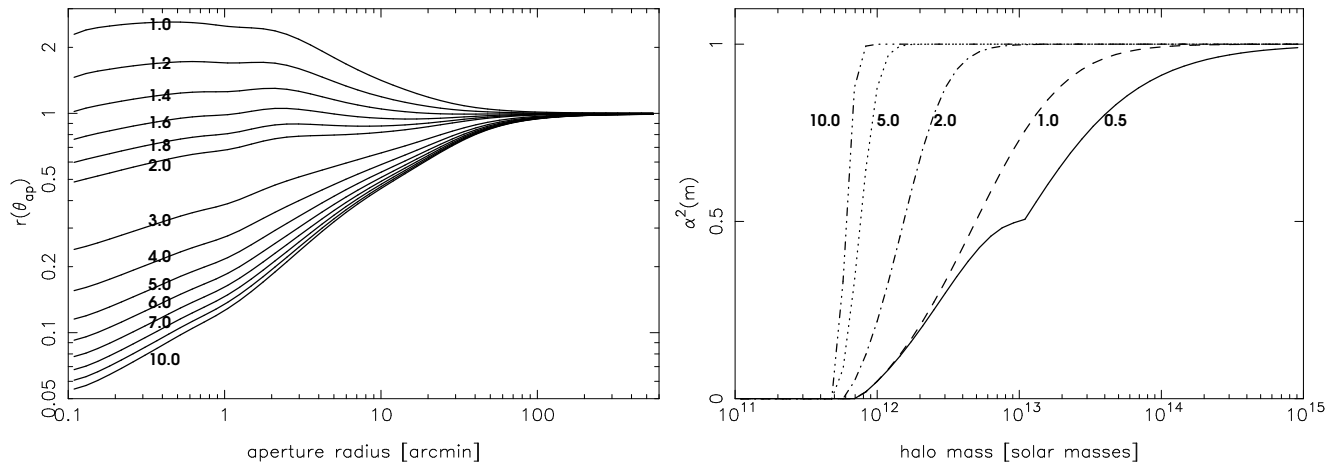,width=17.5cm,angle=0}
  \end{center}
  \caption{\label{fig:variance} Effect of the HOD variance parameter
  $\lambda$ (for both populations equally) on the cross-correlation
  function of two hypothetical populations of galaxies.  The
  parameters for the galaxy populations were chosen arbitrarily from
  \citet{2005ApJ...633..791Z}. As model parameters red galaxies have
  $\{10^{12.69}\msol,0.15,10^{12.94}\msol,10^{13.82}\msol,1.08\}$,
  blue galaxies have
  $\{10^{11.68}\msol,0.15,10^{11.86}\msol,10^{13}\msol,1.02\}$ for
  $\{m_{\rm min},\sigma_{\log{m}},m_0,m^\prime,\epsilon\}$ (see text
  for details). The galaxy concentration is set $f=1$ in both cases,
  both populations have no central galaxy component. The JHOD
  correlation factor is set to $R=+1$. Galaxies are uniformly
  distributed within the redshift range $z\in[0.2,0.4]$. \emph{Left
  panel:} Different lines correspond to a bias parameter
  $r(\theta_{\rm ap})$ for different $\lambda$ (numbers attached to
  solid lines). Top line: a variance of the HOD according to the
  \changed{``Poisson satellite'' model}, $\lambda=1$. Bottom line: a
  variance almost Poisson for all haloes, $\lambda=10$. \emph{Right
  panel:} The corresponding HOD variance parameter,
  $[\alpha^\prime(m)]^2$, of the blue galaxies for some of the lines
  in the left panel.}
\end{figure*}

The HOD variance or number of galaxy pairs of the \changed{``Poisson
satellite'' model} can, using the notation of
\citet{2001ApJ...546...20S}, be written as
\begin{equation}\label{eq:alphazheng}
  \ave{N(N-1)}=\alpha^2\ave{N}^2~\;;~\alpha^2\equiv\frac{\ave{N^{\rm
  sat}}(\ave{N^{\rm sat}}+2)} {(\ave{N^{\rm cen}}+\ave{N^{\rm
  sat}})^2}\;,
\end{equation}
where \mbox{$\ave{N(N-1)}$} is the mean number of galaxy pairs,
regardless of whether they are central galaxies or satellites.  The
variance (or number of galaxy pairs, see Eq. \Ref{eq:varpair}) becomes
Poisson for haloes $m$ with $\alpha=1$, sub-Poisson for $\alpha<1$ and
super-Poisson for $\alpha>1$. In the \changed{``Poisson satellite''
model} $\alpha$ is a function increasing slowly from zero near
$m=m_{\rm min}$ to unity for large $\ave{N(m)}$. See the dashed line
in right panel of Fig. \ref{fig:variance}.

We found, however, that this mean number of galaxy pairs per halo
hardly reproduces the deep decrease in the bias parameter
$r(\theta_{\rm ap})$ of our red and blue galaxy sample (see Section
\ref{sect:relbiasresults}), presumably because it becomes Poisson at
too large $m$. 

For that reason, we relax the assumption of the \changed{``Poisson
  satellite'' model} by introducing another model parameter,
  $\lambda$, that delays the onset of a Poisson variance in $N(m)$,
  $\lambda<1$, or accelerates it, $\lambda>1$. We achieve this by
  keeping the shape of $\alpha(m)$ as in the \changed{``Poisson
  satellite'' model}, Eq. \Ref{eq:alphazheng}, but rescaling the
  (log)mass scale, giving us a new $\alpha^\prime(m)$:
\begin{equation}
  \alpha^\prime(m)\equiv\alpha\!\left(m\times[m/m_{\rm min}]^{\lambda-1}\right)\;.
\end{equation}
Thus, for $\lambda=1$ our model uses exactly the same HOD variance
that is postulated in the original satellite model. Compare this to
the parametrisation of \citet{2001ApJ...546...20S}, where $\alpha(m)$
is postulated to increase linearly with $\ln{m}$ from $m_{\rm min}$
onwards.

The number of satellite pairs, needed for the power spectra integrals,
is then calculated from $\alpha^\prime(m)$ via
\begin{equation}\label{eq:satpairs}
  \ave{N^{\rm sat}(N^{\rm sat}-1)}=
    [\alpha^\prime]^2\ave{N}^2-2\ave{N^{\rm sat}}\;,
\end{equation}
invoking the relation between $\alpha(m)$, now substituted by the
rescaled $\alpha^\prime(m)$, and the mean number of galaxy
pairs. Following from this, the variance of the number of satellites
is consequently
\begin{equation}\label{eq:satvar}
  \sigma^2(N^{\rm sat})=
    [\alpha^\prime]^2\ave{N}^2-
    \ave{N^{\rm sat}}(\ave{N^{\rm sat}}+1)\;.
\end{equation}
The variance of the total number of galaxies, required for models
lacking central galaxies, is simply
\begin{equation}\label{eq:nvar}
  \sigma^2(N)=[\alpha^\prime]^2\ave{N}^2+\ave{N}(1-\ave{N})\;.
\end{equation}

The variance parameter $\alpha(m)$ cannot be reduced arbitrarily,
though, as a too small variance (too sub-Poisson) may be in conflict
with the mean number of galaxies. Put in other words, the number of
pairs, Eqs. \Ref{eq:alphazheng} and \Ref{eq:satpairs}, and the
variances, Eqs. \Ref{eq:satvar} and \Ref{eq:nvar}, have to be larger
or equal than zero, i.e.
\begin{equation}
  [\alpha^\prime]^2\geq
  {\rm max}\!\left\{\frac{2\ave{N^{\rm sat}}}{\ave{N}^2},
  \frac{\ave{N^{\rm sat}}(1+\ave{N^{\rm sat}})}{\ave{N}^2},1-\frac{1}{\ave{N}}\right\}\;.
\end{equation}
If $[\alpha^\prime]^2$ is smaller than the r.h.s., we set
$[\alpha^\prime]^2$ equal to the r.h.s. As $\alpha^\prime$ is usually
an increasing function with halo mass $m$, this resetting is not
necessary for $\lambda\geq1$ but may be important if we try to delay
Poisson statistics compared to the \changed{``Poisson satellite''
model}.

The effect of $\lambda$ on the bias parameter $r(\theta_{\rm ap})$ is
demonstrated in Fig. \ref{fig:variance} for some particular examples.
It is remarkable to see that an early (low $m$) Poisson variance in
the galaxy number suppresses the cross-power towards smaller
$r(\theta_{\rm ap})\lesssim1$, while a late Poisson variance (higher
$m$) yields higher values for $r(\theta_{\rm ap})\gtrsim1$ on small
scales. Therefore, the bias parameter $r(\theta_{\rm ap})$ promises to
be a probe for the variance in the HOD.

Finally, with this parametrisation each galaxy sample has six,
$\{m_{\rm min},\sigma_{\log{m}},m^\prime,\epsilon, f, \lambda\}$,
different parameters. Moreover, we have one more additional parameter
which expresses the correlation within the JHOD of blue and red
galaxies.

\changed{Based on this, we consider three different flavours of the
model -- detailed in the following sections -- which permit red or/and
blue galaxies as central galaxies and always have red and blue
satellites. If a sample has no central galaxy, we still use
$\ave{N(m)}$ as in Eq. \Ref{eq:nmean} and $\ave{N(N-1)}$ as in
Eq. \Ref{eq:alphazheng} (with rescaled $\alpha$) but distribute
\emph{all} $N(m)$ galaxies like ``satellites'' over the halo;
$\ave{N(m)}$ is the HOD of all galaxies in that case. This tests the
idea -- no central galaxy for a galaxy population but everything else
in the model unchanged -- if the data really requires a galaxy sample
to have a central galaxy.}

\begin{figure*}
  \begin{center}
    \epsfig{file=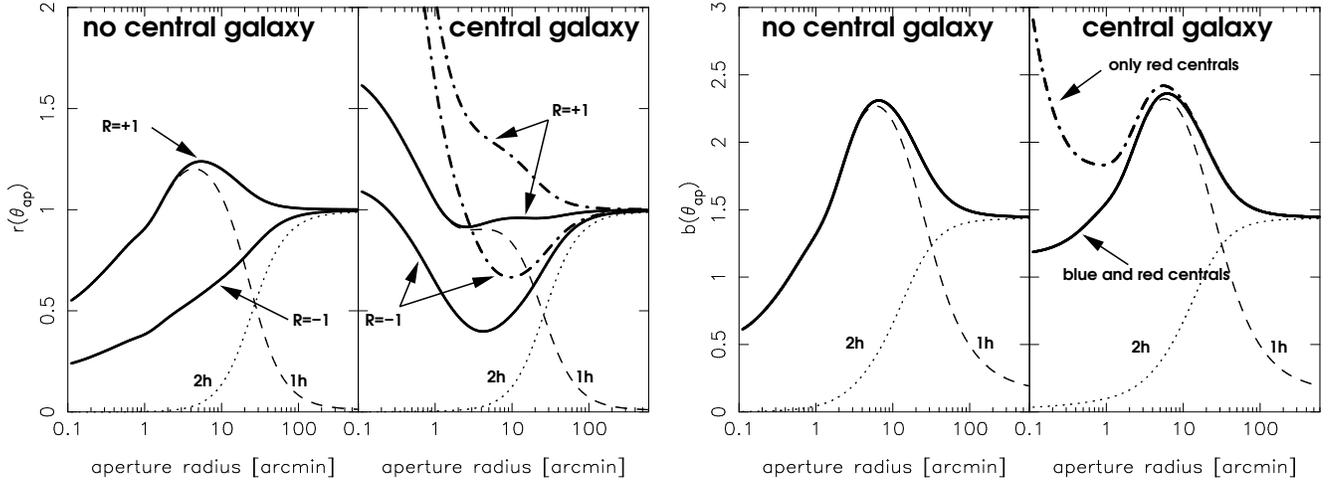,width=17.5cm,angle=0}
  \end{center}
  \caption{\label{fig:split} Example predictions for the relative
  linear stochastic bias of red and blue galaxies for three different
  scenarios (with/without central galaxy component, only red galaxies
  have a central component). The model parameters are chosen as in
  Fig. \ref{fig:variance} except that we fix $\lambda=3.0$ for both
  blue and red galaxies and we vary the JHOD parameter $R$.
  \emph{Left figure}: The bias parameter $r(\theta_{\rm ap})$ is
  plotted (solid lines) for two different JHOD-correlation factors
  (halo-mass invariant), $R=\pm1$. To separate contributions to the
  cross-power originating from the one-halo (1h; dashed line) and
  two-halo (2h; dotted line) term, the cross-power was recalculated
  for $R=+1$ considering only one contribution at a time. The solid
  thick line in the right panel assumes a model where the blue and red
  samples have a central component, whereas the thick dashed-dotted
  line is for a model where only the red sample has a central
  component (no one-halo/two-halo term splitting). \emph{Right
  figure}: Linear bias factor, $b(\theta_{\rm ap})$ (solid lines; red
  galaxies are more strongly clustered). The JHOD-correlation factor
  has no impact here. To estimate the importance of the one- and
  two-halo terms, $b(\theta_{\rm ap})$ was also recalculated with the
  red galaxy power spectrum consisting of only either the one- or
  two-halo term. The thick dashed-dotted line in the right panel
  corresponds to the model where only the red sample has a central
  component(no one-halo/two-halo term splitting).}
\end{figure*}

In Fig. \ref{fig:split} we show a few examples for the linear
stochastic bias between two galaxy populations based on the three
scenarios, detailed below. By splitting the contributions to the
cross-power spectrum, stemming from the one- and two-halo terms, the
figure shows in which regime we can expect the two different terms to
dominate. The transition is roughly at an aperture radius of $20\,\rm
arcmin$ for a mean redshift of $z\sim0.3$. \changed{This corresponds
to an effective projected (proper) scale of $5.6\,h^{-1}\rm Mpc$. For
comparison, the (comoving) virial radius of a NFW halo with
\mbox{$m\approx1.4\times10^{14}\,h^{-1}\msol$} is \mbox{$r_{\rm
vir}\approx0.8\,h^{-1}\rm Mpc$}, or \mbox{$r_{\rm
vir}\approx0.64\,h^{-1}{\rm
Mpc}\,\,(m/10^{14}\,h^{-1}\msol)^{0.33}$}.}  Also shown is the clear
impact of the JHOD correlation factor. Galaxies with tendency of
avoidance have a smaller cross-power.

\newcommand{\nn}[2]{\Ave{N^{\rm #1}_iN^{\rm #2}_j}} 
\newcommand{\n}[2]{\Ave{N^{\rm #1}_{#2}}}

\subsubsection{Number of cross-pairs for red and blue central galaxies}
The statistical cross-moments for a central galaxy scenario (central
galaxies for both or just one population) requires generally more
information on the JHOD than just first- and $2^{\rm nd}$-order
moments of \mbox{$P(N_i,N_j|m)$}. The correlations between satellite
numbers and central galaxy numbers have to be specified as
well. Namely, required are also \mbox{$\ave{N_i(m)|N_j(m)=0}$}, the
mean number of $i$-galaxies for haloes which do not contain any
$j$-galaxy, and \mbox{$P(N_i(m)N_j(m)=0)$}, which is the probability
to find either no $i$-galaxy \emph{or} no $j$-galaxy inside a halo of
mass $m$. We show this in Appendix \ref{sect:moments}.

To not unnecessarily increase the number of free model parameters in a
central-satellite galaxy scenario, we can make a reasonable
approximation, though. For every galaxy population having central
galaxies, we set \mbox{$N^{\rm cen}=1$} for halo masses $m$ where the
mean number of galaxies is \mbox{$\ave{N(m)}\geq1$}, and \mbox{$N^{\rm
sat}=0$} otherwise. This means that if the mean number of galaxies
belonging to a population is less or equal one, it is essentially only
the central galaxy we can find in those haloes. On the other hand, if
the mean galaxy number is larger than one, then there is always at
least a central galaxy.

Applying this approximation allows us to express the cross-moments of
the HOD entirely in terms of known variances, means and the JHOD
correlation factor $R_{ij}(m)$. In a scenario with blue and red
central galaxies, we hence distinguish four cases:
\begin{enumerate}
\item \changed{Case}: \mbox{$\ave{N_i}<1$}, \mbox{$\ave{N_j}<1$}\\
       Here, we have
       \begin{eqnarray}
	 \ave{N_i^{\rm sat}N_j^{\rm cen}}&=&\ave{N_j^{\rm sat}N_i^{\rm
	     cen}}=\ave{N_i^{\rm sat}N_j^{\rm sat}}=0\;.
       \end{eqnarray}
\item \changed{Case}:
      \mbox{$\ave{N_i}\geq1$}, \mbox{$\ave{N_j}<1$}\\ We
      will use:
      \begin{eqnarray}
	\nn{sat}{sat}&=&\nn{cen}{sat}=0\;,\\\nonumber
	\nn{sat}{cen}&=&\!\!\n{sat}{i}\n{cen}{j}+ R_{ij}\sigma(N_i^{\rm
	  sat}) \sigma(N_j^{\rm cen})\;.
      \end{eqnarray}      
\item \changed{Case}: \mbox{$\ave{N_i}\geq1$}, \mbox{$\ave{N_j}\geq1$}\\ Finally
    for richly populated haloes, we will have:
    \begin{eqnarray}
      \nn{cen}{sat}&=&\n{sat}{j}~\;;~\nn{sat}{cen}=\n{sat}{i}\;,\\\nonumber
      \nn{sat}{sat}&=&\!\!\n{sat}{i}\n{sat}{j}+ R_{ij}\sigma(N_i^{\rm
	sat}) \sigma(N_j^{\rm sat})\;.
    \end{eqnarray}
\end{enumerate}

\changed{For the sake of simplicity, this model unrealistically
  also assumes that red and blue galaxies have central galaxies
  simultaneously, which would mean for haloes hosting red and blue
  galaxies that we would find a red and blue galaxy at, or very close,
  to the centre. A more realistic model would add another rule that
  decides when (and with what probability) we have a red or a blue
  central galaxy, but never both at the same time.}

\changed{However, our approximation is fair as long as we rarely find
  haloes with one red \emph{and} one blue galaxies. As seen later on,
  our galaxy samples start to populate haloes at different
  mass-thresholds, blue galaxies from $\sim10^{11}\msol h^{-1}$ and
  red galaxies from $\sim10^{12}\msol h^{-1}$. This means a) haloes
  that have one red galaxy have typically more than one blue galaxy
  and b) haloes with one blue galaxy have typically no red galaxy. In
  the first case a), the auto-power spectrum of blue galaxies is
  dominated by the satellite terms so that placing one blue galaxy at
  the centre does not make much of a difference. The same is true for
  the red/blue cross-power. In the second case b), we have no red
  galaxy that could be placed at the centre together with a blue
  galaxy so that the problem does not arise in the first place. To
  further improve the approximation, we also set for the cross-power
  kernel $K^{\rm 1h}(k,m)$, Table \ref{kernels}, the cross-correlation
  $\ave{N^{\rm cen}_iN^{\rm cen}_j}=0$ which has to vanish if there is
  always only one central galaxy inside a halo.}

\subsubsection{Number of cross-pairs for no central galaxies}
We also explore the possibility that neither the red nor the blue
sample of galaxies have central galaxies. We do this by distributing
the central galaxies of the previously described model with the same
density profile as the satellite galaxies. Therefore, we have a mean
galaxy number according to Eq. \Ref{eq:nmean} with a variance as in
Eq. \Ref{eq:nvar}. The mean number of galaxy pairs is expressed by
$[\alpha^\prime]^2\ave{N}^2$. Moreover, for this scenario the
factorial cross- moments are straightforward, the approximation
described in the forgoing section is not necessary.  The
cross-correlation moment of the JHOD is thus:
\begin{equation}
  \Ave{N_iN_j}=
  \Ave{N_i}\Ave{N_j}+R_{ij}\sigma(N_i)\sigma(N_j)\;.
\end{equation}

\subsubsection{Number of cross-pairs for mixed model with only red central galaxies}
As last scenario we consider the possibility that only the red sample
has central galaxies, while blue galaxies are always satellites.
Within this model, blue galaxies are described according to the ``no
central galaxy'' scenario and red galaxies according to the previous
``central galaxy'' scenario.

Again, for cross-moments in principle we also needed to specify
\mbox{$\ave{N_j(m)|N_i(m)=0}$}, $j$ denotes blue galaxies and $i$ the
red galaxies (Appendix \ref{sect:moments}). One finds by using the
aforementioned approximation:
\begin{eqnarray}\nonumber
  \Ave{N_i^{\rm cen}N_j}&=&
  \Ave{N_i^{\rm cen}}\Ave{N_j}+R_{ij}
  \sigma(N_i^{\rm cen})\sigma(N_j)\;,\\
  \Ave{N_i^{\rm sat}N_j}&=&0\,
\end{eqnarray}
for \mbox{$\ave{N_i}<1$}, and
\begin{eqnarray}
  \Ave{N_i^{\rm cen}N_j}&=&\Ave{N_j}\;,\\
  \nonumber
  \Ave{N_i^{\rm sat}N_j}&=&
  \Ave{N_i^{\rm sat}}\Ave{N_j}+R_{ij}
  \sigma(N_i^{\rm sat})\sigma(N_j)
\end{eqnarray}
otherwise.

\subsection{Fitting the halo model to the data}
\label{sect:jhodresults}

\subsubsection{Method}
\changed{We now use all the results on the clustering of the red and
blue galaxy sample and their correlation to the dark matter density,
gathered on the foregoing pages, to constrain the parameters of our
halo model, see Sect. \ref{sect:jhodmodel}. The number densities of
blue and red galaxies for each redshift bin is added as additional
information (Table \ref{redbluefig}, bottom) to be fitted by the model
as well (Eq. \ref{eq:halonbar}).}

How can we relate the halo model 3D-power spectra to the observables,
which are projections onto the sky?  The mean tangential shear about
the lenses is, for the $i$-th redshift bin, related to the angular
cross-power spectrum between the lens number density and lensing
convergence, $P^{{\rm g}\kappa}_i(\ell)$, by
\citep[e.g.][]{2007A&A...461..861S}
\begin{equation}
  \Ave{\gamma_{{\rm t},i}(\theta)}=
  \int_0^\infty\frac{\d\ell\ell}{2\pi}P^{{\rm
  g}\kappa}_i(\ell)J_2(\ell\theta)\;.
\end{equation}
A similar relation connects the $\caln$-statistics to the angular power
spectrum of galaxy clustering, $P_{ij}(\ell)$, see
Eq. \Ref{nstat}.

To translate the 3D-model power spectra, Eq. \Ref{eq:modelpower}, to
the projected angular power spectra for a given redshift distribution
of lens galaxies, $p_{\rm l}(w)$, and source galaxies, $p_{\rm s}(w)$,
we use Limber's equation in Fourier space \citep{bas01}:
\begin{equation}
  P_{ij}(\ell)=\int_0^\infty\d w\left(\frac{p_{\rm l}(w)}{f_{\rm
  k}(w)}\right)^2P_{ij}\left(\frac{\ell}{f_{\rm k}(w)},w\right)
\end{equation} 
and
\begin{eqnarray}\nonumber
  P_i^{{\rm g}\kappa}(\ell)&=& \frac{3H_0^2}{2c^2}\Omega_{\rm m}
  \int_0^\infty\!\!\!\!\d w\frac{p_{\rm
  l}(w)\overline{W}(w)}{a(w)f_{\rm
  k}(w)}P_{i,j=0}\left(\frac{\ell}{f_{\rm k}(w)},w\right)\;,\\
  \overline{W}(w)&\equiv&\int_w^\infty\d w^\prime p_{\rm s}(w^\prime)
  \frac{f_{\rm k}(w^\prime-w)}{f_{\rm k}(w^\prime)}\;.
\end{eqnarray} 
We denote the Hubble constant, the vacuum speed of light, the
cosmological scale factor by $H_0$, $c$ and $a(w)$, respectively.  The
second argument $w$ in the 3D-power spectrum, $P_{ij}(k,w)$, is used
to express a possible time-evolution of the power as function of
comoving radial distance.  

We expect the evolution of the 3D-power spectra within the lens galaxy
redshift bins to be moderate so that we neglect the time-evolution
over the range of a redshift bin. The 3D-power spectra are computed
for a radial distance $w=(w_1+w_2)/2$, where $w_1$ and $w_2$ are the
distance limits of the redshift bin. \changed{This particular $w$ is
justified by the redshift probability distributions inside the
$z$-bins, Fig. \ref{fig:pzlenses}, which are relatively symmetric
about the mean.}

Again, we employ the MCMC-method to trace out the posterior likelihood
function of our halo-model parameters. The $\caln$-statistics and the
GGL signal are put together to constrain the model.  In general, the
JHOD correlation factor $R(m)$ (we have just one: between the red and
blue sample) is a function of the halo mass. As the effect of $R(m)$
becomes negligible for small relative fluctuations in the halo
occupation number of galaxies (large $m$), we assume the same
correlation parameter for all $m$, which is, consequently, mainly
constrained by haloes with a small number of galaxies ($m\sim m_{\rm
min}$). For the scope of this paper, where we have relatively small
galaxy samples and relatively large uncertainties in the clustering
statistics (compared to SDSS, for example) this is an acceptable
approximation. For future studies investigating a mass-dependence of
the JHOD correlation parameter may be interesting and feasible.

We confine the parameter space of the model (flat priors) to
\mbox{$10^{6}\msol\le (m_{\rm min},m^\prime)\le10^{16}\msol$},
\mbox{$10^{-3}\le\sigma_{{\rm log}m}\le1$}, \mbox{$0\le\epsilon\le2$},
\mbox{$|R|\le1$}, \mbox{$10^{-1}\le f\le 3$} and
\mbox{$0.9\le\lambda\le10$}. Three model scenarios are fitted
separately: i) both the blue and red populations have central
galaxies, ii) no central galaxies, and iii) only the red sample has
central galaxies.

In order to possibly discriminate between the three scenarios we
estimate the Bayesian evidence for each scenario in every redshift bin
(Appendix \ref{sect:bayesian}). This method is a more sophisticated
approach for model discrimination than a ``simple'' $\chi^2$
comparison of a best-fit because, rather than looking merely at the
height of the (posterior) likelihood function, the width of the
likelihood function is taken into account as well.

\begin{figure*}
  \begin{center}
    \epsfig{file=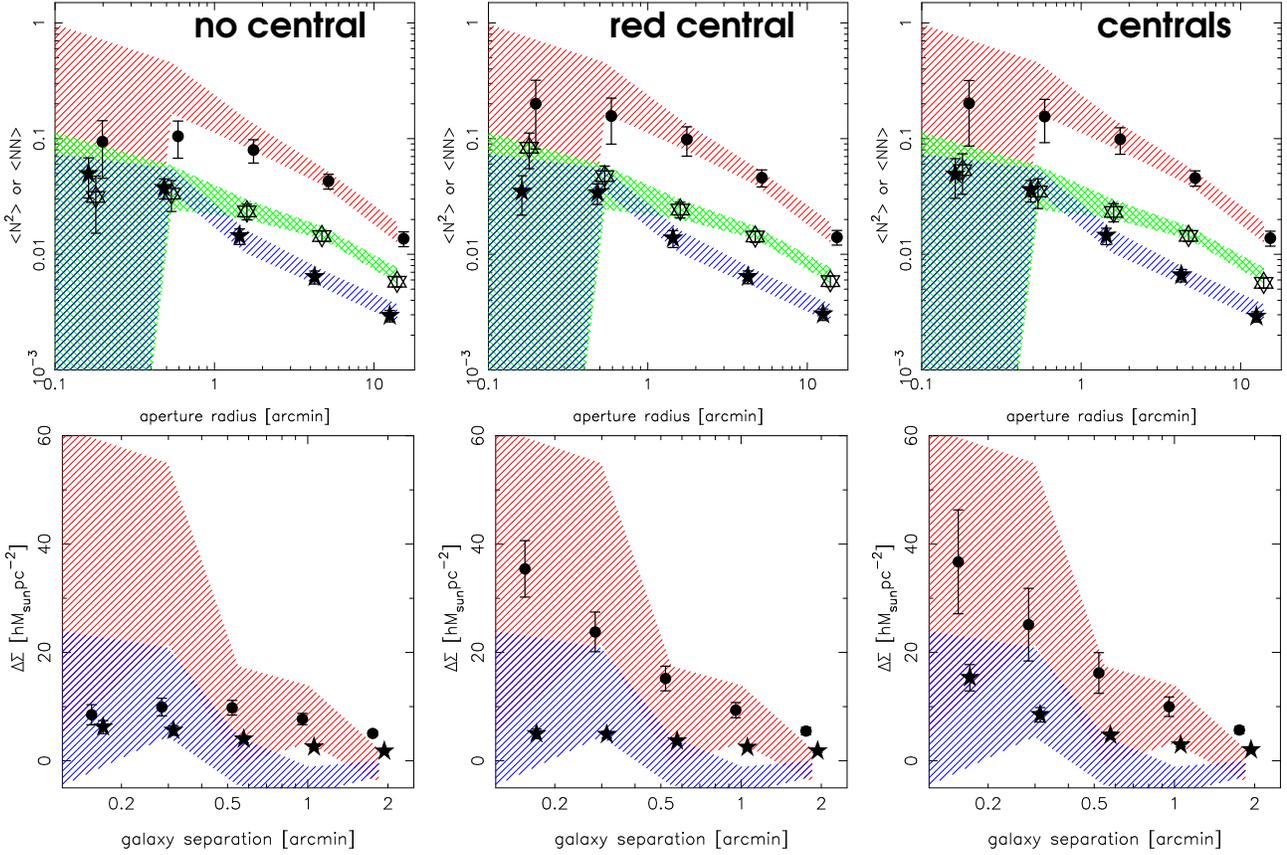,width=17cm,angle=0}
  \end{center}
  \caption{\label{fig:scenariosbin3} \changed{Average fit, compared
  to the observation, of the $\caln$-statistics (\emph{first row};
  points: variance of red galaxies, filled stars: variance of blue
  galaxies, open stars: cross-correlation) and GGL (\emph{second row};
  filled stars: blue galaxies, points: red galaxies) as predicted by
  the three different halo-model scenarios (first column: no central
  galaxies, second column: red central galaxies, third column: blue
  and red central galaxies) for the redshift bin $\bar{z}=0.7$. The
  model fits (average over the MCMC tracks) and their
  $1\sigma$-variance are denoted by points and errorbars, the shaded
  areas bracket the $1\sigma$-range in \combo. For the GGL panels
  shaded areas are the combined constraints from all redshift
  bins. The corresponding model predictions for the number density of
  red galaxies are (in units of $h^3\rm Mpc^{-3}$)
  $11\pm8\times10^{-3}$ (no central), $8\pm4\times10^{-3}$ (red
  central), $9\pm6\times10^{-3}$ (red and blue centrals) and for blue
  galaxies $51\pm15\times10^{-3}$ (no central), $57\pm18\times10^{-3}$
  (red central), $65\pm21\times10^{-3}$ (red and blue centrals).}}
\end{figure*}

\subsubsection{Results}
Fig. \ref{fig:scenariosbin3} shows the model fits to the data
belonging to the redshift bin \mbox{$\bar{z}=0.7$}. This redshift bin
was chosen since it has the best signal-to-noise in our data.

\begin{table}
  \begin{center}
    \caption{\label{tab:mcmc2} Quality of the halo model fits and a
    Bayesian comparison of the three different models describing the
    same data. The $\chi^2/{\rm dof}$ per degree of freedom (${\rm
    dof}=26-13=13$) is for the maximum-likelihood fit to the data. By
    $\Delta\ln{E}$ we mean the difference of the Bayesian
    (log)evidence for a particular scenario relative to the scenario
    with the highest evidence at the same redshift.}
    \begin{tabular}{c|cc|r}
      $\bar{z}$&$\chi^2/{\rm dof}$&$\Delta\ln{E}$&model type\\
      \hline
      0.3&0.51&0.00&no central galaxies  \\
      0.3&0.56&0.44&red central galaxies \\
      0.3&0.45&0.57&blue and red centrals\\
      \hline
      0.5&1.25&0.00&no central galaxies  \\
      0.5&1.20&0.86&red central galaxies \\
      0.5&1.24&0.88&blue and red centrals\\
      \hline
      0.7&0.57&0.00&no central galaxies  \\
      0.7&0.50&1.02&red central galaxies \\
      0.7&0.60&1.19&blue and red centrals\\
      \hline
      0.9&0.59&0.00&red central galaxies \\
      0.9&0.55&0.07&blue and red centrals\\
      0.9&0.62&0.31&no central galaxies  \\
    \end{tabular}
  \end{center}
\end{table}

Note that the points (model prediction) in this figure are the model
averages along the MCMC tracks and not the best-fitting models
(minimum $\chi^2$). The $1\sigma$-standard deviation of the fits is
denoted by error bars.  The MCMC-average is useful to study how well a
model fits the data in general and where the main problems of the
model in describing the data are located.

We gather from this figure that, in general, qualitatively all three
model scenarios appear to provide a pretty good description of the
data. The main differences appear for $\theta_{\rm
ap}\lesssim1^\prime$ ($\caln$-statistics) and
$\theta\lesssim0^\prime\!\!.5$ (GGL); the differences in the $\cal
N$-statistics are small, though.  The scenario with no central
galaxies at all is bound to systematically smaller values for the
$\caln$-statistics and, thus, has small difficulties in explaining the
clustering statistics of red galaxies for smaller scales.  Since
statistical uncertainties grow large for small scales, those
difficulties are not too significant for our data.

Three other issues can be identified:
\begin{itemize}
  \item \changed{Although the galaxy clustering and the GGL is well
  described by the models, there is only moderate agreement between
  the predicted and observed mean number densities of blue galaxies:
  the predicted numbers are higher. The measured value (Table
  \ref{redbluefig}) is, however, still within the $2\sigma$-scatter of
  $\bar{n}_{\rm blue}$ of the Markov chains. For example, for
  \mbox{$\bar{z}=0.7$} the model predicts consistently for all
  scenarios \mbox{$\bar{n}_{\rm blue}\sim
  (57\pm18)\times10^{-3}\,h^{3}\rm Mpc^{-3}$}, whereas the data
  estimate is \mbox{$\bar{n}_{\rm blue}=(18\pm3)\times10^{-3}\,h^3\rm
  Mpc^{-3}$}. This could point towards an inaccuracy of the halo model
  or/and to a systematic underestimate of $\bar{n}_{\rm blue}$
  provided by the $V_{\rm max}$-estimator. The observed numbers of red
  galaxies are always within the $1\sigma$-scatter of the Markov
  chains although the means of the chains are always larger than the
  observed values, which again could be indicative of an
  overprediction by our halo-model. Tensions between the halo model
  predictions for galaxy clustering and number densities were also
  noticed by other studies, such as \citet{2008ApJ...685L...1Q}.}
  \item The GGL-signal of the blue galaxies always conflicts the data
    beyond \mbox{$\theta\gtrsim0^\prime\!.5$} (too high) because the
    observed $\Delta\Sigma$ quickly drops to essentially negative in
    that regime (if the GGL of all redshift bins is
    combined). Considering that most of the other model predictions
    fit quite well and that, to the knowledge of the authors, no
    negative GGL-signal for these galaxy separations has been found in
    the literature, this conflict could very well be a hint towards
    systematics in the lensing data undiscovered so far.
  \item The model without red central galaxies is in mild conflict with
    the GGL measurement for the smallest separations: a galaxy centred
    on the halo centre boosts the GGL signal, which is arguably favoured
    by the data (when the GGL-signal of all redshift bins is combined).
\end{itemize}

Table \ref{tab:mcmc2} lists the (reduced) $\chi^2$ of the maximum
likelihood fits and the estimated Bayesian evidence of the three
different model scenarios. All scenarios at all redshifts give good
fits to the data. The worst fits, $\chi^2/{\rm dof}\sim1.25$, are for
the redshift bin $\bar{z}=0.5$ which may be related to the sudden drop
of the $\caln$-statistics for the largest aperture radius (upper right
panel in Fig. \ref{combonsqd}). Since this $\chi^2$ still has a
probability of $\sim20\%$, we can consider it as a statistical fluke.

\changed{Except for $\bar{z}=0.7$, where we find weak evidence for a
  model without central galaxies, the Bayesian model discrimination
  does not prefer any particular model. Combining the evidence of all
  redshift bins (the Bayesian evidence just sums up), assuming independent
  statistical information, however yields substantial evidence
  (\mbox{$\Delta\ln{E}\approx2$}) for a model without central
  galaxies. On the other hand, by looking at
  Fig. \ref{fig:scenariosbin3} we concluded that a model without
  central galaxies has some problems describing the data on the
  smallest scales. Moreover, our central galaxy models have always,
  except for $\bar{z}=0.3$, a better $\chi^2$. All of which taken
  together makes us suspicious, if the Laplacian approximation that is
  used to compute the Bayesian evidence is really accurate enough for
  our work. Our model comparison therefore seems to be inconclusive
  and an improved statistical analysis or a larger data set will have
  to revisit this question in the future.}

Table \ref{tab:mcmc} (after the bibliography) lists the constraints on
the halo-model parameters for all redshifts and all scenarios. For an
parameter average over all redshifts (bottom block of table), we
combine the 1D-probability density functions (PDFs), $P_i(p_j)$, as
estimated from the MCMCs, of all parameters, $p_j$, for all redshifts
$i$ to obtain a total 1D-posterior of $p_j$:
\begin{equation}
  P(p_j)=\prod_iP_i(p_j)\;.
\end{equation}
From this total posterior we derive the mean and r.m.s.-variance of
every parameter. Note that for top-hat and equal $P_i$'s one has
$P=P_i$, meaning combining information does not improve anything in
this case.

First of all, we find that the fits seem to be robust with respect to
the three different model scenarios, \changed{with $R$, $f$ and maybe
$m^\prime$ being the only possible exceptions.}

For all aforementioned parameters, we do not see a clear trend with
redshift, \changed{maybe with $m_{\rm min}$ being an exception which
(the median) is increasing slightly in $\bar{z}=0.9$ but still is
consistent with the rest. The redshift-combined result of this
parameter is \mbox{$m_{\rm min}=10^{12.1\pm0.2}h^{-1}\msol$} for red
and \mbox{$m_{\rm min}=10^{11.2\pm0.1}h^{-1}\msol$} for blue
galaxies. Therefore, blue-cloud galaxies clearly populate smaller
haloes than red-sequence galaxies. In fact, this is the halo-model
explanation for the different clustering strengths of red and blue
galaxies.}

\changed{The combined result for the parameter pair $m^\prime$ and $\epsilon$,
describing the galaxy occupancy as function of halo mass beyond
$m_{\rm min}$, is roughly for both red and blue galaxies
\mbox{$m^\prime=10^{13.0\pm0.4}h^{-1}\msol$} and
\mbox{$\epsilon=1.1\pm0.2,1.3\pm0.2$} for red and blue galaxies,
respectively.}

The least constrained parameter in our analysis is clearly
$\sigma_{\log{m}}$ which has for all redshifts
$0.5\pm0.2$. \changed{This is essentially} what one would expect from
a top-hat PDF, non-vanishing between $0\ldots1$. Therefore, our data
does not add information that significantly improves our
prior. Combining all redshifts shrinks the $1\sigma$-confidences
somewhat, though, because the individual 1D-PDFs are not completely
flat.

Apparently only little more information is added to $\lambda$, the HOD
variance parameter, by the data. With a top-hat prior between
$0.9\ldots10$ we would expect as constraint $\lambda=5.5\pm2.5$, which
is roughly what we find for the individual redshift bins,
\changed{excluding $\lambda$ of blue galaxies for the model with red
central galaxies only}. However, the 1D-PDFs are not completely flat
preferring some regions in parameter space, too. In particular, they
exclude values near $\lambda\sim1$ which is the HOD variance in the
\changed{``Poisson satellite'' model}. As discussed earlier, this is
because $\lambda=1$ cannot explain the observed deep drop in
$r(\theta_{\rm ap})$ for small scales. Quantitatively, we infer from
the redshift-combined PDF of $\lambda$ that we can decisively exclude
values less than $\lambda=2$ for red galaxies and $\lambda=3$ for blue
galaxies with $95\%$ confidence. \changed{This fits to the findings of
\citet{2005MNRAS.361..415C} that found no indications of a
sub-Poissonian variance, $\alpha(m)<1$, in their analysis of red and blue
galaxies.}

\changed{For the combined concentration parameter of red and blue
galaxies we find values that depend slightly on the adopted
scenario. In a model with no central galaxies, red galaxies require a
concentration larger than that of dark matter, $f_{\rm
red}=1.9\pm0.5$, which, however drops to $f_{\rm red}\approx1.3\pm0.5$
if a central red galaxy is allowed. Our conclusion is that the data
demands a centrally concentrated spatial distribution of red galaxies
either by a larger $f_{\rm red}$ or by a central galaxy. Conversely,
for blue galaxies we find for no central galaxies, $f_{\rm
blue}=1.0\pm0.4$, consistent with the dark matter distribution, but
distributions flatter than dark matter, $f_{\rm blue}=0.6\pm0.3$, if
central galaxies are present.}

\changed{The JHOD correlation factor, $R$, turns out to be hard to
measure. For the individual redshift bins we find little improvement
compared to the flat prior. For all $z$-bins combined, we find
$R=+0.1\pm0.2$ (no central galaxies/only red central galaxies) and
$R=+0.5\pm0.2$ (blue and red central galaxies). This implies that the
number of red and blue galaxies are uncorrelated in the first case and
slightly positively correlated in the latter case.}

Inside Fig. \ref{redbluebias} we plotted the bestfit solutions to the
relative linear stochastic bias of the red and blue population found
along the MCMC tracks for a mean redshift of $\bar{z}=0.7$. The halo
model reproduces the scale-dependence of bias factor and correlation
factor at that redshift well (note that the shaded confidence regions
are combinations of all redshift bins). All three scenarios reveal
very similar trends with most differences for very small scales which,
however due to measurement noise (galaxy shot noise), are not well
constrained. If we use the halo model and bestfit parameters to
extrapolate the bias parameters to large scales we find a bias factor
between \mbox{$b\sim1.46-1.58$} depending on the particular
scenario. Those values reconcile our measurements at relatively small
scales with results from various other studies that measured galaxy
biasing on larger scales (see
Sect. \ref{sect:relbiasresults}). Furthermore, it underlines the need
for both small and large scale measurements of galaxy clustering in
order to discriminate different halo-model scenarios. In that context,
we also would like to point to Fig. \ref{fig:scenariosbin3}, lower
row. Here we can easily see that the three halo-model scenarios
predict clearly different GGL for galaxy separations smaller than
\mbox{$\sim12^\pprime$}, \changed{corresponding to a physical scale of
$\sim60\,h^{-1}\rm kpc$. This shows that GGL has an important model
discriminating power on those scales. However, the halo model outlined
here may be inaccurate on exactly those scales as it does not include
the effect of lenses hosting individual haloes inside their parent
halo \citep{2003MNRAS.345..529S}.} Unfortunately, the statistical
uncertainties in our GGL measurements do not allow to fully exploit
this potential.

\section{Summary}

For this paper we studied the clustering and, in particular, the
relative clustering of red sequence galaxies and blue-cloud galaxies
in \combo~ (fields: S11, A901, CDFS) inside four redshift bins up to a
redshift of $z\sim1$. The two samples were separated by applying a
redshift-dependent cut along the red-sequence. An additional cut was
applied to assure that red and blue galaxies at all redshifts have
roughly the same colour-dependent $M_V$-limits. The red sample has
\mbox{$\ave{M_V}=-20.0\pm0.1\,\rm mag$}, the blue sample has
\mbox{$\ave{M_V}=-18.8\pm0.1\,\rm mag$} ($h=1.0$).

\changed{By looking at the spatial correlation function of the
samples, we found for all redshifts combined a correlation length of
\mbox{$r_0=5.5\pm0.9,3.0\pm0.4\,h^{-1}\rm Mpc$} for the red and blue
galaxies, respectively. The corresponding power-law indices of the
spatial correlation function were \mbox{$\delta=0.85\pm0.10,
0.65\pm0.08$}. A significant evolution of these parameters with
redshift was not found.}

Parameterising the relative biasing of the red and blue sample in
terms of the linear stochastic bias, we measured for all redshifts
combined that the bias factor, $b(\theta_{\rm ap})$, is slightly
scale-dependent within a range of aperture radii of \mbox{$\theta_{\rm
ap}=6^\pprime-20^\prime$}, varying between $b\sim1.7-2.2$. We found
that the second parameter, $r(\theta_{\rm ap})$ -- quantifying the
correlation of galaxy number density fluctuations as function of scale
--, is scale-dependent, too. It drops from a value close to unity at
larger scales of \mbox{$\theta_{\rm ap}\sim20^\prime$} to
\mbox{$r\sim0.6\pm0.15$} at \mbox{$\theta_{\rm ap}\sim6^\pprime$}. An
aperture radius of \mbox{$\theta_{\rm ap}=10^\prime$} corresponds to a
proper spatial scale of $2.8,3.8,4.5,4.8\,h^{-1}\rm Mpc$ at the
redshifts $z=0.3,0.5,0.7,0.9$, respectively. The measurements
emphasise the different clustering of the red and blue sample at all
redshifts, but do not exhibit a clear evolution with redshift beyond
the statistical uncertainties.

We also looked at the mean tangential ellipticity of a population of
faint background galaxies as function of separation from red and blue
galaxies (GGL). For this analysis, shear catalogues with a mean source
redshift of $\bar{z}=0.78$ from the GaBoDS were taken. The GGL-signal
detected corresponds to a projected differential surface mass density
of \mbox{$\Delta\Sigma=35\pm25\,h\rm\msol pc^{-2}$}, red galaxies, and
\mbox{$\Delta\Sigma=16\pm10\,h\rm\msol pc^{-2}$}, blue galaxies, at a
galaxy-galaxy separation of $\sim12^\pprime$ (roughly
\mbox{$\sim60\,h^{-1}\rm kpc$}). This indicates that the red galaxies are
either typically more massive than blue galaxies, or are residing
inside a matter richer environment than the blue galaxies.

A large part of the paper discussed a dark-matter halo based model
that was employed to describe the angular clustering, including the
cross-correlation function of clustering, the GGL-signal and the
(ratio of) number densities of the red and blue galaxy sample
simultaneously. Due to large statistical errors for the GGL, the
GGL-signal mainly served as an upper limit for a model predicted
signal. 

We used three different variants of our halo-model, all having the
same number of free parameters, to fit the data: i) neither the red
nor the blue population have central galaxies, ii) only the red sample
has central galaxies, and iii) both the red and blue sample can have
central galaxies and there is always one central galaxy. A Bayesian
method of model discrimination was performed to decide which model at
which redshift may be most suitable in explaining the
data. \changed{The model discrimination was inconclusive, a more
accurate treatment or a larger data set is required.}  In this
context, we pointed out that GGL at small separations is most
sensitive to the presence or absence of a central galaxy.

Describing the cross-correlation function required the extension of
the halo-model descriptions currently available in the literature by
at least one additional parameter \changed{(see
\citet{2003MNRAS.339..410S, 2002MNRAS.332..697S} which is equivalent
to our approach only if $R(m)=0$)}. The extension is necessary for a
full parametrisation of the $2^{\rm nd}$-order joint HOD of two galaxy
populations. We called this parameter the correlation factor of the
joint HOD of two galaxy populations, $R(m)$. It expresses the tendency
of different galaxy types to avoid ($R(m)$ close to minus unity) or
\changed{attract} each other ($R(m)$ close to plus unity) inside/into
a same dark matter halo of mass $m$. A vanishing $R(m)$ indicates
uncorrelated halo-occupation numbers. The $2^{\rm nd}$-order
cross-correlation function of two galaxy samples, or equivalently
$r(\theta_{\rm ap})$, is most sensitive to $R(m)$ for small
separations where the one-halo term is dominating ($\theta_{\rm
ap}\lesssim20^\prime$ for $\bar{z}=0.3$). In principle, $R(m)$ is
halo-mass dependent but its impact on the cross-correlation function
becomes negligible for haloes with large galaxy occupation numbers
(large $m$). Therefore, observations mainly constrain $R(m)$ for
smaller haloes. \changed{The measurements of $R$ for the individual
$z$-bins yielded only little constraints. Combining all redshifts we
found \mbox{$R=+0.5\pm0.2$} (positive correlation of galaxy numbers)
for iii) and \mbox{$R=+0.1\pm0.2$} for i) and ii) (no correlation).}

We found it necessary to add another new degree of freedom, $\lambda$,
to the model in order to explain the observed, towards smaller scales
declining sub-unity bias parameter $r(\theta_{\rm ap})$. The parameter
$\lambda$ regulates how quickly a HOD of galaxy numbers populating a
dark-matter halo assumes a Poisson variance, \changed{see
Sect. \ref{sect:lambdadef} for an explanation}. We demonstrated that
the cross-correlation function of galaxy clustering is quite sensitive
to $\lambda$: $r(\theta_{\rm ap})$ seems to reach values substantially
smaller than unity only for $\lambda>1$.  With all models i)-iii) we
decisively excluded, for red and blue galaxies, a value of
$\lambda\lesssim2$ (red) and $\lambda\lesssim3$ (blue) with $95\%$
confidence. \changed{This rules out with high confidence, at least for
our red and blue galaxy samples, a model like that of
\cite{2004ApJ...609...35K} (``Poisson satellites'') for the mean
number of galaxy pairs inside a halo.}

\changed{The halo model parameters of the red and blue sample mostly
differ for $m_{\rm min}$. We concluded that the
average of the mass-scale for \mbox{$z=0\ldots1$} is \mbox{$m_{\rm
min}=10^{12.1\pm0.2}h^{-1}\msol$} for red and \mbox{$m_{\rm
min}=10^{11.2\pm0.1}h^{-1}\msol$} for blue galaxies.}

\changed{Finally, we presented evidence that red galaxies are more
concentrated towards the halo centre than the dark matter, which
either has to be achieved by a central red galaxy or a larger
concentration parameter \mbox{$f_{\rm red}=1.9\pm0.5$}. The
distribution of blue galaxies is consistent with the dark matter
distribution inside a halo for i), but flatter with \mbox{$f_{\rm
blue}=0.6\pm0.3$} for ii) or iii).}

As an outlook, we would like to mention that the joint HOD of two
galaxy populations is also probed by a new statistics, third-order
galaxy-galaxy lensing (red and blue galaxies as lenses), that are
outlined in \citet{2005A&A...432..783S} and have been recently
detected for the first time in \citet{2008A&A...479..655S}. Combining
those statistics with the second-order cross-correlation function in
forthcoming surveys promises to set better constraints on the JHOD
correlation factor.
\section*{Acknowledgements}

We would like to thank Michael J.I. Brown, Peder Norberg and Alan
Heavens for very helpful discussions. \changed{We are also grateful to
the anonymous referee of this paper for his many useful suggestions.}
This work was supported by the European DUEL Research-Training Network
(MRTN-CT-2006-036133). Patrick Simon further acknowledges support by
PPARC, Hendrik Hildebrandt also received funding from the
Deutsche-Forschungsgemeinschaft (DFG) under the project
ERB327/2-1. This work was further supported by the DFG under the
project SCHN 342/6--1 and by the Priority Programme SPP 1177 `Galaxy
evolution' of the Deutsche Forschungsgemeinschaft under the project
SCHN 342/7--1.

\bibliographystyle{mn2e}
\bibliography{combobias}

\begin{landscape}
  \begin{table}
  \begin{center}
    \vspace{3cm}
    \caption{\label{tab:mcmc}\changed{Compilation of MCMC results for the halo
    model parameters (see Section \ref{sect:jhodmodel}), obtained from
    fitting three different model scenarios to the $\caln$-statistics,
    relative galaxy numbers and GGL-signal. The three scenarios are
    denoted by ``S'' (simple, no central galaxies), ``M'' (mixed, only
    red central galaxies) and ``C'' (central, red and blue central
    galaxies), respectively. The parameters of $m_{\rm min}$ and
    $m^\prime$ are in units of solar masses, $h^{-1}\msol$, the
    parameter $R$ represents the JHOD correlation factor between the
    red and blue sample. Quoted parameter values stand for median and
    $1\sigma$ confidence limits as derived from the MCMCs.  The last
    block of values at the bottom of the table contains constraints
    (mean and r.m.s. variance) from all redshifts combined. The
    fiducial cosmological model uses WMAP3 parameters.}}
  \begin{tabular}{c|cccccc|cccccc|c|c}
    &\multicolumn{6}{c|}{\textbf{RED
	SAMPLE}}&\multicolumn{6}{c|}{\textbf{BLUE
	SAMPLE}}&&\\
    $\bar{z}$&
    $[\log_{10}{m_{\rm min}}$&$\sigma_{{\rm log}m}$&$\log_{10}{m^\prime}$&$\epsilon$&$f$&$\lambda$]&
    $[\log_{10}{m_{\rm min}}$&$\sigma_{{\rm
	log}m}$&$\log_{10}{m^\prime}$&$\epsilon$&$f$&$\lambda$]&$R$&type
    \\\hline\hline&&&&&&&&&&&&&\\
    $0.3$&
    $12.1^{+0.5}_{-0.4}$&$0.5^{+0.3}_{-0.3}$&$13.9^{+1.5}_{-1.5}$&$1.0^{+0.5}_{-0.4}$&$1.7^{+0.9}_{-0.9}$&$4.3^{+2.9}_{-2.0}$&
    $11.0^{+0.3}_{-0.3}$&$0.5^{+0.3}_{-0.3}$&$13.6^{+1.6}_{-0.9}$&$1.1^{+0.5}_{-0.5}$&$1.5^{+1.0}_{-0.9}$&$4.5^{+2.6}_{-2.0}$&
    $+0.0^{+0.6}_{-0.6}$&S\\
    $0.3$&
    $11.9^{+0.3}_{-0.2}$&$0.5^{+0.3}_{-0.3}$&$12.6^{+1.6}_{-1.4}$&$0.8^{+0.5}_{-0.4}$&$1.3^{+1.0}_{-0.8}$&$5.0^{+3.1}_{-2.7}$&
    $10.9^{+0.2}_{-0.2}$&$0.5^{+0.3}_{-0.3}$&$13.0^{+1.7}_{-0.6}$&$1.1^{+0.4}_{-0.5}$&$0.6^{+0.8}_{-0.4}$&$5.5^{+2.8}_{-2.5}$&
    $+0.0^{+0.5}_{-0.5}$&M\\
    $*0.3$&
    $11.8^{+0.3}_{-0.3}$&$0.5^{+0.3}_{-0.3}$&$12.9^{+0.9}_{-0.8}$&$1.0^{+0.4}_{-0.3}$&$1.4^{+0.9}_{-0.9}$&$5.3^{+3.0}_{-2.8}$&
    $10.8^{+0.3}_{-0.2}$&$0.5^{+0.3}_{-0.3}$&$12.8^{+0.7}_{-0.5}$&$1.2^{+0.4}_{-0.3}$&$1.2^{+1.0}_{-0.8}$&$3.9^{+3.0}_{-2.0}$&
    $+0.3^{+0.5}_{-0.6}$&C
    \\&&&&&&&&&&&&&\\
    \hline&&&&&&&&&&&&&\\
    $0.5$&
    $12.2^{+0.2}_{-0.2}$&$0.5^{+0.3}_{-0.3}$&$13.9^{+1.3}_{-1.7}$&$0.7^{+0.6}_{-0.4}$&$1.9^{+0.7}_{-0.9}$&$5.5^{+2.6}_{-2.6}$&
    $11.3^{+0.2}_{-0.2}$&$0.5^{+0.3}_{-0.3}$&$13.1^{+1.8}_{-0.8}$&$1.0^{+0.4}_{-0.4}$&$0.9^{+0.8}_{-0.5}$&$4.9^{+3.0}_{-2.4}$&
    $+0.2^{+0.5}_{-0.5}$&S\\
    $0.5$&
    $12.0^{+0.2}_{-0.2}$&$0.5^{+0.3}_{-0.3}$&$12.3^{+1.5}_{-1.0}$&$0.6^{+0.3}_{-0.3}$&$1.3^{+0.9}_{-0.7}$&$5.2^{+2.9}_{-2.7}$&
    $11.2^{+0.2}_{-0.1}$&$0.5^{+0.3}_{-0.3}$&$12.7^{+1.0}_{-0.6}$&$0.9^{+0.4}_{-0.4}$&$0.5^{+0.4}_{-0.2}$&$6.0^{+2.4}_{-2.6}$&
    $+0.2^{+0.5}_{-0.5}$&M\\
    $0.5$&
    $12.0^{+0.2}_{-0.2}$&$0.5^{+0.3}_{-0.3}$&$12.6^{+1.0}_{-0.8}$&$0.7^{+0.3}_{-0.3}$&$1.4^{+0.9}_{-0.8}$&$5.2^{+2.9}_{-2.7}$&
    $11.2^{+0.2}_{-0.2}$&$0.5^{+0.3}_{-0.3}$&$12.6^{+0.7}_{-0.5}$&$1.0^{+0.3}_{-0.3}$&$0.4^{+0.5}_{-0.2}$&$4.7^{+3.2}_{-2.2}$&
    $+0.6^{+0.3}_{-0.5}$&C
    \\&&&&&&&&&&&&&\\
    \hline&&&&&&&&&&&&&\\
    $0.7$&
    $12.0^{+0.3}_{-0.2}$&$0.5^{+0.3}_{-0.3}$&$13.2^{+0.3}_{-0.3}$&$1.6^{+0.3}_{-0.3}$&$1.8^{+0.8}_{-0.8}$&$5.3^{+2.8}_{-2.7}$&
    $11.1^{+0.1}_{-0.1}$&$0.5^{+0.3}_{-0.3}$&$13.0^{+0.2}_{-0.3}$&$1.7^{+0.2}_{-0.3}$&$1.2^{+0.8}_{-0.5}$&$6.9^{+1.9}_{-2.1}$&
    $+0.1^{+0.5}_{-0.5}$&S\\
    $0.7$&
    $12.0^{+0.2}_{-0.2}$&$0.5^{+0.3}_{-0.3}$&$13.1^{+0.3}_{-0.4}$&$1.5^{+0.3}_{-0.3}$&$1.6^{+0.8}_{-0.8}$&$5.3^{+2.7}_{-2.8}$&
    $11.1^{+0.1}_{-0.1}$&$0.5^{+0.3}_{-0.3}$&$12.9^{+0.2}_{-0.2}$&$1.6^{+0.2}_{-0.2}$&$0.8^{+0.6}_{-0.4}$&$7.7^{+1.5}_{-1.9}$&
    $-0.2^{+0.5}_{-0.4}$&M\\
    $0.7$&
    $12.1^{+0.2}_{-0.2}$&$0.5^{+0.3}_{-0.3}$&$13.1^{+0.3}_{-0.4}$&$1.5^{+0.3}_{-0.2}$&$1.5^{+0.9}_{-0.8}$&$5.1^{+3.1}_{-3.1}$&
    $11.1^{+0.1}_{-0.1}$&$0.5^{+0.3}_{-0.3}$&$12.7^{+0.3}_{-0.3}$&$1.4^{+0.2}_{-0.2}$&$0.9^{+0.7}_{-0.5}$&$6.5^{+2.2}_{-2.3}$&
    $+0.3^{+0.5}_{-0.6}$&C
    \\&&&&&&&&&&&&&\\
    \hline&&&&&&&&&&&&&\\
    $0.9$&
    $12.5^{+0.3}_{-0.3}$&$0.5^{+0.3}_{-0.3}$&$14.5^{+1.0}_{-1.8}$&$0.9^{+0.6}_{-0.5}$&$1.7^{+0.8}_{-0.9}$&$4.5^{+3.0}_{-2.2}$&
    $11.5^{+0.2}_{-0.2}$&$0.5^{+0.3}_{-0.3}$&$14.5^{+1.0}_{-1.2}$&$1.2^{+0.5}_{-0.5}$&$1.2^{+1.0}_{-0.7}$&$4.5^{+2.3}_{-1.9}$&
    $-0.1^{+0.6}_{-0.5}$&S\\
    $0.9$&
    $12.5^{+0.3}_{-0.3}$&$0.5^{+0.3}_{-0.3}$&$14.4^{+1.1}_{-2.1}$&$0.9^{+0.6}_{-0.5}$&$1.5^{+0.9}_{-0.9}$&$4.4^{+3.3}_{-2.3}$&
    $11.4^{+0.2}_{-0.2}$&$0.5^{+0.3}_{-0.3}$&$14.4^{+1.0}_{-1.2}$&$1.1^{+0.5}_{-0.6}$&$0.6^{+0.7}_{-0.4}$&$5.7^{+2.3}_{-2.2}$&
    $+0.1^{+0.6}_{-0.6}$&M\\
    $0.9$&
    $12.5^{+0.3}_{-0.3}$&$0.4^{+0.3}_{-0.3}$&$14.2^{+1.2}_{-1.6}$&$0.8^{+0.6}_{-0.5}$&$1.1^{+1.2}_{-0.9}$&$4.7^{+2.9}_{-2.4}$&
    $11.7^{+0.2}_{-0.3}$&$0.5^{+0.3}_{-0.3}$&$14.5^{+1.1}_{-1.4}$&$1.2^{+0.5}_{-0.5}$&$1.0^{+1.1}_{-0.7}$&$3.9^{+2.1}_{-1.7}$&
    $+0.5^{+0.3}_{-0.6}$&C
    \\&&&&&&&&&&&&&\\
    \hline\hline&&&&&&&&&&&&&\\
    ALL&
    $12.2^{+0.2}_{-0.2}$&$0.5^{+0.2}_{-0.2}$&$13.4^{+0.7}_{-0.7}$&$1.3^{+0.3}_{-0.3}$&$1.9^{+0.5}_{-0.5}$&$4.3^{+1.6}_{-1.6}$&
    $11.2^{+0.1}_{-0.1}$&$0.5^{+0.2}_{-0.2}$&$13.0^{+0.2}_{-0.2}$&$1.4^{+0.2}_{-0.2}$&$1.0^{+0.4}_{-0.4}$&$5.3^{+1.4}_{-1.4}$&
    $+0.1^{+0.2}_{-0.2}$&S\\
    ALL&
    $12.1^{+0.1}_{-0.1}$&$0.5^{+0.2}_{-0.2}$&$12.9^{+0.4}_{-0.4}$&$1.1^{+0.2}_{-0.2}$&$1.3^{+0.5}_{-0.5}$&$4.4^{+1.9}_{-1.9}$&
    $11.2^{+0.1}_{-0.1}$&$0.5^{+0.2}_{-0.2}$&$12.9^{+0.2}_{-0.2}$&$1.4^{+0.2}_{-0.2}$&$0.5^{+0.2}_{-0.2}$&$7.0^{+1.3}_{-1.3}$&
    $+0.1^{+0.2}_{-0.2}$&M\\
    ALL&
    $12.1^{+0.1}_{-0.1}$&$0.5^{+0.2}_{-0.2}$&$13.1^{+0.3}_{-0.3}$&$1.1^{+0.2}_{-0.2}$&$1.4^{+0.5}_{-0.5}$&$5.1^{+1.4}_{-1.4}$&
    $11.1^{+0.1}_{-0.1}$&$0.5^{+0.2}_{-0.2}$&$12.8^{+0.2}_{-0.2}$&$1.3^{+0.2}_{-0.2}$&$0.6^{+0.3}_{-0.3}$&$4.8^{+1.2}_{-1.2}$&
    $+0.5^{+0.2}_{-0.2}$&C
  \end{tabular}
  \end{center}
\end{table}
\end{landscape}

\appendix

\section{Dark matter halo properties of model}
\label{sect:dmhaloes}
\changed{For $n(m)$, we use the semi-analytical peak-background split
prescription of \citet{1999MNRAS.308..119S}:}
\begin{equation}
  n(m)\propto \frac{1}{m^2}\frac{\sqrt{a}\nu}{1+[a\nu^2]^{-p}}{\rm
  e}^{-\frac{a\nu^2}{2}}\frac{\d\ln{\nu^2}}{\d\ln{m}}
\end{equation}
with \mbox{$\nu=\delta_{\rm c}/(D(z)\sigma(m))$},
$a=0.707$ and $p=0.3$. The expression 
\begin{equation}
  \sigma(m)=\frac{9}{2\pi^2}
  \int_0^\infty\d k k^2\,\frac{[\sin{(rk)}-rk\cos{(rk)}]^2}{(rk)^6}P_{\rm lin}(k)
\end{equation}
 denotes the top-hat variance of the linear dark matter fluctuations,
 within spheres of radius \mbox{$r=\left(\frac{3m}{4\pi\rho_{\rm
 crit}\Omega_{\rm m}}\right)^{1/3}$}. The function $D(z)$ denotes the
 linear growth factor, to be normalised to one for $z=0$
 \citep{peebles80}:
\begin{equation}
  D(z)\propto H(a(z))\int_0^{(1+z)^{-1}}\d a \frac{1}{[aH(a)]^3}\;,
\end{equation}
where $a(z)=(1+z)^{-1}$ is the cosmological scale factor and $H(a)$
the Hubble parameter. The constant $\delta_{\rm c}$, the over-density
of virialised haloes undergoing linear spherical collapse at $z=0$, is
chosen according to \citet{1997PThPh..97...49N}:
\begin{equation}
\delta_{\rm c}=\frac{3(12\pi)^{2/3}}{20}
\left[1+0.0123\log_{10}{\Omega_{\rm m}}\right]\;.
\end{equation}
For the cosmology adopted in this paper, one finds \mbox{$\delta_{\rm
c}=1.674$}.  \changed{For the linear dark matter power spectrum the
prescription of \citet{1998ApJ...496..605E} is employed.} It is
evolved to redshift $z$ using the linear growth factor $D(z)$.

\changed{The halo bias function we implemented into our model is from
\citet{2005ApJ...631...41T} which is a modification of the original
function of \citet{2001MNRAS.323....1S}}:
\begin{eqnarray}
  &&b(m)=1+\frac{1}{\sqrt{a}\delta_{\rm c}}\times\\\nonumber
  &&\left[\sqrt{a}(a\nu^2)+\sqrt{a}b(a\nu^2)^{1-c}-
  \frac{(a\nu^2)^c}{(a\nu^2)^c+b(1-c)(1-c/2)}\right]
\end{eqnarray}
with $a=0.707$, $b=0.35$, $c=0.8$ and $\nu$ as previously defined. To
guarantee that, when using this prescription for $b(m)$, the two-halo
term asymptotically fits the linear dark matter power spectrum $P_{\rm
lin}(k)$ on large-scales (small $k$) we artificially normalise all
$P^{\rm 2h}_{ij}(k)$ in Eq. \Ref{eq:twohalo} by \citep[see
e.g.][]{2005MNRAS.362.1451M}
\begin{equation}
  \bar{b}^2\equiv
  \left[\frac{1}{\overline{m}}
    \int\d m\,n(m)\,b(m)\right]^2\;.
\end{equation}
This normalisation is $\bar{b}=1.02,1.08$ for $z=0.0,1.0$.

The dark matter density profile of haloes is in our model a truncated
NFW \citep{1996ApJ...462..563N}:
\begin{equation}
  u_0(r,m)\propto \left\{
  \begin{array}{ll}
    \frac{1}{\frac{r}{r_{\rm s}}\left(1+\frac{r}{r_{\rm
    s}}\right)^2}&{\rm for~}r\le r_{\rm vir}\\
    0&{\rm otherwise}
  \end{array}\right.\;,
\end{equation}
with \mbox{$r_{\rm s}\equiv\frac{r_{\rm vir}}{c(m,z)}$} with
    \mbox{$r_{\rm vir}^3=\frac{3m}{4\pi\rho_{\rm crit}\Omega_{\rm
    m}\Delta_{\rm vir}(z)}$} and $c(m,z)$ being the so-called viral
    radius and concentration parameter, respectively. For the
    over-density within the virial radius we use
    \citep{2001MNRAS.321..559B}:
\begin{equation}
  \Delta_{\rm vir}(z)=
  \frac{18\pi^2+82(\Omega_{\rm m}(z)-1)-39(\Omega_{\rm m}(z)-1)^2}{\Omega_{\rm m}(z)}\;,
\end{equation}
$\rho_{\rm crit}$ (at redshift $z=0$) and $\Omega_{\rm m}(z)$ are the
 critical matter density and matter density parameter, respectively;
 \mbox{$\Omega_{\rm m}\equiv\Omega_{\rm m}(z=0)$}. The concentration
 parameter is calculated according to \citet{2000MNRAS.318..203S}:
\begin{equation}
  c(m,z)=\frac{10}{1+z}\left(\frac{m}{m_\ast}\right)^{-0.20}\;,
\end{equation}
where the present day non-linear mass scale, $m_\ast$, is defined by
\mbox{$\delta_{\rm c}=\sigma(m_\ast)$}. For our adopted cosmology we
find \mbox{$m_\ast=1.9\times10^{12}h^{-1}\msol$}.

\section{Exact statistical moments for central galaxy models}
\label{sect:moments}

As can be seen in Table \ref{kernels}, in a scenario where we split
the halo occupation number of one population (mixed scenario) or both
populations (pure central scenario) into one central galaxy and
satellite galaxies one has to specify $\ave{N_i^{\rm sat}N_j^{\rm
cen}}$, $\ave{N_i^{\rm sat}N_j^{\rm sat}}$ or $\ave{N_i^{\rm
sat}N_j}$, $\ave{N_i^{\rm cen}N_j}$ as function of halo mass $m$,
respectively. Here, we give general expressions in terms of the JHOD,
$P(N_i,N_j|m)$, of these moments. A detailed derivation for one of the
moments is given, the others can be obtained in a similar manner. Note
that we skip the arguments ``$(m)$'' and ``$|m$'' in the JHOD for the
derivation; all following equations are for haloes of a fixed mass.

The central galaxy-satellite splitting is done in such a way that we
always have $N^{\rm cen}=1$ if $N\geq1$, $N^{\rm cen}=0$ for $N=0$,
$N^{\rm sat}=N-1$ for $N>1$ and $N^{\rm sat}=0$ for $N\leq1$. This is
to say that (${\rm H}(x)$ is the Heaviside step function):
\begin{eqnarray}
  N_i^{\rm sat}&=&
  {\rm H}\left(N_i-1\right)\left(N_i-1\right)\;,\\
  N_i^{\rm cen}&=&
  {\rm H}\left(N_i-1\right)\;.
\end{eqnarray}

\changed{We focus on $\ave{N_i^{\rm sat}N_j^{\rm cen}}$. Let us substitute the
previous expressions and write down explicitly the ensemble average in
the statistical moment as sum over the JHOD (the states, halo
occupation numbers, are integers), taking into account the Heaviside
step functions:}
\begin{eqnarray}\nonumber
  &&\!\!\!\!\!\!\!\!\!\Ave{N_i^{\rm sat}N_j^{\rm cen}}
  =\sum_{N_i=1}^\infty\sum_{N_j=1}^\infty
  \left(N_i-1\right)P(N_i,N_j)\\\nonumber
  &=& \sum_{N_i=0}^\infty\sum_{N_j=0}^\infty
  \left(N_i-1\right)P(N_i,N_j)\\\nonumber
  &-&
  \sum_{N_i=0}^\infty\sum_{N_j=0}^0
  \left(N_i-1\right)P(N_i,N_j)\\\nonumber
  &-&
  \sum_{N_i=0}^0\sum_{N_j=0}^\infty
  \left(N_i-1\right)P(N_i,N_j)\\\nonumber
  &+&
  \sum_{N_i=0}^0\sum_{N_j=0}^0
  \left(N_i-1\right)P(N_i,N_j)\\\nonumber
  &=&\Ave{N_i}-1-\sum_{N_i=0}^\infty
  \left(N_i-1\right)P(N_i|N_j=0)\\\nonumber
  &+&
  \sum_{N_j=0}^\infty P(N_j|N_i=0)-\sum_{N_i,N_j=0}^0P(N_i,N_j)\\\nonumber
  &=&\Ave{N_i}-1-\Ave{N_i|N_j=0}+P(N_j=0)\\\nonumber
  &+&P(N_i=0)-P(N_i=0,N_j=0)\\
  &=&\Ave{N_i}-1-\Ave{N_i|N_j=0}+P(N_iN_j=0)\;,
\end{eqnarray}
where we denote by
\begin{equation}
  \Ave{N_i|N_j=0}\equiv
  \sum_{N_i=0}^\infty N_iP(N_i,N_j=0)
\end{equation}
the conditional halo-average of $N_i$ for which \mbox{$N_j=0$},
whereas \mbox{$P(N_iN_j=0)$} denotes the probability to find a halo
which has either $N_i=0$ \emph{or} $N_j=0$. The conditional mean is
zero, \changed{if all haloes have at least one $j$-galaxy.} We have
used in the foregoing calculus the relations (also with $i$ and $j$
interchanged)
\begin{eqnarray}
  P(N_i)&\!\!=\!\!&\sum_{N_j=0}^\infty\,P(N_i,N_j)\;,\\\nonumber
  P(N_iN_j\!=\!0)&\!\!=\!\!&P(N_i\!=\!0)+P(N_j\!=\!0)-P(N_i\!=\!0,N_j\!=\!0)
\end{eqnarray}
and
\begin{equation}
  \sum_{N_i=0}^0
  \sum_{N_j=0}^\infty\,N_i\,P(N_i,N_j)=0\;.
\end{equation}

Analogues to this little exercise in probability theory, we find:
\begin{eqnarray}
  &&\!\!\!\!\!\!\!\!\!\Ave{N_i^{\rm sat}N_j^{\rm
  sat}}=\\\nonumber
  &=&
  \left[\Ave{N_i}-1\right]\left[\Ave{N_j}-1\right]+
  R_{ij}\sigma(N_i)\sigma(N_j)\\\nonumber
  &+&\Ave{N_i|N_j=0}+\Ave{N_j|N_i=0}-P(N_iN_j=0)\;,
\end{eqnarray}
and for the mixed scenario:
\begin{equation}
  \Ave{N_i^{\rm cen}N_j}=
  \Ave{N_j}-\Ave{N_j|N_i=0}\;,
\end{equation}
and
\begin{eqnarray}
  &&\!\!\!\!\!\!\!\!\!\Ave{N_i^{\rm sat}N_j}=\\\nonumber
  &&
  \left[\Ave{N_i}-1\right]\Ave{N_j}
  +R_{ij}\sigma(N_i)\sigma(N_j)+\Ave{N_j|N_i=0}\;.
\end{eqnarray}

In conclusion, we find that the statistical moments for the
$2^{\rm nd}$-order clustering statistics are surprisingly complex in a
central galaxy scenario. Only in very simple cases, such as \emph{no
correlation} between distinct galaxy populations, i.e. for the pure
central scenario
\begin{eqnarray}
 \Ave{N_i^{\rm sat}N_j^{\rm sat}}&=&
 \Ave{N_i^{\rm sat}}\Ave{N_j^{\rm sat}}~{\rm and}\\
 \Ave{N_i^{\rm cen}N_j^{\rm sat}}&=&
 \Ave{N_i^{\rm cen}}\Ave{N_j^{\rm sat}}\;,
\end{eqnarray}
we find simple statistical cross-moments. Another trivial case is
given, if the haloes, for a fixed mass, always contain ``$i$''-
\emph{and} ``$j$''-galaxies: the conditional means and $P(N_iN_j=0)$
in the foregoing equations vanish. Simply providing the fluctuations
in the satellite number, the mean number of satellites and central
galaxies, and the JHOD correlation factor, $R_{ij}$, does not suffice
in general, though. We have to make additional assumptions about the
JHOD in order to work out all factorial moments of second order.

\section{Estimation of Bayesian evidence}
\label{sect:bayesian}

In applying Bayesian statistics to the problem of model fitting
\citep{2003Book...MACKAY} one assesses the posterior probability
function of the model parameters given the observed data, $\vec{d}$:
\begin{equation}
  P(\vec{p}|\vec{d})=
  \frac{P(\vec{d}|\vec{p})P(\vec{p})}{P(\vec{d})}\propto
  P(\vec{d}|\vec{p})P(\vec{p})\;;
\end{equation}
$P(\vec{p})$ is a prior on the model parameters, possibly obtained
from other observations, or motivated
theoretically. $P(\vec{d}|\vec{p})$ is the probability of the data
given a set of model parameters (likelihood function); this involves
essentially the specification of a PDF for the noise in the
observation. In this paper, we tackle the problem of estimating the
posterior likelihood, $P(\vec{p}|\vec{d})$, by using the MCMC
technique.

For the mere constraints in model parameter space the probability of
the observed data, $P(\vec{d})$, is unimportant because it is just
some constant. It can be used in statistics, however, to discriminate
between a set of models trying to explain the data simultaneously,
since $P(\vec{d})$ is actually the probability to obtain the data
\emph{under the assumption} that a certain model is correct and taking
into account the remaining uncertainties in $\vec{p}$. As PDFs are
normalised to one, one finds for a particular model that
\begin{equation}
  P(\vec{d})=\int\d\vec{p}\,P(\vec{d}|\vec{p})P(\vec{p})\;,
\end{equation}
where the integral is over the full parameter space. 

Evaluating this integral requires, compared to the posterior, its own
way of treatment \citep[e.g.][]{2007MNRAS.380.1029H}. We will not go
into that here. Instead, we are using the approximation mentioned in
\citet{2003Book...MACKAY}, by saying that the posterior likelihood
around the maximum-likelihood point can satisfyingly well described by
a multivariate Gaussian; the inverse covariance of the parameter
estimates -- the Hessian of the (log)posterior -- at this point is the
Fisher matrix \citep{2003Book...MACKAY}, $\mat{F}$, which can be
estimated from the MCMC tracks:
\begin{eqnarray}
  &&\ln{E}\equiv\ln{P(\vec{d})}\\\nonumber
  &&=-\frac{1}{2}
  \left[
    \ln{\det{(2\pi\mat{C})}}+\chi^2+\ln{\det{(\mat{F}/2\pi)}}-2\ln{P(\vec{p}_{\rm ml})}
    \right]\;,
\end{eqnarray}
where $P(\vec{p}_{\rm ml})$ is the prior likelihood of the
maximum-likelihood parameters $\vec{p}_{\rm ml}$ of the model,
$\mat{C}$ is the covariance of the data noise (Gaussian) and $\chi^2$
is evaluated at the maximum-likelihood point. By $\ln{E}$ we mean the
Bayesian evidence of the data for a given model.

A model ``A'' with larger $\ln{E_a}$ is more likely to reproduce the
observed data than a model ``B'' with smaller evidence
$\ln{E_b}$. Depending on the ratio of evidence probabilities, or
difference $\Delta\ln{E}=\ln{E_a}-\ln{E_b}$, we would favour model
``A'' over model ``B'' ($\Delta\ln{E}=1$ is a relative probability of
$37\%$). \citet{1961Book...JEFFREYS} suggests that the difference of
two models is ``substantial'' for \mbox{$1\le\Delta\ln{E}<2.5$},
``strong'' for \mbox{$2.5\le\Delta\ln{E}<5$} and ``decisive''
otherwise.

\bsp

\label{lastpage}

\end{document}